%% file: main.tex
\pdfoutput=1

\documentclass[12pt, a4paper]{article}
\usepackage[margin=2.5cm]{geometry}
\usepackage{graphicx}
\usepackage{amsmath}
\usepackage{amssymb}
\usepackage{theorem}
\usepackage{natbib}
\usepackage[]{algorithm}
\usepackage{algpseudocode}
\usepackage{subfig}
\usepackage{bm}
\usepackage{color}
\usepackage{url}

\title{Bayesian inference for stochastic differential equation mixed effects models of a tumor xenography study}

\author{Umberto Picchini${}^{a,b}$, \bigskip Julie Lyng Forman${}^{c}$
  \\
  ${}^{a}${\small Department of Mathematical Sciences},\\
  {\small Chalmers University of Technology and the University of Gothenburg}\\ 
  {\small Email:} {\small {\tt picchini@chalmers.se}}  \\
  ${}^{b}${\small Centre for Mathematical Sciences},
  {\small Lund University}\\  
  ${}^{c}${\small Section of Biostatistics, Department of Public Health},
  {\small University of Copenhagen}\\ 
  {\small Email:} {\small {\tt jufo@biostat.ku.dk}}  \\  
  }

\date{}

\begin{document}
\maketitle

\begin{abstract}
We consider Bayesian inference for stochastic differential equation mixed effects models (SDEMEMs) exemplifying tumor response to treatment and regrowth in mice. We produce an extensive study on how a SDEMEM can be fitted using both exact inference based on pseudo-marginal MCMC and approximate inference via Bayesian synthetic likelihoods (BSL). We investigate a two-compartments SDEMEM, these corresponding to the fractions of tumor cells killed by and survived to a treatment, respectively. Case study data considers a tumor xenography study with two treatment groups and one control, each containing 5-8 mice. Results from the case study and from simulations indicate that the SDEMEM is able to reproduce the observed growth patterns and that BSL is a robust tool for inference in SDEMEMs.
Finally, we compare the fit of the SDEMEM to a similar ordinary differential equation model. Due to small sample sizes, strong prior information is needed to identify all model parameters in the SDEMEM and it cannot be determined which of the two models is the better in terms of predicting tumor growth curves. In a simulation study we find that with a sample of 17 mice per group BSL is able to identify all model parameters and distinguish treatment groups.
\end{abstract}

\section{Introduction}
Pre-clinical cancer trials aim at understanding the dynamics of tumor
growth and evaluate the effect of treatments such as radio- and
chemotherapies in delaying this. A typical trial involves repeated
measurements of the volume of solid tumors grown in mice. Tumors are
grown until a critical size is reached, in case of which the mouse must
be sacrificed for ethical reasons, or until a planned end of study.
Data from these trials pose a statistical challenge due to 
the missing data caused by the sacrifice and the substantial variation in
growth patterns between subjects. Even within the same treatment
group, it occurs that some tumors are eliminated following treatment,
others continue to grow unaffected, and yet others display a
decrease in volume followed by regrowth \citep{laajala2012improved}.

\cite{heitjan1993statistical} review traditional approaches to
analysing tumor xenography experiments, including ANOVA, MANOVA, and
linear mixed models for tumor volumes. An overall drawback of
the linear models is that inference targets the mean log-volume of the tumor, which
is problematic as all large volumes are censored due to sacrifice. 
It has been suggested to avoid bias by instead comparing
time to sacrifice, tumor doubling time, or a similar survival outcome, see
\cite{stuschke1990methods}, \cite{wu2009assessing} and
\cite{wu2011confidence}. However, this limits statistical efficiency as the the full information in the data is not used. Moreover, delays and doubling times may in practice be hard to
measure accurately due to day-to-day perturbations in growth, measurement
error, and discrete time follow-up. 
Rank-based comparisons of composite tumor volume/time to sacrifice 
outcomes is a more robust and powerful approach \citep{peron2016}. However, this does not offer much insight on the tumor growth dynamics and the effect of the treatment. Also, a fully specified parametric model would be required for power calculations and optimal design.

\cite{demidenko2013mixed} reviews non-linear mixed models for tumor
growth and re-growth following treatment. In this paper we will focus on the double exponential model
which identifies two latent compartments corresponding to the fraction of the
tumor which is killed by the treatment and the one that survives. This
model offers a phenomenological explanation for the variation in
individual tumor growth patterns, by recognizing a) the proportion of
the tumor killed by the treatment, b) the rate of elimination of the dead
tumor cells, and c) the growth rate of the surviving part of the tumor.    
Clinically relevant quantities such as tumor
doubling times, tumor growth delay and surviving fraction of tumor
cells can be deduced from the double exponential model (\citealp{demidenko2006assessment}, \citealp{demidenko2010three}).
More recent non-linear mixed model approaches specify the individual growth curves
semi-parametrically \citep{xia2013model}, or as splines (\citealp{kong2011modeling}, \citealp{zhao2011bayesian}). These models allow for much flexibility in
individual growth curves but do not share the biological interpretation
of the double exponential or delayed double exponential models.

A drawback of the double exponential, and other classical non-linear
mixed models, is that the only source of intra-subject variation is given by
independent identically distributed measurement errors.
This may not be realistic, as growth rates are subject to day-to-day variation,
due to biological processes not easily accounted for, e.g. mutations in the cancer and the response of the immune system. 
In recent years, a number of works have promoted the use of stochastic differential equation mixed effects models (SDEMEMs) as a more realistic alternative to the classical
nonlinear mixed  models. For instance, \cite{donnet2010bayesian} find that a
stochastic differential equation (SDE) version of the Gompertz growth model
is superior to its nonlinear deterministic mixed model counterpart, for prediction of the body weight of growing chicken. \cite{donnet2013review} report similar findings from pharmacokinetic experiments and \cite{whitaker2015bayesian} provide more recent references to inference for SDEMEMs. 
What prevents more widespread application of SDEMEMs is that inference for nonlinear SDE models is overall challenging, even when not considering random effects and measurement errors, because SDEs generally have intractable likelihood functions \citep{fuchs2013inference}. In case measurement error occurs, the observed process is no longer Markovian and a state-space type model must be considered \citep{cappe2006inference}. This paper considers inference for  state-space models with latent dynamics given by an SDE, which is performed using state-of-the-art likelihood-based methods employing sequential Monte Carlo (SMC) filters. More specifically, one of our attempts to estimate model parameters applies a pseudo-marginal method (PMM, \citealp{andrieu2009pseudo}) which returns exact Bayesian inference, despite employing SMC approximations. We further consider a methodology which is able to target more general models beyond the state space class, namely a Bayesian version of the synthetic likelihoods (SL) approach \citep{price2016bayesian}. SL was initially proposed in \cite{wood2010statistical} and does not impose any assumption on the complexity of the model, the only requirement being the ability to simulate artificial datasets from the model. However, SL returns approximate inference, unlike PMM. To the best of our knowledge
this is the first application of synthetic likelihoods to SDEMEMs.
%and construct summary statistics for the generated datasets. 
We show that Bayesian SL (BSL) applied to SDEMEMs produces results qualitatively similar to the exact Bayesian methodology, both with experimental data and with data produced in a simulation study. Further, we find that BSL is easier to calibrate than PMM and returns results that are less sensitive to calibration setup (see supplementary material). Also, BSL is able to demonstrate a difference in treatment efficacy in a simulation study with seventeen subjects in each treatment group, while PMM is not.
Importantly, when considering applications to experimental data, we find that having only 5-8 subjects per treatment group is not sufficient to identify all model parameters accurately. Nevertheless, Bayesian inference still offers an opportunity to perform exploratory data analyses in small scale experiments, since informative priors based on subject-matter expertise may compensate for otherwise too small sample sizes.

Although SDEMEMs are complex models with three layers of randomness, we conclude that BSL should be considered an additional tool for inference in this class of models. Moreover, in future studies BSL would allow us to investigate models beyond the state-space type,  where SMC methods might not be applicable.

The structure of the paper is as follows: Section \ref{s:models}
contrasts the classical nonlinear mixed models with
the stochastic differential equation mixed effects models (SDEMEMs).
Section \ref{sec:inference-sdemem} explains the pseudo-marginal method
for exact Bayesian inference in SDEMEMs. Section \ref{sec:synlik} considers inference using synthetic likelihoods.  In section
\ref{sec:case-study} we analyze data from a tumor xenography study. In section \ref{s:odefit} results from exact Bayesian inference for ordinary differential equations mixed-effects models are given. In section \ref{sec:simulation-study} we run two simulation studies on artificial data. Data and software to run our experiments are available at \url{https://github.com/umbertopicchini/sdemem-tumor}. Further results and methodological considerations are available as supplementary material.

\section{Mixed effects models of tumor growth}
\label{s:models}

\subsection{Ordinary mixed effects models}\label{sec:ordinary-models}
Denote with $M$ the number of subjects in a given treatment group. Assume that tumor volumes from subject $i$ are measured at discrete time points 
$t_{i0}<\ldots<t_{in_i}$, $i=1,\ldots,M$. In a planned experiment, such as the one
considered in section \ref{sec:case-study}, time points will usually be
the same for all subjects, i.e.\ $t_{ij}=t_j$, while the number of observations
$n_i$ may differ between subjects as the mice get sacrificed when their tumor volume exceeded a critical size prescribed by the ethical guidelines. Denote with $V_{ij}=V_i(t_j)$ the exact tumor volume for subject $i$
at time $t_j$, $j=1,...,n_i$. We model the observations as

\begin{equation}
\label{e:obsmodel}
Y_{ij} = \log(V_{ij})+\varepsilon_{ij},\qquad i=1,...,M;j=1,...,n_i
\end{equation}
where the $\varepsilon_{ij}$'s are i.i.d.\ normally distributed measurements errors with
$\varepsilon_{ij}\sim\mathcal{N}(0,\sigma_\varepsilon^2)$. 
This means that we assume tumor volumes to be measured with
multiplicative log-normal measurement errors. In experimental practice the length and width of the tumor is measured on the skin surface and the volume is approximated by that of an ellipsoid, resulting in a measurement accuracy that is typically within $\pm 20\%$ of the true volume.
  
In regulated experiments the mice are sacrificed long before the tumor
volumes reach steady state. Hence, unperturbed growth in the control
group is adequately described by a simple exponential growth
model. Let $\beta_1,\ldots,\beta_M$ denote the random subject-specific growth
rates, then the growth curves for the control group are given by
\begin{equation}
\label{e:ode0}
\frac{dV_i(t)}{dt}=\beta_iV_i(t),\quad\quad V_{i}(0)=v_{i,0},\qquad i=1,...,M.
\end{equation}
Distributional assumptions for the $\beta_i$'s are in section \ref{sec:sde-model}.
Of course, equation \eqref{e:ode0} is solved explicitly by
$V_i(t)=v_{i,0}e^{\beta_i t}$. Note that
with the further assumption that growth rates are normally distributed
and initial tumor volumes log-normally distributed across the
population, the observation model (\ref{e:obsmodel}) is merely a
standard linear mixed model with a random intercept and a random
slope.

If tumor volumes are observed post treatment, then the double
exponential model in \cite{demidenko2013mixed} describes the total volume
in terms of surviving tumor cells $V^{\textnormal{surv}}$ and cells killed by the treatment $V^{\textnormal{kill}}$ as
\begin{equation}
\begin{cases}
V_i(t)&=V^{\textnormal{surv}}_i(t)+V^{\textnormal{kill}}_i(t),\\
\label{e:ode1}
\frac{dV^{\textnormal{surv}}_i(t)}{dt}&=\beta_iV^{\textnormal{surv}}_i(t),
\quad V_{i}^{\textnormal{surv}}(0)=(1-\alpha_i)v_{i,0},\qquad i=1,...,M\\
\frac{dV^{\textnormal{kill}}_{i}(t)}{dt}&=-\delta_iV^{\textnormal{kill}}_{i}(t),
\quad V_{i}^{\textnormal{kill}}(0)=\alpha_i v_{i,0}.
\end{cases}
\end{equation}
Here $\alpha_i\in [0,1]$ denotes the proportion of the tumor that has
been killed by the treatment in subject $i$, while $\delta_i$ denotes the elimination rate for the dead tumor cells in subject $i$. 
Equation (\ref{e:ode1}) has the explicit solution $V_i(t)=(1-\alpha_i)v_{i,0}e^{\beta_i t}+\alpha_iv_{i,0}e^{-\delta_i t}$. Distributional assumptions for $\delta_i$ and $\alpha_i$ are in section \ref{sec:sde-model}.

\subsection{Stochastic differential equation mixed effects model}\label{sec:sde-model}
The assumption of time constant growth and elimination rates in the
ordinary mixed models is usually not realistic, since growth is
affected by various biological processes that are not easily accounted
for (see \citealp{donnet2010bayesian} for a motivating example).
We therefore suggest to replace the ordinary differential equation model specified by (\ref{e:ode0}) with a SDE model such as the geometric Brownian motion,
\begin{equation}
\label{e:sde0}
dV_{i}(t)=(\beta_i+\gamma^2/2)V_{i}(t)dt + \gamma V_{i}(t) dB_{i}(t),
\quad V_{i,0}=v_{i,0}.
\end{equation}
Here the $\{B_{i,t}\}_{t\geq 0}$'s are independent standard Brownian
motions and $\gamma^2$ denotes the intra-subject growth rate
variance. This means that the instantaneous growth rate is not exactly $\beta_i$ but deviates from this by a random normal perturbation. The motivation for including the term $\gamma^2/2$ in the
drift of the SDE is that the individual growth process is then given
by $V_{i}(t)=v_{i,0}e^{\beta_i t + \gamma B_i(t)}$ which is a log-normally
distributed stochastic process 
with geometric mean $v_{i,0}e^{\beta_i t}$, which coincides
with the ordinary exponential growth model
(\ref{e:ode0}). With the further assumption that growth rates are distributed as
$\beta_i\sim\mathcal{N}(\bar{\beta}_0,\sigma_\beta^2)$ and initial tumor volumes as
$\log(v_{i,0})\sim\mathcal{N}(\bar{v}_0,\sigma_0^2)$ across the
population, volumes at time $t_{ij}$ would follow
a log-normal distribution with geometric mean
$\bar{v}_0e^{\bar{\beta}_0t_j}$ which is the same as in the ordinary log-linear mixed model. 
However, in our case studies we assume $v_{i0}$ to be fixed known mathematical constants, as detailed in section \ref{sec:results-pmm}.

The ordinary double exponential model (\ref{e:ode1}) can similarly be
replaced by a SDEMEM with the following specification 
\begin{equation}
\begin{cases}
Y_{ij} &= \log(V_{ij})+\varepsilon_{ij},\qquad i=1,...,M;\quad j=1,...,n_i\\
V_i(t)&=V^{\textnormal{surv}}_i(t)+V^{\textnormal{kill}}_i(t),\\
\label{e:sde1}
dV^{\textnormal{surv}}_{i}(t)&=(\beta_i+\gamma^2/2)V^{\textnormal{surv}}_{i}(t)dt
+\gamma V^{\textnormal{surv}}_{i}(t)dB_{i}(t),
\quad V^{\textnormal{surv}}_{i}(0)=(1-\alpha_i)v_{i,0}\\
dV^{\textnormal{kill}}_{i}(t)&=(-\delta_i+\tau^2/2)V^{\textnormal{kill}}_{i}(t)dt
+\tau V^{\textnormal{kill}}_{i}(t)dW_{i}(t),
\quad V^{\textnormal{kill}}_{i}(0)=\alpha_i v_{i,0}
\end{cases}
\end{equation}
with random effects $\beta_i\sim\mathcal{N}(\bar{\beta},\sigma^2_\beta)$, $\delta_i\sim\mathcal{N}(\bar{\delta},\sigma^2_\delta)$ and $\alpha_i\sim\mathcal{N}_{[0,1]}(\bar{\alpha},\sigma^2_\alpha)$ where here and in the following $\mathcal{N}_{[0,1]}$ denotes a Gaussian distribution truncated to the interval [0,1] (for example, this means that $\alpha_i\sim \mathcal{N}(\bar{\alpha},\sigma^2_\alpha|0\leq \alpha_i\leq 1)$).
The $\{W_{i}(t)\}_{t\geq 0}$'s are additional standard Brownian
motions assumed mutually independent and independent of the $\{B_{i}(t)\}_{t\geq 0}$'s, of the $\varepsilon_{ij}$, of the (fixed) system initial conditions and of the random effects. Here
$\tau^2$ denotes the intra-subject elimination rate variance. The SDEs in \eqref{e:sde1} have explicit solutions given by $V^{\textnormal{surv}}_{i}(t) = V^{\textnormal{surv}}_{i}(0)e^{\beta_i t+\gamma B_i(t)}$ and $V^{\textnormal{kill}}_{i}(t) = V^{\textnormal{kill}}_{i}(0)e^{-\delta_it+\tau W_i(t)}$ respectively. It is easy to simulate paths from \eqref{e:sde1} due to the independent normal increments of the Brownian motions. 
Please note that, while processes $\log V^{\textnormal{surv}}_i(t)$ and $\log V^{\textnormal{kill}}_i(t)$ both have Gaussian transition densities, instead $\log V_i(t)$ is not Gaussian distributed, and this prevents an analytic expression for the likelihood function to be found (the integrals in equation \eqref{eq:likelihood} cannot be solved analytically). We choose a truncated Gaussian distribution for the individual
treatment effect $\alpha_i$, as this assigns strictly
positive probabilities densities to the values zero and one. This is to anticipate that an effective treatment could have the
effect that tumors are completely eliminated, while an inefficient
treatment might not kill any tumor cells.

We stress that the models considered above by no means are the only possibilities for specifying dynamical models for tumor growth. Additional random effects could be added to the subject specific growth curves, e.g.\ to describe a delay in the treatment effect, a different diffusion term could replace $\gamma V_i(t)dB_i(t)$, or
a so-called stochastic growth rate model (SGRM) could be specified by $V_i(t)=\exp\{\int_0^t\beta_i(u)du\}$, where the time-varying growth rate $\{\beta_i(t)\}_{t\geq 0}$ could be modelled by e.g. an Ornstein-Uhlenbeck process
%$d\beta_i(t)=-\rho_i(\beta_i(t)-\bar{\beta}_i)dt+\gamma_idB_{i}(t)$ 
(a continous time autoregressive process evolving around its mean $\bar{\beta}_i$). 
Note that, since the integrated diffusion process in the SGRM is not a Markov process, the more general methodology from section \ref{sec:synlik} would be needed to analyze this type of model.

\section{Likelihood-based inference for SDEMEMs}
\label{sec:inference-sdemem}
In this section we discuss likelihood inference for SDEMEMs such as
model (\ref{e:sde0}) and (\ref{e:sde1}) and generalizations
thereof. Both models can be viewed as instances of the general state-space SDEMEM

\begin{equation}
\begin{cases}
\bm{Y}_{ij} &= \bm{g}(\bm{X}_{i}(t_{ij}),\bm{\varepsilon}_{ij}), \qquad \bm{\varepsilon}_{ij}\sim_{i.i.d.} \mathcal{N}(\bm{0},\sigma^2_{\varepsilon}\mathrm{I}_{d_y})\\
d\bm{X}_{i}(t) &= \bm{\mu}(\bm{X}_{it},t,\bm{\phi}_i)dt+\bm{\sigma}(\bm{X}_{it},t,\bm{\kappa})d\bm{B}_{i}(t), \qquad \bm{X}_{i}(t_{0})\sim \pi_0(\bm{x}_{i}(t_{0})|\bm{\phi}_i)\\
\bm{\phi}_i &\sim_{i.i.d.} p(\bm{\phi}_i|\bm{\eta}).\\
\end{cases}
\label{eq:sdemem-statespace}
\end{equation}
where each $\bm{Y}_{ij}$ has dimension $\dim(\bm{Y}_{ij})=d_y$, $\dim(\bm{\varepsilon}_{ij})=d_y$, $\mathrm{I}_{d_y}$ is the $d_y\times d_y$ identity matrix, and $\bm{X}_{i}(t)$ has dimension $\dim(\bm{X}_{i}(t))=d_x$, with $d_x\geq d_y$ at every $t$. 
Model \eqref{eq:sdemem-statespace} has the following interpretation: for each subject
$i$, $\{\bm{X}_{i}(t)\}_{t\geq 0}$ represents the hidden (unobservable) biological process of interest, with dynamics governed by the drift and diffusion functions $\bm{\mu}(\cdot)$ and $\bm{\sigma}(\cdot)$ which
are assumed known, save from the the subject specific parameters (random effects) $\bm{\phi}_i$ and the common model parameters $\bm{\kappa}$. 
In case of model \eqref{e:sde1}, the latent process is $\bm{X}_i(t)=(V_i^{\textnormal{kill}},V_i^{\textnormal{surv}})$, $\bm{\kappa}=(\gamma,\tau)$, $\bm{\phi}_i=(\log\alpha_i,\log\beta_i,\log\delta_i)$, $\bm{\eta}=(\bar{\alpha},\bar{\beta},\bar{\delta},\sigma_\alpha,\sigma_\beta,\sigma_\delta)$ and $\bm{X}_i(t_0)=(x_{i,0}^{\mathrm{surv}},x_{i,0}^{\mathrm{kill}})$ with $\bm{x}_{i,0}^{\mathrm{surv}}=(1-\alpha_i)v_{i,0}$ and $x_{i,0}^{\mathrm{kill}}=\alpha_iv_{i,0}$. Regularity conditions for the existence and uniqueness of a solution to the stochastic differential equation can be found e.g.\ in \cite{fuchs2013inference}.
Observations $\{\bm{Y}_{ij}\}$ are assumed to consist of discrete time measurements of the latent process $\{\bm{X}_{i}(t)\}$, perturbed with measurement error $\bm{\varepsilon}_{ij}$ via a known function $\bm{g}(\cdot)$. 
E.g.\ model \eqref{e:sde1} is specified with $\bm{g}(\bm{v},\bm{\varepsilon})=\log(v^{\textnormal{kill}}+v^{\textnormal{surv}})+\varepsilon$. 
Finally, the subject specific random effects $\bm{\phi}_i$ are assumed distributed with a density $p(\cdot|\bm{\eta})$ parametrized by the ``population parameter'' $\bm{\eta}$. 
The aim of our analysis is to perform inference for the vector parameter $\bm{\theta}=(\bm{\eta},\bm{\kappa},\sigma_{\varepsilon})$.
A main feature of the SDEMEM  \eqref{eq:sdemem-statespace} is its ability to discriminate between the temporal intra-subject variability ($\bm{\kappa}$), the inter-subjects variability ($\mathrm{Var}(\bm{\phi_i})$), and the measurement error variance ($\sigma_{\varepsilon}^2$). Knowledge of these distinct sources of variation will be valuable when planning experiments and performing power calculations.

It is important to notice that measurements $\bm{Y}_{ij}$ are conditionally independent given the latent states $\bm{X}_{ij}:=\bm{X}_{i}(t_j)$ and $\bm{\phi}_i$, implying that
\eqref{eq:sdemem-statespace} is a state-space model \citep{cappe2006inference}. This, as well as the Markov property of the latent process $\{\bm{X}_i(t)\}$, is essential for the inference methods described in this section to work. However, these  are not required properties for the methodology in section \ref{sec:synlik}. 

Denote with $\bm{y}_i=\{\bm{y}_{ij}\}_{j=1,...,n_i}$ the collection of observations for subject $i$ and with $\bm{X}_i=\{\bm{X}_{ij}\}_{j=1,...,n_i}$ the corresponding values
of the latent process. Let $\bm{y}=(\bm{y}_1,...,\bm{y}_M)\in\mathcal{Y}$ denote the full set of measurements for all subjects in a certain experimental group. Standard methods for frequentist as well as Bayesian estimation
of the model parameters $\bm{\theta}$
require the evaluation of the likelihood function
$p(\bm{y}|\bm{\theta})=\prod_{i=1}^M p(\bm{y}_i|\bm{\theta})$. The hidden Markov structure
implies the following derivation
\begin{eqnarray}
\nonumber
p(\bm{y}_i|\bm{\theta})&=&\int p(\bm{y}_i|\bm{\phi}_i;\bm{\theta})p(\bm{\phi}_i|\bm{\theta})d\bm{\phi}_i\\
\label{eq:likelihood}
&=&\int\biggl(\int p(\bm{y}_i|\bm{X}_i;\bm{\theta})p(\bm{X}_i|\bm{\phi}_i;\bm{\theta})d\bm{X}_i\ \biggr) p(\bm{\phi}_i|\bm{\theta})d\bm{\phi}_i\\
\nonumber
&=&\int\biggl(\int\biggl\{\prod_{j=1}^{n_i}p(\bm{y}_{ij}|\bm{X}_{ij},\bm{\theta})p(\bm{X}_{i,j}|\bm{X}_{i,j-1},\bm{\phi}_i;\bm{\theta})\biggr\} p(\bm{X}_{i0}|\bm{\phi}_i,\bm{\theta})d\bm{X}_i\biggr) p(\bm{\phi}_i|\bm{\theta})d\bm{\phi}_i.
\end{eqnarray}
Note that the term $p(\bm{X}_{i0}|\bm{\phi}_i,\bm{\theta})$ vanishes in either of the
cases where $\bm{X}_{i0}$ is included among the random effects or is assumed
to be a known constant, $\bm{X}_{i0}:=\bm{x}_{i0}$.

Function (\ref{eq:likelihood}) is not
analytically tractable. Thus, inference for SDEMEMs relies on either
more specific model assumptions, such as the latent processes being
Gaussian, or the use of computationally intensive 
methods. \cite{delattre2013coupling} show how to conduct likelihood
inference in SDEMEMs using the stochastic approximate EM
algorithm (SAEM) coupled with an extended Kalman
filter. \cite{donnet2014using} propose a particle MCMC
algorithm to perform the S-step in SAEM. In either case the use of
SAEM requires explicit specification of sufficient summary statistics
for the augmented likelihood $p(\bm{y}_i,\bm{X}_i,\bm{\phi}_i|\bm{\theta})$. While
providing fast and accurate inference in models with a latent Gaussian
structure, the derivation of the summary statistics is a tedious if
not impossible task for more complex models of realistic interest. See \cite{picchini2016likelihood} for a likelihood-free version of SAEM. 

Bayesian inference targets the parameter posterior distribution
$\pi(\bm{\theta}|\bm{y})\propto p(\bm{y}|\bm{\theta})\pi(\bm{\theta})$ where $\pi(\bm{\theta})$ is
the corresponding prior distribution. Bayesian methodology for SDEMEMs was first
studied by \cite{donnet2010bayesian} who implemented a Gibbs sampler
that applies to the case where the SDE has an explicit solution, and
which can be extended to the more general state-space model by using an
Euler-Maruyama discretization. A recent review of Bayesian inference
methods for SDEMEMs can be found in \cite{whitaker2015bayesian}. It is
important to notice that MCMC algorithms can be constructed to sample from the exact posterior of $\bm{\theta}$, for models admitting a non-negative unbiased estimator of $p(\bm{y}|\bm{\theta})$ (\citealp{beaumont2003estimation}, \citealp{andrieu2009pseudo}). In this spirit, we exemplify a \textit{pseudo-marginal} method
(PMM, \citealp{andrieu2009pseudo}) using sequential Monte Carlo (SMC). The key idea is to substitute the intractable
$p(\bm{y}|\bm{\theta})$ with an unbiased non-negative estimate $\hat{p}(\bm{y}|\bm{\theta})$, and plug this
in an otherwise standard Metropolis-Hastings algorithm (see algorithm
\ref{alg:MCMC}).
\begin{algorithm}
\scriptsize
\caption{A pseudo-marginal MCMC algorithm}
\begin{algorithmic}
\State 1.  \textbf{Input:} a positive integer $R$. Fix a starting value $\bm{\theta}^*$ or generate it from its prior $\pi(\bm{\theta})$ and set $\bm{\theta}_1:=\bm{\theta}^*$. Set a kernel $q(\bm{\theta}'|\bm{\theta})$. Use algorithm \ref{alg:smc} or the APF to obtain an unbiased estimate $\hat{p}(\bm{y}|\bm{\theta}^*)$ of $p(\bm{y}|\bm{\theta}^*)$. Set $r=1$.
\State \textbf{Output:} $R$ correlated draws from $\pi(\bm{\theta}|\bm{y})$ (possibly after a burnin).
\State 2. Generate a $\bm{\theta}^{\#}\sim q(\bm{\theta}^{\#}|\bm{\theta}^*)$. Use algorithm \ref{alg:smc} or the APF to obtain an unbiased estimate $\hat{p}(\bm{y}|\bm{\theta}^\#)$ of $p(\bm{y}|\bm{\theta}^\#)$. 
\State 3. Generate a uniform random draw $u\sim U(0,1)$, and calculate the acceptance probability 
\begin{eqnarray*}
\alpha=\min\biggl[1,\frac{\hat{p}(\bm{y}|\bm{\theta}^\#)}{\hat{p}(\bm{y}|\bm{\theta}^*)}
\times  \frac{q(\bm{\theta}^*|\bm{\theta}^{\#})}{q(\bm{\theta}^{\#}|\bm{\theta}^{*})} \times \frac{\pi(\bm{\theta}^{\#})}{\pi(\bm{\theta}^*)} \biggr]. 
\end{eqnarray*}
If $u>\alpha$, set $\bm{\theta}_{r+1}:=\bm{\theta}_{r}$ otherwise set $\bm{\theta}_{r+1}:=\bm{\theta}^{\#}$, $\bm{\theta}^*:=\bm{\theta}^{\#}$ and $\hat{p}(\bm{y}|\bm{\theta}^*):=\hat{p}(\bm{y}|\bm{\theta}^\#)$. Set $r:=r+1$ and go to step 4. 
\State 4. If $r\leq R$ repeat steps 2--3 otherwise stop.
\end{algorithmic}
\label{alg:MCMC}
\end{algorithm} 
An unbiased estimate of the likelihood function can be obtained using SMC filters, of which two popular examples are the bootstrap filter (BF, \citealp{gordon1993novel}), adapted in algorithm \ref{alg:smc} below, and the auxiliary particle filter (APF, \citealp{pitt1999filtering}, \citealp{pitt2012some}), with a version suitable for our case studies detailed in the supplementary material. Here we describe the BF, as it is more approachable for a general audience and sufficient to convey the methodology. The interested reader is referred to the supplementary material which contains useful notes on how to implement either a BF or an APF for model \eqref{e:sde1}, as well as a comparison between BF and APF. Note that in most cases of practical interest, the forward propagation step in both BF and APF requires a numerical scheme, such as Euler-Maruyama (see for example \citealp{golightly2011bayesian}), though this is not the case with model \eqref{e:sde1} as the analytic solutions for $V_i^{\mathrm{surv}}(t)$ and $V_i^{\mathrm{kill}}(t)$ are known (section \ref{sec:sde-model}).  
The approximated likelihood is
\begin{equation}
\hat{p}(\bm{y}|\bm{\theta}) = \prod_{i=1}^M \hat{p}(\bm{y}_i|\bm{\theta}),\label{eq:unbiased-all}
\end{equation}
and when using BF we have
\begin{equation}
\hat{p}(\bm{y}_i|\bm{\theta}) = \hat{p}(\bm{y}_{i1}|\bm{\theta})\prod_{j=2}^{n_i} \hat{p}(\bm{y}_{ij}|\bm{y}_{i,1:j-1},\bm{\theta})=\prod_{j=1}^{n_i}\biggl(\frac{1}{L}\sum_{l=1}^L w_{ij}^l\biggr ),
\label{eq:unbiased-individual}
\end{equation}
where $L$ is the number of particles used to propagate the latent state forward and the $w_{ij}^l$'s are importance weights. Strategies to tune the value of $L$ can be found in \cite{doucet2015efficient} and \cite{sherlock2015efficiency}. In the context of our case study, results using different values of $L$ are compared in the supplementary material. We performed the resampling step using the stratified method of
\cite{kitagawa1996monte}. Also, note that in sections \ref{sec:case-study}--\ref{sec:simulation-study} we used a
Gaussian kernel $q(\cdot|\cdot)$ to propose parameters via the
adaptive Gaussian random walk algorithm of \cite{haario2001adaptive}.

\begin{algorithm}
\scriptsize
\caption{SMC bootstrap filter (BF) for mixed-effects state-space models}
\begin{algorithmic}
\State \textbf{Input:} a positive integer $L$, a starting value for $\bm{\theta}$
and a starting value $\bm{x}_0$. Set time $t_{0}=0$ and corresponding starting states
$\bm{X}_{i0}=\bm{x}_{i0}$. We use the convention that all steps involving the index $l$ must be
performed for all $l\in\{1,\ldots,L\}$. \\

\textbf{Output:} all the $\hat{p}(\bm{y}_{ij}|\bm{y}_{i,1:j-1})$, $i=1,...,M$; $j=1,...,n_i$.
\For{$i=1,...,M$}
\State draw $\bm{\phi}_i^l\sim p(\bm{\phi}_i|\bm{\theta})$
  \If{$j=1$}
  \State Sample $\bm{x}_{i1}^l  \sim p(\bm{x}_{i1}|\bm{x}_{i0},\bm{\phi}_i^l;\bm{\theta})$.
   \State Compute  $w_{i1}^l =p(\bm{y}_{i1}|\bm{x}_{i1}^l)$
      and $\hat{p}(\bm{y}_{i1})=\sum_{l=1}^L w_{i1}^l/L$.
      \State Normalization:  $\tilde{w}_{i1}^l:={w}_{i1}^l/\sum_{l=1}^L {w}_{i1}^l$. Interpret $\tilde{w}_{i1}^l$ as a probability associated to $\bm{x}_{i1}^l$.
\State Resampling: sample $L$ times with replacement from the probability distribution $\{\bm{x}_{i1}^l,\tilde{w}_{i1}^l\}$. Denote the sampled particles with $\tilde{\bm{x}}_{i1}^l$.
  \EndIf
 \For{$j=2,...,n_i$}

\State Forward propagation: sample $\bm{x}_{ij}^l\sim p(\bm{x}_{ij}|\tilde{\bm{x}}_{i,j-1}^l,\bm{\phi}_i^l;\bm{\theta})$.
\State Compute $w_{ij}^l=p(\bm{y}_{ij}|\bm{x}_{ij}^l)$ and normalise $\tilde{w}_{ij}^l:={w}_{ij}^l/\sum_{l=1}^L {w}_{ij}^l$
\State Compute $\hat{p}(\bm{y}_{ij}|\bm{y}_{i,1:j-1})=\sum_{l=1}^L w_{ij}^l/L$
\State Resample $L$ times with replacement from $\{\bm{x}_{ij}^l,\tilde{w}_{ij}^l\}$. Sampled particles are $\tilde{\bm{x}}_{ij}^l$. 
\EndFor
\EndFor
\end{algorithmic}
\label{alg:smc}
\end{algorithm}

The main distinction between model \eqref{eq:sdemem-statespace} and other state-space models
is that latent states are subject specific and can be further
decomposed into a time-dependent component $\bm{X}_i$ and a time-independent
component $\bm{\phi}_i$. Therefore, when applying SMC we first draw $\bm{\phi}_i$ and then, conditionally on such draw, we propagate forward particles corresponding to the states
$\bm{X}_i$. As proven by \cite{del2004genealogical} and \cite{pitt2012some}, each
individual estimate \eqref{eq:unbiased-individual} produced by
BF and APF is unbiased (where the expectation is taken with respect to the
distribution used to generate all the random variables employed in the SMC approximation). Since measurements from different subjects are assumed independent, it follows that $\hat{p}(\bm{y}|\bm{\theta})$ in \eqref{eq:unbiased-all} is an unbiased
estimator for ${p}(\bm{y}|\bm{\theta})$. Due to this, parameters drawn according to algorithm \ref{alg:MCMC} have  stationary distribution $\pi(\bm{\theta}|\bm{y})$ (after a burn-in), \textit{for any number of particles} $L$ (\citealp{beaumont2003estimation}, \citealp{andrieu2009pseudo}).

\section{Approximate inference for SDEMEMs using synthetic likelihoods}\label{sec:synlik}

In this section we discuss approximate Bayesian inference for SDEMEMs using synthetic likelihoods \citep{wood2010statistical}. Similarly to approximate Bayesian computation (ABC, see \citealp{marin2012approximate} for a review), the synthetic likelihoods methodology is a black-box approach solely relying on simulations from the assumed data-generating model. It is therefore a tool suitable for models having an otherwise intractable likelihood. It is important to notice that both ABC and SL do
not require the model to have a state-space representation. 
Compared to exact methods, the drawbacks of the approximate methodologies is the loss of statistical efficiency and a need to validate
their performance on a case to case basis. 

Similarly to ABC, SL relies on a set of
carefully selected summary statistics for the data $\bm{s}:=\bm{s}(\bm{y})$. However, while in ABC no assumption is made for the distribution of $\bm{s}$, SL assumes
that summary statistics follow a multivariate normal distribution, $\bm{s} \sim
\mathcal{N}(\bm{\mu}(\bm{\theta}),\bm{\Sigma}(\bm{\theta}))$ (see \citealp{fasiolo2016extended} and \citealp{an2018robust} on relaxing this assumption). If this holds true, and if parameters in $\bm{\theta}$ can be identified from $\bm{\mu}(\bm{\theta})$
and  $\bm{\Sigma}(\bm{\theta})$, then inference for $\bm{\theta}$ can be based on the
Gaussian likelihood of $\bm{s}$ instead of the intractable likelihood of
$\bm{y}$. Most often
$\bm{\mu}(\bm{\theta})$ and $\bm{\Sigma}(\bm{\theta})$ are unknown functions which have to be approximated with simulations. Hence synthetic likelihoods can be viewed as an instance of the simulated method of moments \citep{mcfadden1989smm}. 

The implementation of SL is straightforward. For a given $\bm{\theta}$, $N$
synthetic datasets $\bm{y}^{*1},...,\bm{y}^{*N}$ are generated independently from the model, hence each entry in the vector $\bm{y}^{*n}$ belongs to the space $\mathcal{Y}$ (see the notation in section \ref{sec:inference-sdemem}) and $\dim(\bm{y}^{*n})=\dim(\bm{y})$, $n=1,...,N$. Summary statistics
$\bm{s}^{*n}=\bm{s}(\bm{y}^{* n})$ are computed for each simulated dataset and from these
we obtain the estimates:

\[
\hat{\bm{\mu}}_N(\bm{\theta}) = \frac{1}{N}\sum_{n=1}^{N}\bm{s}^{*n},\qquad  
\hat{\bm{\Sigma}}_N(\bm{\theta}) = \frac{1}{N-1}\sum_{n=1}^{N}(\bm{s}^{*n}-\hat{\bm{\mu}}_N(\bm{\theta}))(\bm{s}^{*n}-\hat{\bm{\mu}}_N(\bm{\theta}))'.
\]
%The only tuning parameter is $N$.
When applying SL to SDEMEMs it is important that summary statistics reflect
the hierarchical structure of the model. That is, in order to anticipate the intra- and
inter-individual variation of the experimental data we construct 
subject-specific summaries $\bm{s}_i:=\bm{s}^{\mathrm{intra}}(\bm{y}_i)$ for $i=1,\ldots,M$, as well as summaries that represent inter-individuals variation between all subjects $\bm{s}^{\mathrm{inter}}:=\bm{s}^{\mathrm{inter}}(\bm{y}_1,...,\bm{y}_M)$,   
so that $\bm{s}=(\bm{s}_1,...,\bm{s}_M,\bm{s}^{\mathrm{inter}})$. The summaries used in our case study are described in section \ref{sec:results-synlik}.

Approximate normality for summary statistics can often be argued theoretically, by
appealing to the central limit theorem (CLT). If the sample size is small or the summary
statistics do not admit a CLT, then the normal assumption would have
to be verified empirically using simulations, see \cite{wood2010statistical} for details.

Here we follow \cite{price2016bayesian}, who proposed a fully Bayesian approach, henceforth referred to as Bayesian SL (BSL). A BSL procedure samples from the exact posterior $\pi(\bm{\theta}|\bm{s})$ without incurring any bias caused by a finite $N$ (note that ``exact'' sampling is ensured only if the distribution of $\bm{s}$ is really Gaussian). The key feature exploits the idea underlying the pseudo-marginal
method discussed in section \ref{sec:inference-sdemem}, where an
unbiased estimator is used in place of the unknown likelihood function.
\cite{price2016bayesian} noted that plugging-in the estimates
$\hat{\bm{\mu}}_N(\bm{\theta}) $ and $\hat{\bm{\Sigma}}_N(\bm{\theta}) $ into the Gaussian
likelihood  $p(\bm{s}|\bm{\theta})$ in general results in a biased estimator $p_N(\bm{s}|\bm{\theta})$ of $p(\bm{s}|\bm{\theta})$. However, this can be avoided
by instead adopting the unbiased estimator of \cite{ghurye1969unbiased}: 

\begin{align}
\hat{p}(\bm{s}|\bm{\theta}) &= (2\pi)^{-d/2}\frac{c(d,N-2)}{c(d,N-1)(1-1/N)^{d/2}}|(N-1)\hat{\bm{\Sigma}}_N(\bm{\theta})|^{-(n-d-2)/2}\nonumber\\
& \times \biggl\{\psi\bigl((N-1)\hat{\bm{\Sigma}}_N(\bm{\theta}) - (\bm{s}-\hat{\bm{\mu}}_N(\bm{\theta}))(\bm{s}-\hat{\bm{\mu}}_N(\bm{\theta}))'/(1-1/N) \bigr)\biggr\}^{(N-d-3)/2}.
\label{eq:synlik}
\end{align}
Here $\pi$ denotes the mathematical constant (not the prior), $d=\dim(\bm{s})$, $N$ is assumed to satisfy $N>d+3$, and for a square matrix $\bm{A}$ the function $\psi(\bm{A})$ is defined as $\psi(\bm{A})=|\bm{A}|$ if $\bm{A}$ is positive definite and $\psi(\bm{A})=0$ otherwise, where $|\bm{A}|$ is the determinant of $\bm{A}$. Finally $c(k,v)=2^{-kv/2}\pi^{-k(k-1)/4}/\prod_{i=1}^k\Gamma(\frac{1}{2}(v-i+1))$.

To produce inference for SDEMEMs with BSL we can use Algorithm \ref{alg:synlikMCMC} below, in analogy with algorithm \ref{alg:MCMC} in section \ref{sec:inference-sdemem}. 
The former merely uses SL to draw from the posterior $\pi(\bm{\theta}|\bm{s})$ instead of
$\pi(\bm{\theta}|\bm{y})$, assuming that $\bm{s}$ follows a Gaussian distribution. 
\begin{algorithm}
\scriptsize
\caption{Bayesian synthetic likelihoods (BSL)}
\begin{algorithmic}
\State   \textbf{Input:} a positive integer $R$. The observed summary
statistics $\bm{s}$. Fix a starting value $\bm{\theta}^*$ or generate it from the
prior $\pi(\bm{\theta})$. Set $\bm{\theta}_1=\bm{\theta}^*$. Choose a kernel $q(\bm{\theta}'|\bm{\theta})$. Set $r=1$. \\
\textbf{Output:} $R$ correlated samples from $\pi(\bm{\theta}|\bm{s})$. 
\State 1. Conditionally on $\bm{\theta}^*$ generate independently $N$
summaries $\bm{s}^{* 1},...,\bm{s}^{* N}$, compute moments $\hat{\bm{\mu}}_N(\bm{\theta}^*)$, $\hat{\bm{\Sigma}}_N(\bm{\theta}^*)$  and
$\hat{p}(\bm{s}|\bm{\theta}^*)$ from \eqref{eq:synlik}.
\State 2. Generate a $\bm{\theta}^{\#}\sim
q(\bm{\theta}^{\#}|\bm{\theta}^*)$. Conditionally on $\bm{\theta}^\#$ generate
independently $\bm{s}^{\# 1},...,\bm{s}^{\# N}$, compute $\hat{\bm{\mu}}_N(\bm{\theta}^\#)$,
$\hat{\bm{\Sigma}}_N(\bm{\theta}^\#)$  and $\hat{p}(\bm{s}|\bm{\theta}^\#)$. 
\State 3. Generate a uniform random draw $u\sim U(0,1)$, and calculate
the acceptance probability 
\begin{eqnarray*}
\alpha=\min\biggl[1,\frac{\hat{p}(\bm{s}|\bm{\theta}^\#)}{\hat{p}(\bm{s}|\bm{\theta}^*)}
\times  \frac{q(\bm{\theta}^*|\theta^{\#})}{q(\bm{\theta}^{\#}|\bm{\theta}^{*})} \times \frac{\pi(\bm{\theta}^{\#})}{\pi(\bm{\theta}^*)} \biggr]. 
\end{eqnarray*}
If $u>\alpha$, set $\bm{\theta}_{r+1}:=\bm{\theta}_{r}$ otherwise set $\bm{\theta}_{r+1}:=\bm{\theta}^{\#}$, $\bm{\theta}^*:=\bm{\theta}^{\#}$ and $\hat{p}(\bm{s}|\bm{\theta}^*):=\hat{p}(\bm{s}|\bm{\theta}^\#)$. Set $r:=r+1$ and go to step 4. 
\State 4. Repeat steps 2--3 as long as $r\leq R$.
\end{algorithmic}
\label{alg:synlikMCMC}
\end{algorithm} 
In our case studies, we use
algorithm \ref{alg:synlikMCMC} to estimate the model parameters of the
SDEMEMs (\ref{e:sde0}) and (\ref{e:sde1}). To the best of our
knowledge, this is the first application of the synthetic likelihood
methodology to SDEMEMs. 
%The methodology has also been tested with success in the simulation study in section \ref{sec:simulation-study}. 
We refer to section \ref{sec:results-synlik} and the supplementary material for further considerations on the implementation of BSL. 

Note that one of the main conclusion we get from analyzing experimental data  
(section \ref{sec:case-study}) is that the sample sizes are too small to obtain accurate inference for all model parameters. Most importantly, it is difficult to identify the mean treatment efficacy $\bar{\alpha}$ between different treatment groups both with BSL and PMM. Interestingly, our simulation study (section \ref{sec:simulation-study}) suggests that with a moderately larger sample size, BSL is able to identify $\bar{\alpha}$ while PMM is not.

\section{Case study}
\label{sec:case-study}

We consider data from a tumor xenography study originally including four
treatment groups (1: chemo therapy, 2: radiation therapy, 3:
combination therapy I, 4: combination therapy II), and one untreated
control (group 5). Each group consists of 7-8 mice. Mice were followed
up on Mondays, Wednesdays, and Fridays for six consecutive weeks or
until their tumor volume exceeded 1,000 mm$^3$, in which case the mouse
was sacrificed as prescribed by the Danish legislation for the use of animals in scientific research. In groups 2
and 4 about half of the mice were sacrificed within the treatment
period or shortly after, and due to the reduced sample sizes these will not be considered any further.

\begin{figure}[ht]
\centering
\includegraphics[height=4cm,width=6.5cm]{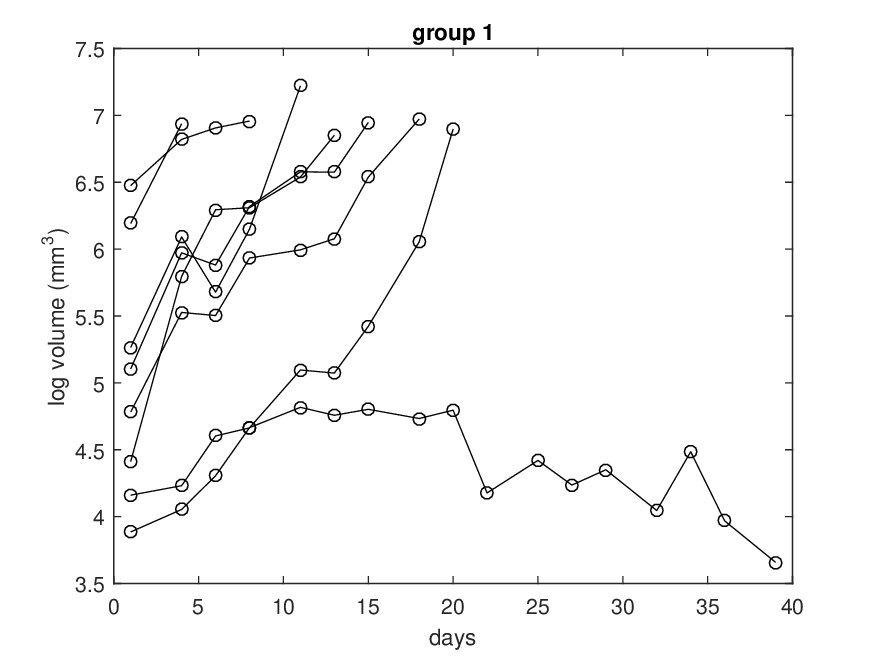}\quad
\includegraphics[height=4cm,width=6.5cm]{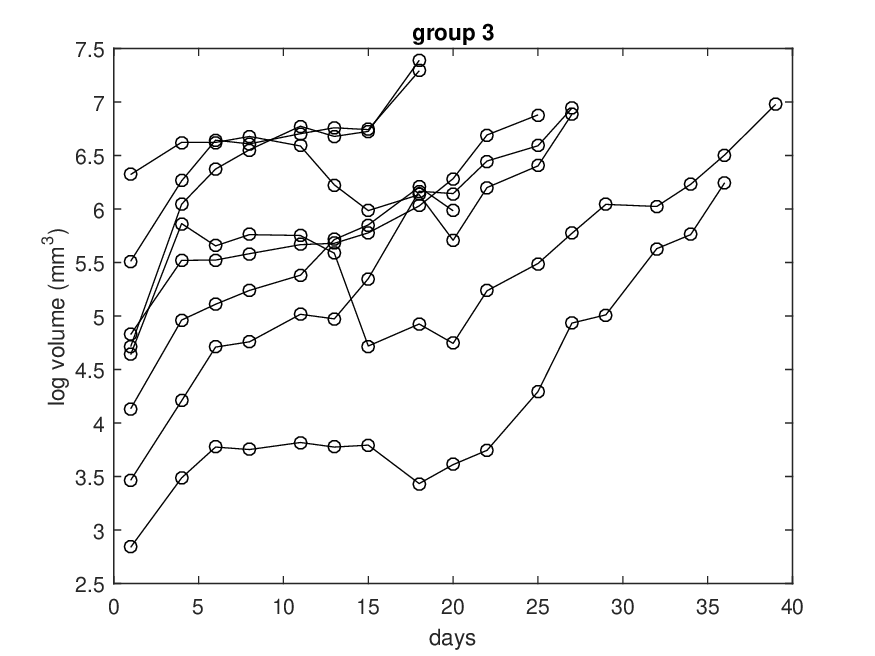}\\
\includegraphics[height=4cm,width=6.5cm]{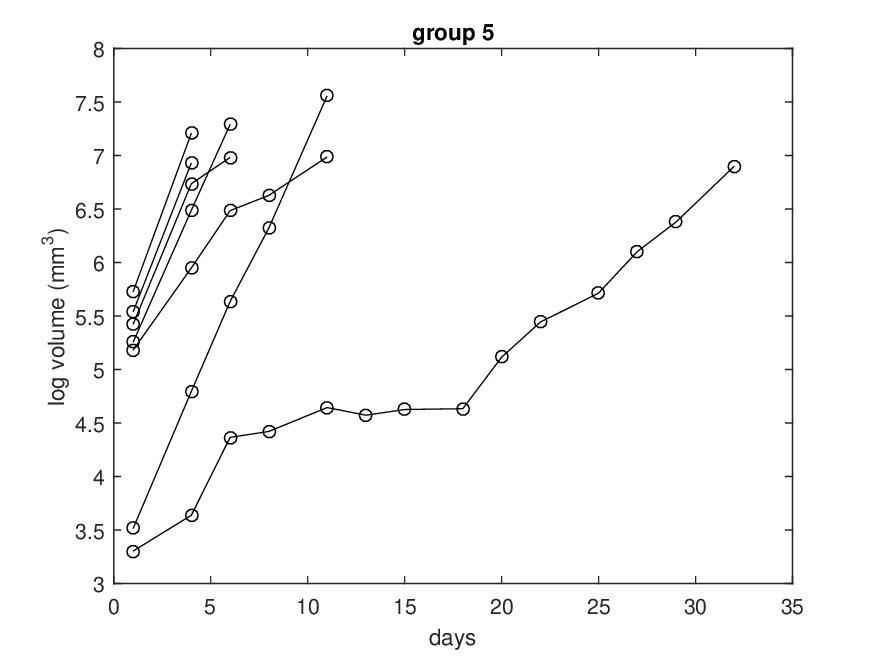}
\caption{\footnotesize{Data of log-volumes ($mm^3$) for three treatment groups.}}
\label{fig:logdata}
\end{figure}

Treatment in
equal size doses was applied on days 1, 4, and 6 of the study. Afterwards
no treatment was administered. The repeated measurements of tumor
volumes in the three remaining groups are shown in Figure \ref{fig:logdata}. It is obvious that growth patterns vary substantially between
subjects. In the untreated control group a single mouse with a slowly
growing tumor survived for 32 days before sacrifice while all other
untreated mice were sacrificed within 10 days. In the active treatment
groups we see patterns of decay followed by regrowth which match the
characteristic shape of the double exponential curve. In the same
groups we also see tumors that appear to grow continuously, unaffected
by the treatment. An outlying mouse in group 1 appears to present a slowly
vanishing tumor. Most likely the implanted tumor cells never grew to form a tumor in this mouse in the first place, hence its data was excluded from the analyses. 
Several mice display tumor volumes that are stable
over shorter durations of time. These stable periods deviate from the
growth patterns of the ordinary simple and double exponential mixed
models, but can be explained by the random variations in growth and
decay rates which are modeled in the double exponential SDEMEM.  
We applied the double exponential SDEMEM (\ref{e:sde1}) to model the post-treatment
log-volumes, i.e. starting from (and including) day 6. Separate
model fits were obtained for treatment groups 1 and 3.  Tumor growth in the untreated controls (group 5) was modeled with the simple exponential SDEMEM (\ref{e:sde0})
starting from day 1 of the study.

For Bayesian analysis of the double exponential SDEMEM
(\ref{e:sde1}) we choose a truncated Gaussian prior on the average
treatment effect $\bar{\alpha}\sim \mathcal{N}_{[0,1]}(0.6,0.2^2)$. Note that this assigns strictly positive probabilities to the values zero and one. This is to anticipate that an effective treatment could have the effect that tumors are completely eliminated, while an inefficient treatment might not kill any tumor cells at all. For the remaining parameters in (\ref{e:sde1}) we choose priors $\log\bar{\beta}\sim
\mathcal{N}(0.7,0.6^2)$, $\log\bar{\delta}\sim
\mathcal{N}(0.7,0.6^2)$, $\sigma_{\beta}\sim \mathrm{InvGam}(4,2)$,
$\sigma_{\delta}\sim \mathrm{InvGam}(4,2)$, $\sigma_{\alpha}\sim
\mathrm{InvGam}(5,1.5)$, $\gamma\sim \mathrm{InvGam}(5,7)$, $\tau\sim \mathrm{InvGam}(5,7)$, and
$\sigma_{\varepsilon}\sim \mathrm{InvGam}(2,1)$, where $\mathrm{InvGam}(a,b)$ denotes the inverse-Gamma distribution with shape $a$ and scale $b$. Note that positive model
parameters have been reparametrised by their logarithms. We refer to the supplementary material for BSL results obtained using less informative priors. To enhance numerical stability 
%when simulating the SDEs solutions, prior to performing inference for a given group of subjects we scale 
the observational times were scaled as $t_{ij}=t_{ij}/t_{\textnormal{max}}\in[0,1]$, since $t_\textnormal{max}=$39 days was the maximum time of follow-up after which the remaining mice were sacrificed.
Parameter estimates should be interpreted accordingly. 
Software and data are available at \url{https://github.com/umbertopicchini/sdemem-tumor}.

\subsection{Results using exact Bayesian inference}\label{sec:results-pmm}
We fitted model \eqref{e:sde1} separately for groups 1 and 3 using the pseudo-marginal (PMM) algorithm, as described in Section
\ref{sec:inference-sdemem}. Unbiased estimates of the likelihood function were obtained via the auxiliary particle filter (APF). For each subject, $v_{i0}=y_{i1}$ was considered a known constant. Recall, however, that in model \eqref{e:sde1} initial states
$v_{i0}^{\textnormal{surv}}$ and $v_{i0}^{\textnormal{kill}}$ depend also on $\alpha_i$. Algorithm \ref{alg:MCMC} was initialized at $\log
\bar{\beta}=1.6$, $\log\bar{\delta}=1.6$,  $\log\bar{\alpha}=-0.36$,
$\log\gamma=0$, $\log\tau=0$,   $\log\sigma_{\beta}=-0.7$,
$\log\sigma_{\delta}=-0.7$,   $\log\sigma_{\alpha}=-2.3$,
$\log\sigma_{\varepsilon}=0$. We used $L=2,000$ particles and $L_2=5$ (the number of particles propagated from each of the $L$ particles to compute first stage weights, see the supplementary material). Chains of length $R=20,000$ were produced and the computation took about 53 minutes for treatment group 1 ($M=5$), and 167 minutes for treatment group 3 ($M=8$) with a \textsc{Matlab} code running on a Intel Core i7-4790 3.60 GHz. For
both groups, average acceptance rates observed during the execution of
algorithm \ref{alg:MCMC} were equal to 30\%.
The chains convergence were verified using the scale reduction factor $\hat{R}$ \citep{gelman1992inference} as implemented in R's \texttt{coda} package \citep{coda}. We considered three chains initialized at very dispersed values compared to the marginal posteriors. All values for $\hat{R}$ were below 1.1 except for $\sigma_\alpha$ ($\hat{R}=1.2$), hence the chains appear to have converged.

Results are shown in Table \ref{tab:realdata-estimates} (the initial 10,000 draws were discarded as burn-in).
The treatment efficacy is estimated at $\bar{\alpha}=60\%$ in group 3
and at $\bar{\alpha}=52\%$ in group 1, however the corresponding posteriors are very wide in both groups. Thus it is not possible to draw conclusions on differences in treatment efficacy between the two groups (posterior marginal of the difference of the two efficacies not shown). In section \ref{sec:larger-simulated-data} we show that having larger sample sizes enables a much better identification of the treatments efficacy $\bar{\alpha}$ if inference is made with BSL.
Note that posteriors for $\log\bar{\beta}$,
$\gamma$, $\tau$ and $\sigma_\varepsilon$ are informative when compared to their priors. Also, the estimate for $\bar{\beta}$ is higher in group 1 than in group 3, as it should be by looking at Figure \ref{fig:logdata}
(recall that in group 1 the decaying growth curve was excluded prior to analysis).
It is reassuring that the measurement error variance is estimated
consistently as $\hat{\sigma}_{\varepsilon}\simeq 0.1-0.2$ in all
groups. This means that tumor volumes were measured with a relative
accuracy approximately within $\pm 20\%$, which is realistic for an
experiment of this type.
On the other hand, we found the marginal posterior for $\sigma_\alpha$ to be
highly sensitive to  the choice of its prior; that is the posterior distribution
followed the shape of the prior, regardless of the choice of hyperparameters.

The one-compartment model \eqref{e:sde0} was fitted to the untreated
controls (group 5). The priors were the same as for
the corresponding parameters in the two-compartments model \eqref{e:sde1}. 
Parameter estimates are shown in Table \ref{tab:realdata-estimates}.
Estimates of the mean population growth rate $\bar{\beta}$ are higher than for groups 1--3, and the diffusion coefficient $\gamma$ is also higher than the corresponding stochasticities for the two-compartment models, but credibility intervals are wide. The measurement error standard deviation $\sigma_\varepsilon$ is compatible with the previous model fits, which is reassuring.

\begin{figure}[ht]
\centering
\includegraphics[height=3cm,width=6cm]{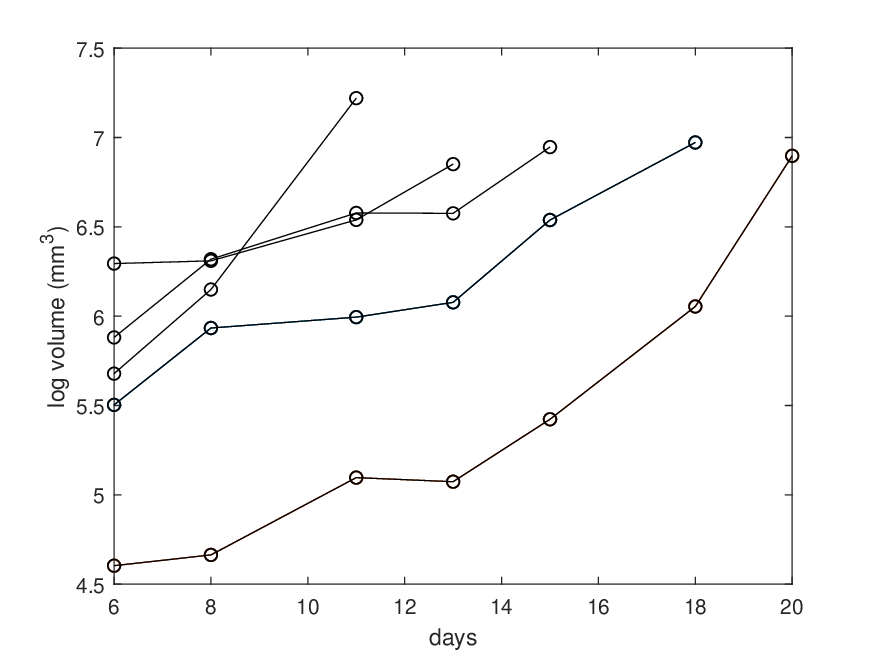}\quad
\includegraphics[height=3cm,width=6cm]{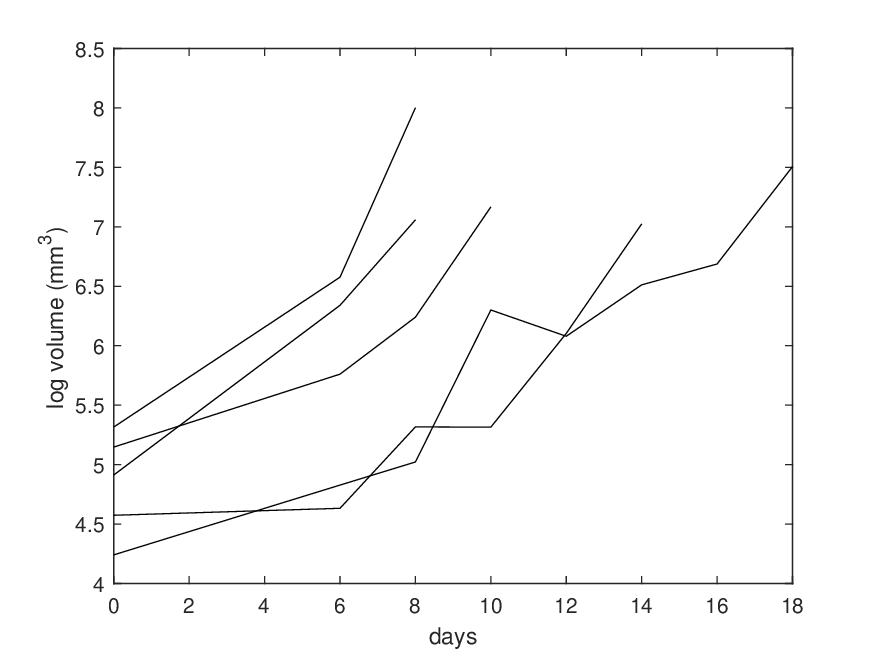}\\
\includegraphics[height=3cm,width=6cm]{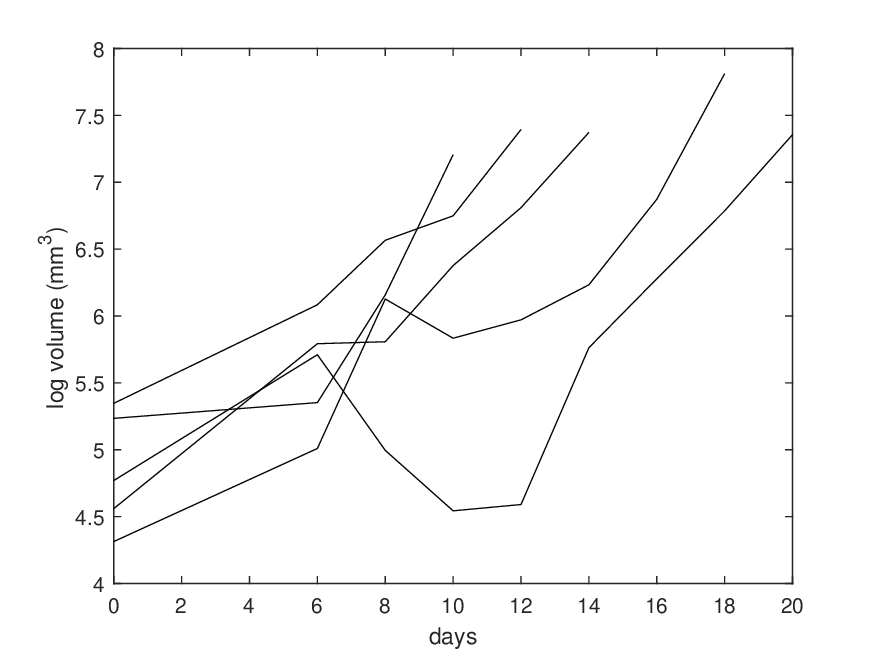}\quad
\includegraphics[height=3cm,width=6cm]{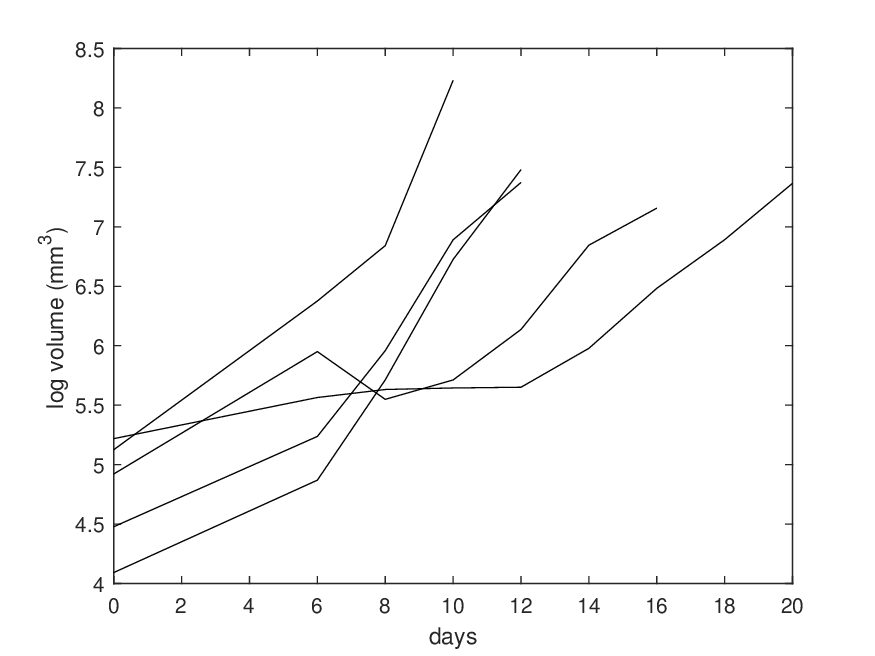}\\
\caption{\footnotesize{Fitted data in group 1 (top left) and three realizations from model \eqref{e:sde1} estimated with exact Bayesian methodology (remaining plots).  Top left panel does not report data for one excluded mouse. Recall for this group measurements at days 1 and 4 were disregarded during estimation, hence times on abscissas start at day 6.}}
\label{fig:group1-simdata}
\end{figure}

\begin{figure}[ht]
\centering
\includegraphics[height=3cm,width=6cm]{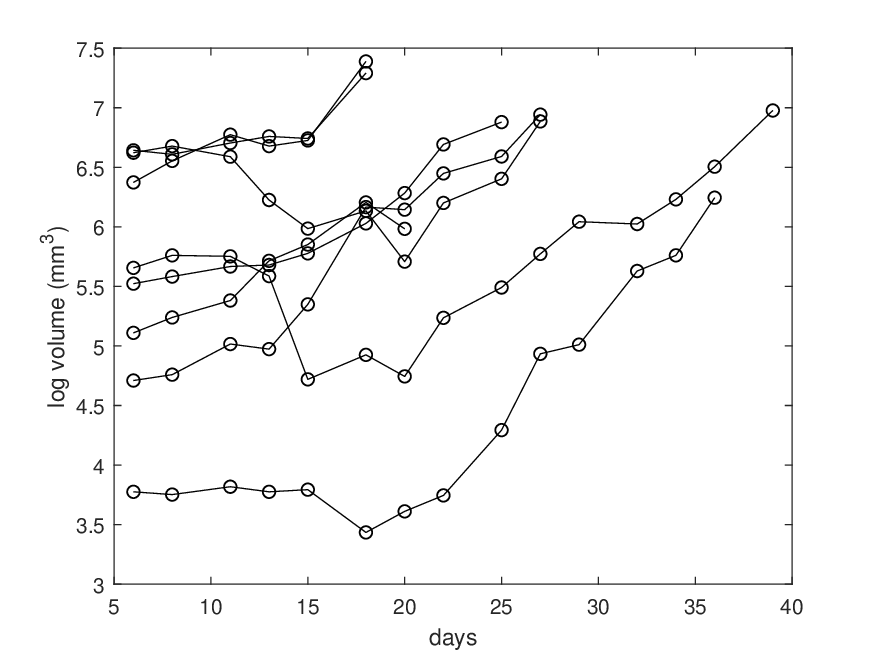}\quad
\includegraphics[height=3cm,width=6cm]{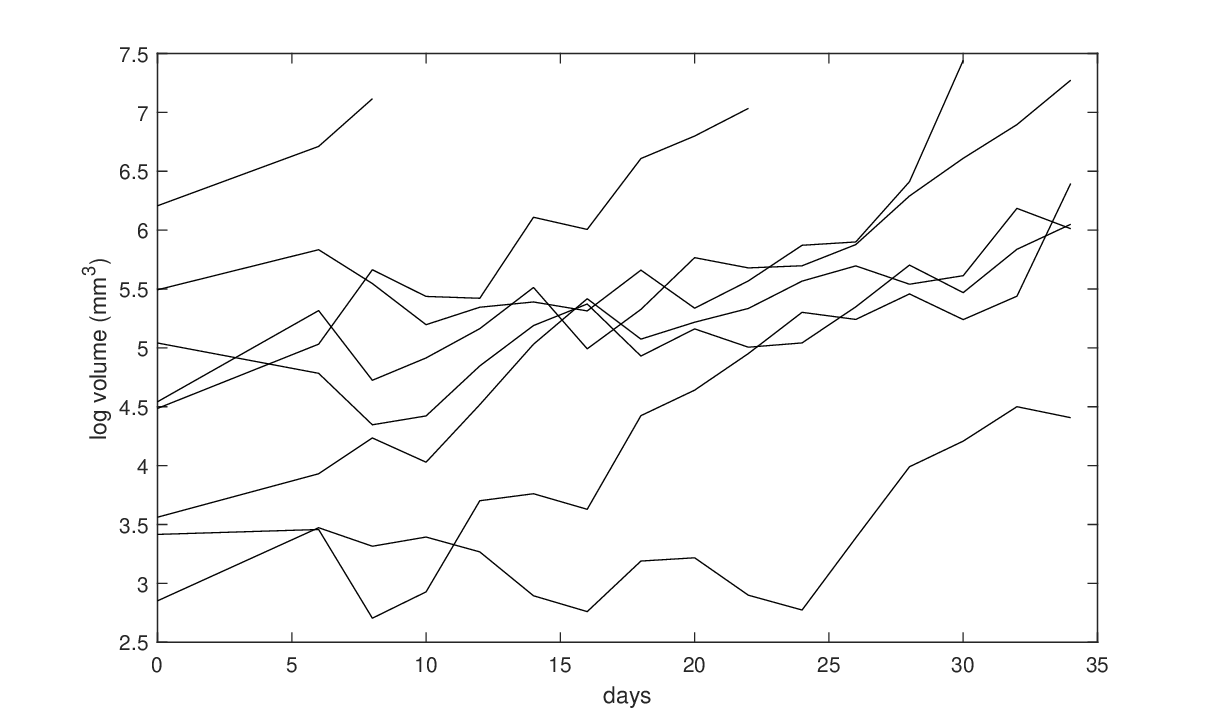}\\
\includegraphics[height=3cm,width=6cm]{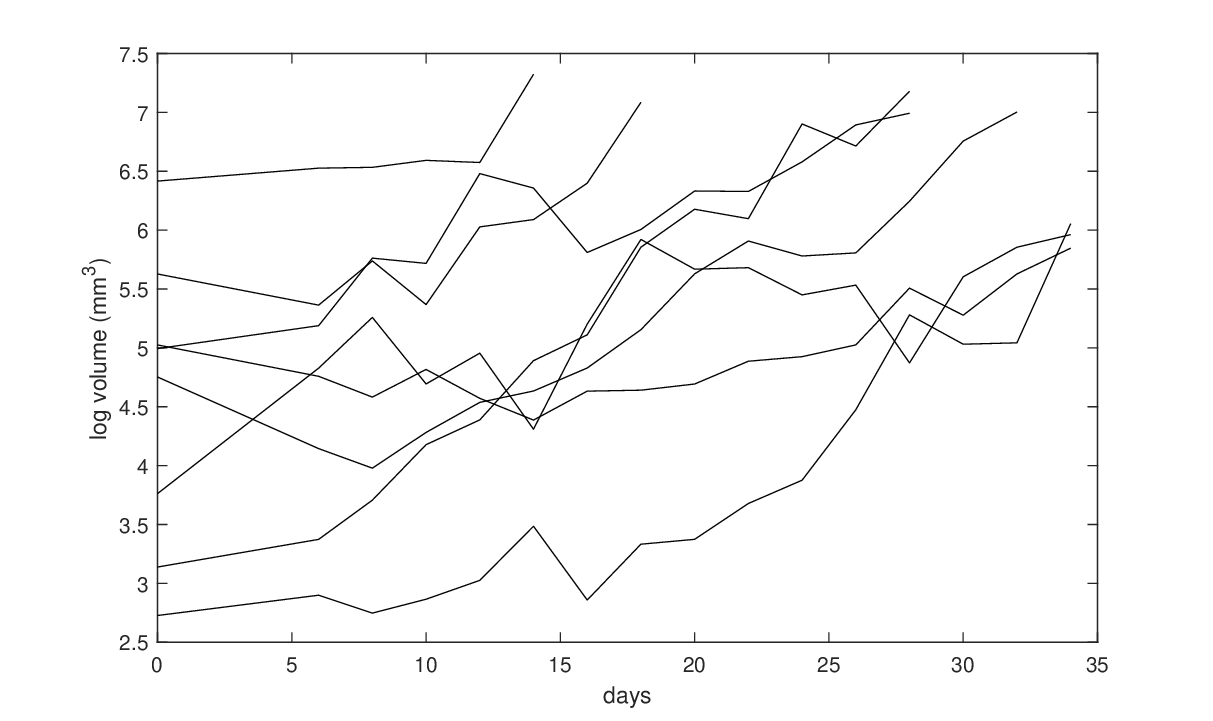}\quad
\includegraphics[height=3cm,width=6cm]{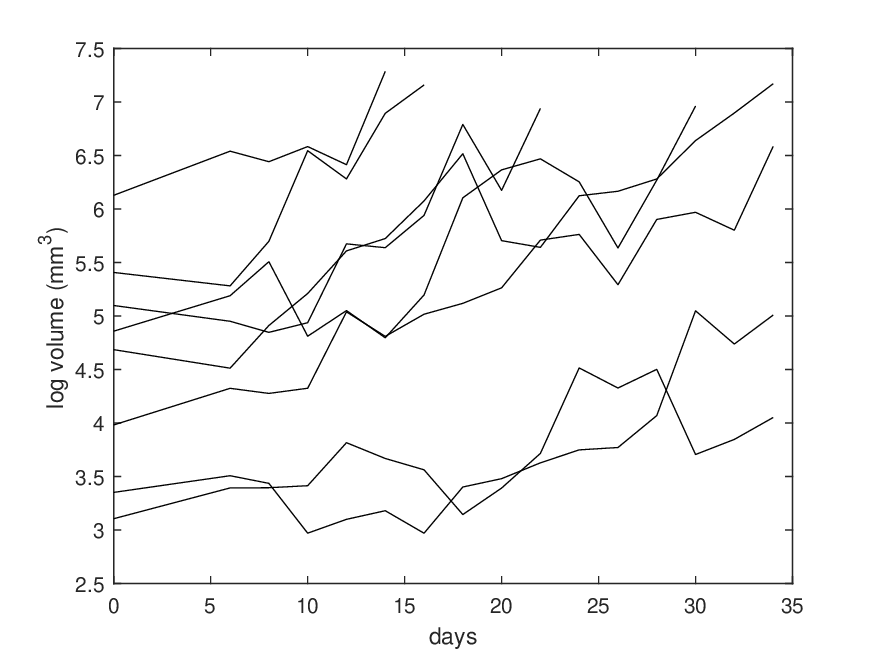}\\
\caption{\footnotesize{Fitted data in group 3 (top left) and three realizations from model \eqref{e:sde1} estimated with exact Bayesian methodology (remaining plots). Recall for this group measurements at days 1 and 4 were disregarded during estimation, hence times on abscissas start at day 6.}}
\label{fig:group3-simdata}
\end{figure}

To make a rough assessment of whether model \eqref{e:sde1} is
realistic compared to the data, we simulated growth curves using the posterior means estimated with PMM (Table
\ref{tab:realdata-estimates}). The simulations are shown in Figures
\ref{fig:group1-simdata} and \ref{fig:group3-simdata}. The overall impression is that model \eqref{e:sde1} is capable of generating growth dynamics that
are similar to the experimental data. A more thorough comparison is produced using posterior predictive checks in section \ref{sec:bsl-posterior-predictive}.

A comparison between PMM using the bootstrap and auxiliary particle filters (BF vs APF)
is presented in the supplementary material. We found that results obtained with APF are more stable. This is no surprise as BF is known to degenerate when the measurements error is small (in our case $\sigma_\varepsilon$ is more than an order of magnitude smaller than the typical log-volumes). Moreover we compared PMM to BSL (see in addition next section) finding that results from PMM seem sensitive to the number of particles $L$, whereas BSL is less affected by the number of simulated datasets $N$.

\begin{table}[ht]
\caption{\footnotesize{Posterior means and 95\% posterior intervals: for each parameter we first report exact Bayesian inference using the pseudo-marginal method PMM and then approximate inference using synthetic likelihoods estimation BSL.}}
\centering \scriptsize
\begin{tabular}{rrrr}
\hline
{} & group 1 & group 3 & group 5\\
\hline\\
$\bar{\beta}$ & 5.81 [3.82,7.83] & 3.33 [2.07,4.64] & 6.70 [4.09,8.90]\\
& 6.59 [4.90,8.75] & 3.93 [2.93,5.04] & 7.48 [6.20,8.98]\\ 
$\bar{\delta}$& 1.84 [0.68,4.59] & 1.14 [0.40,2.32]& --\\
& 1.90 [0.53,4.69] & 1.52 [0.43,3.68]\\
$\bar{\alpha}$& 0.52 [0.24,0.84] & 0.60 [0.31,0.91]  & --\\
& 0.41 [0.13,0.74] & 0.47 [0.17,0.84]\\
$\gamma$& 1.13 [0.66,1.71] & 1.09 [0.70,1.52] & 1.49 [1.07,2.09]\\
& 1.03 [0.63,1.42] & 0.92 [0.56,1.32] & 1.64 [1.26,2.15]\\
$\tau$& 1.50 [0.68,2.98] & 1.82 [1.02,2.63]  &--\\
& 1.51 [0.71,2.77] & 1.75 [1.03,2.64]\\
$\sigma_{\beta}$& 0.61 [0.23,1.37] & 0.51 [0.19,1.67] &  0.68 [0.23,1.96]\\
&  0.55 [0.25,1.19] & 0.59 [0.23,1.28] & 0.40 [0.27,0.60]\\
$\sigma_{\delta}$& 0.67 [0.24,1.68]  & 0.76 [0.26,2.23] & --\\
& 0.66 [0.23,1.74] & 0.71 [0.25,1.91]\\
$\sigma_{\alpha}$ & 0.37 [0.16,0.74] & 0.29 [0.15,0.48] & --\\
& 0.30 [0.14,0.68] & 0.43 [0.14,1.16]\\
$\sigma_{\varepsilon}$& 0.22 [0.19,0.30] & 0.20 [0.19,0.23] & 0.23 [0.20,0.31]\\
& 0.17 [0.10,0.28] & 0.11 [0.07,0.17] & 0.18 [0.11,0.29]\\
\hline
\end{tabular}
\label{tab:realdata-estimates}
\end{table}

\begin{figure}[ht]
\centering
\subfloat[$\log\bar{\beta}$]{\includegraphics[height=2.9cm,width=5cm]{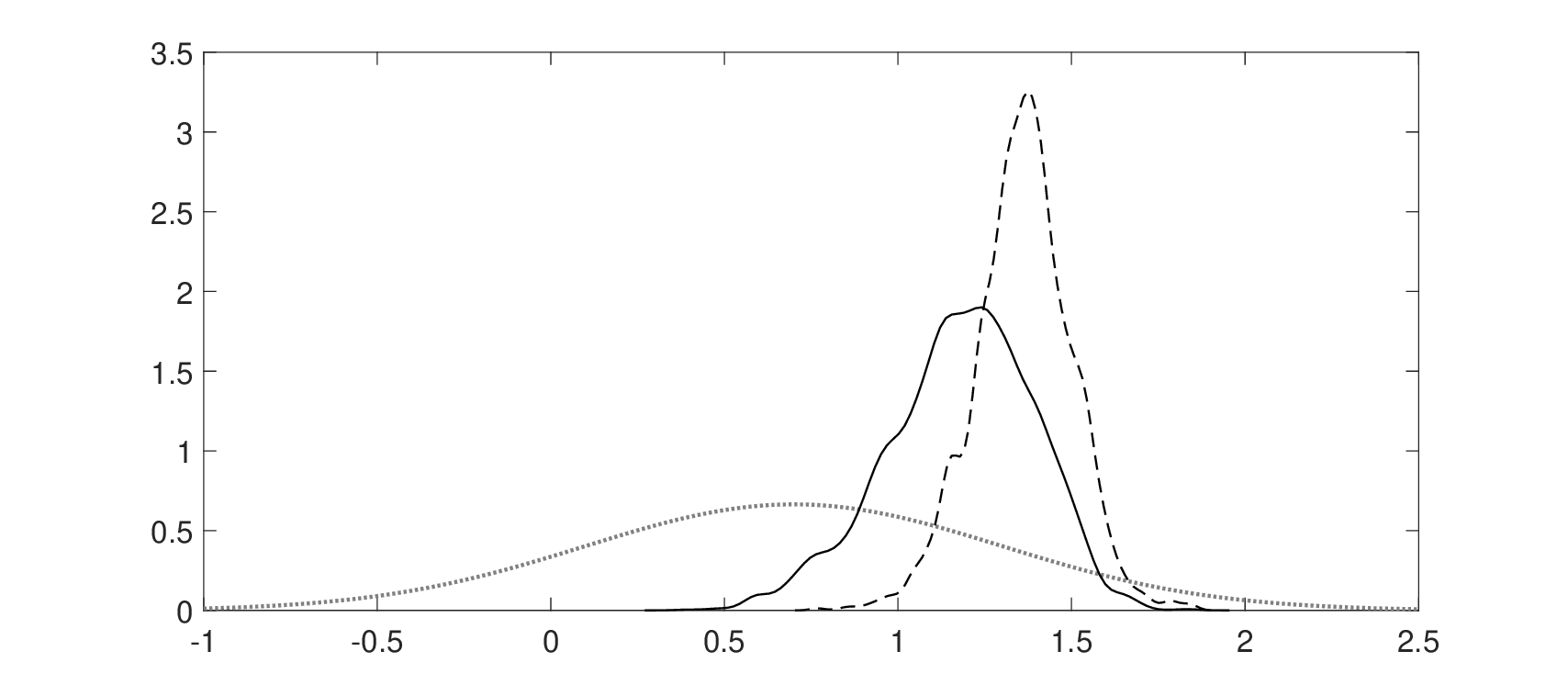}}
\subfloat[$\log\bar{\delta}$]
{\includegraphics[height=2.9cm,width=5cm]{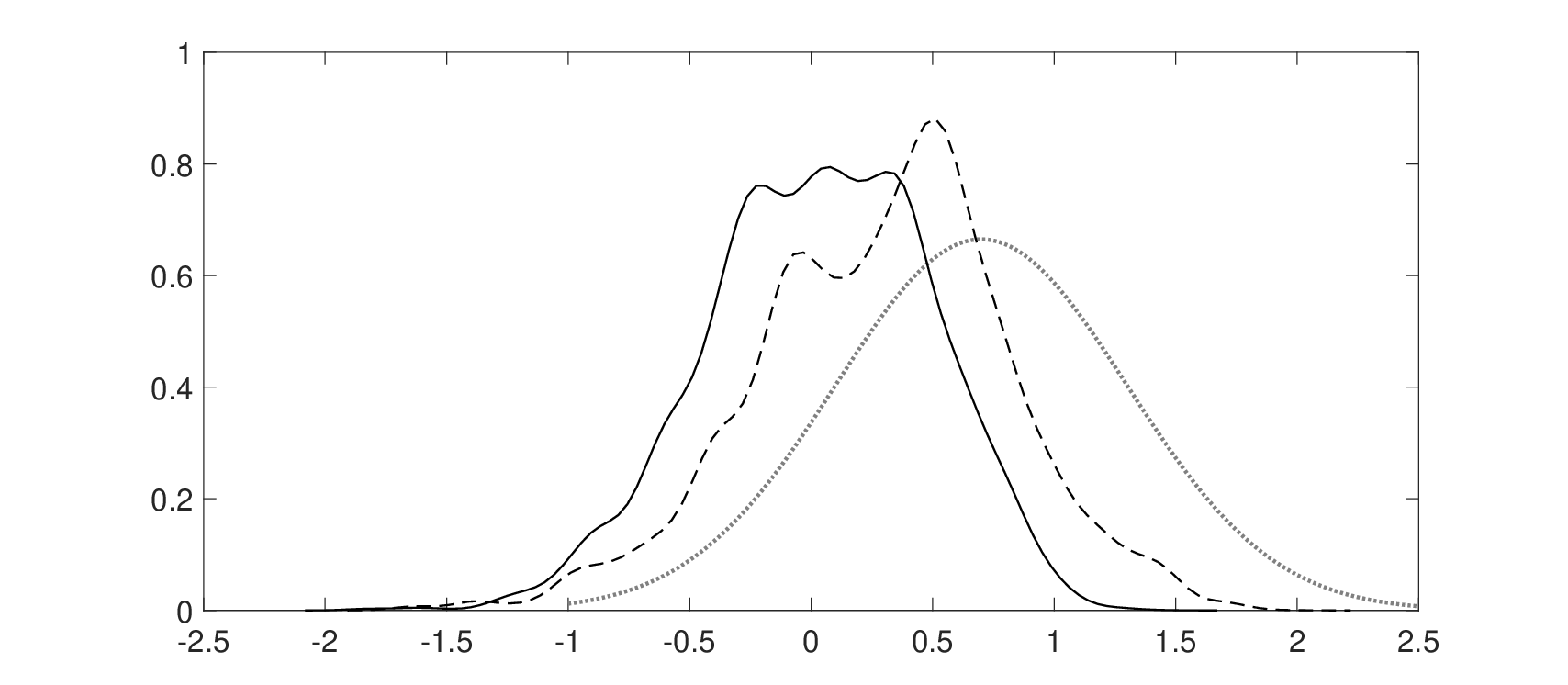}}
\subfloat[$\bar{\alpha}$]{
\includegraphics[height=2.9cm,width=5cm]{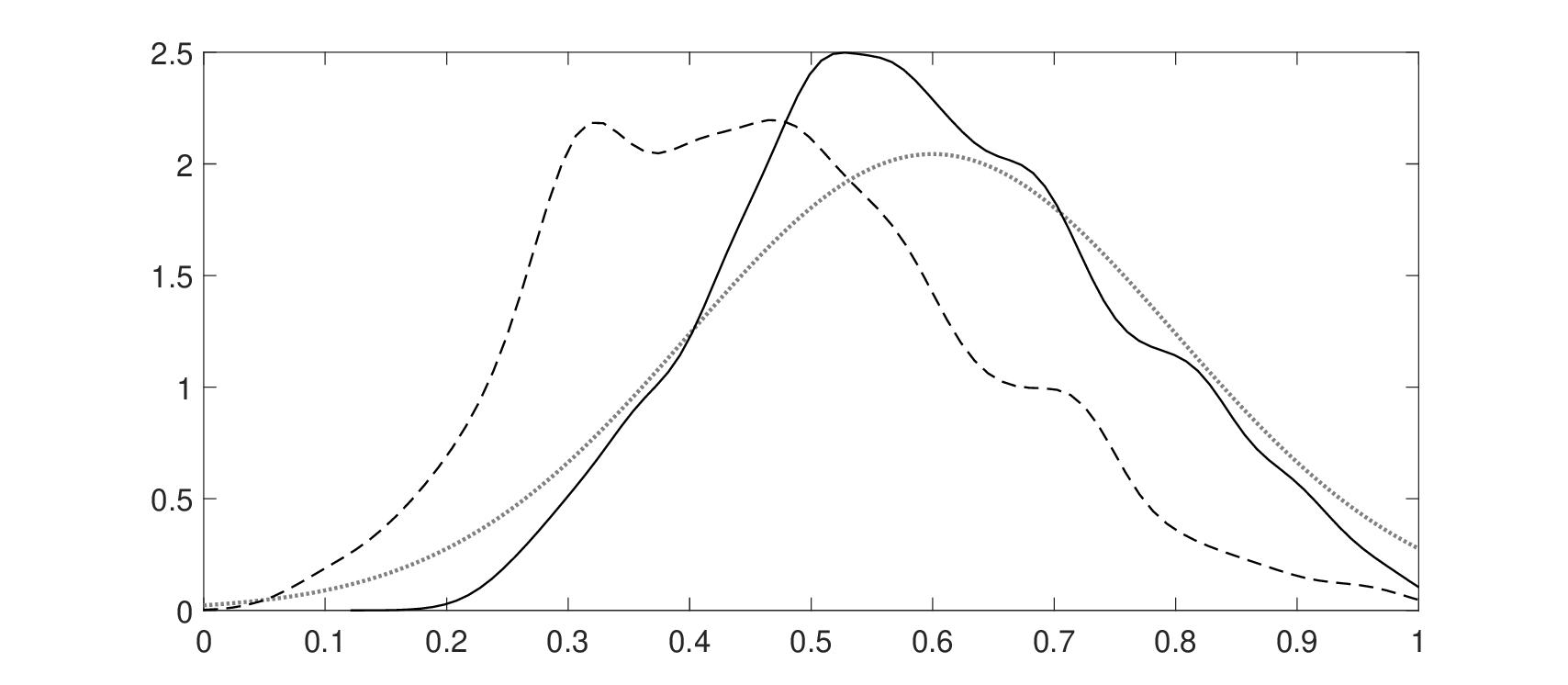}}\\
\subfloat[$\gamma$]{
\includegraphics[height=2.9cm,width=5cm]{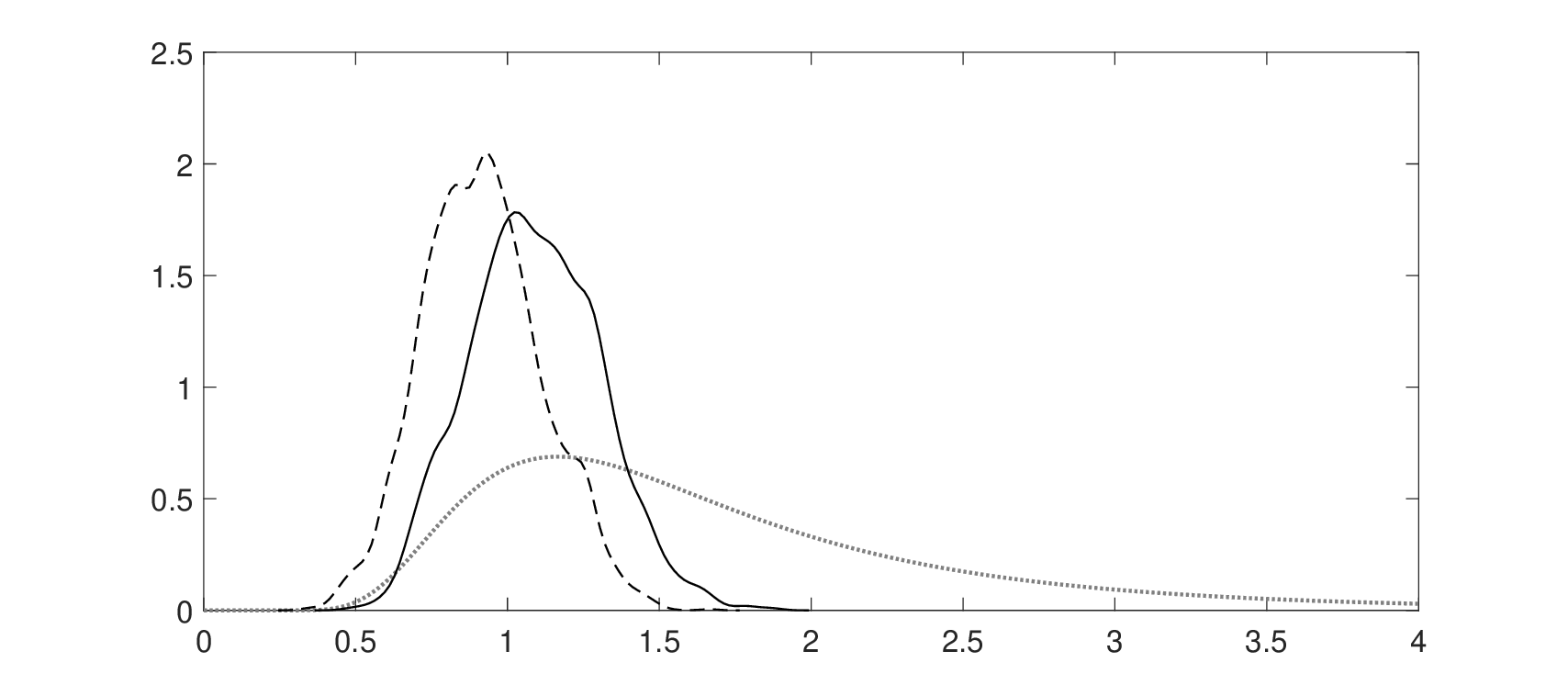}}
\subfloat[$\tau$]{
\includegraphics[height=2.9cm,width=5cm]{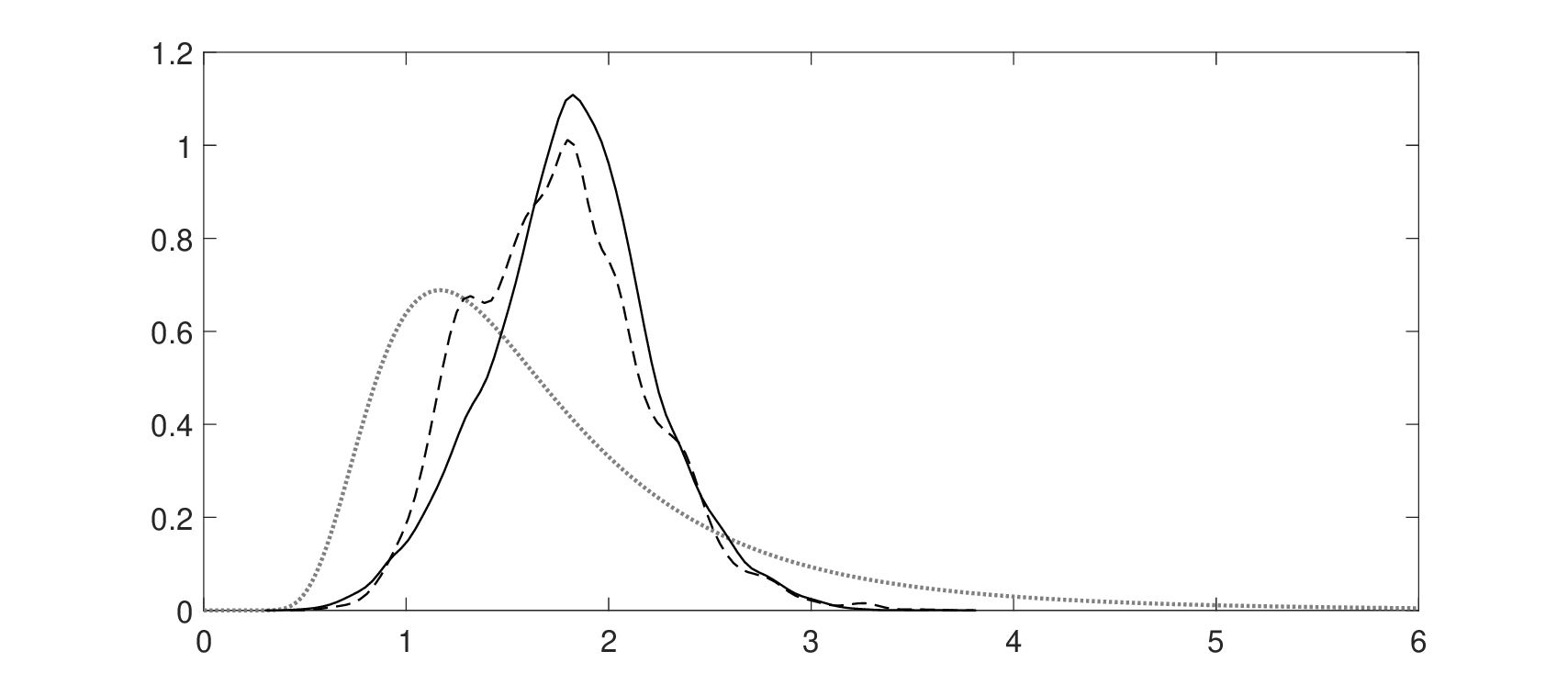}}
\subfloat[$\sigma_{\beta}$]{
\includegraphics[height=2.9cm,width=5cm]{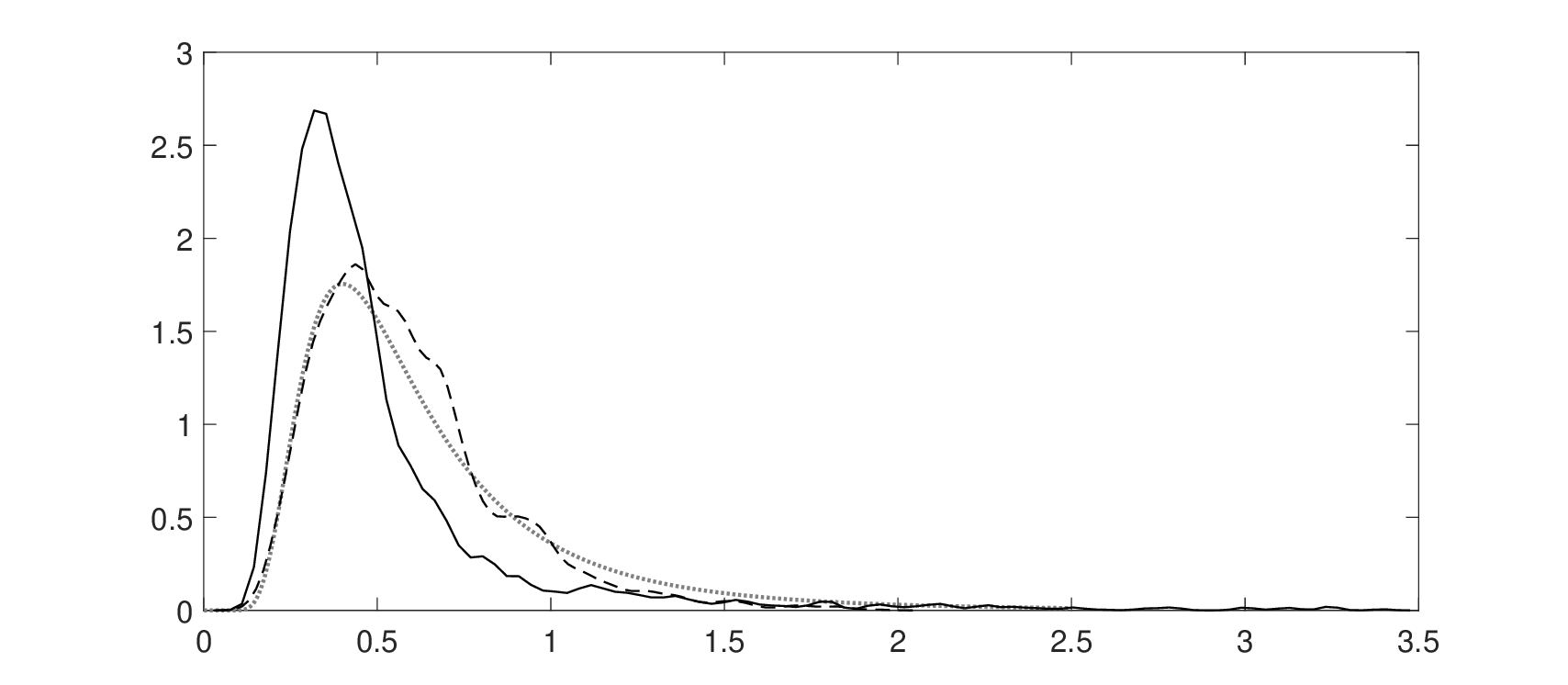}}\\
\subfloat[$\sigma_{\delta}$]{
\includegraphics[height=2.9cm,width=5cm]{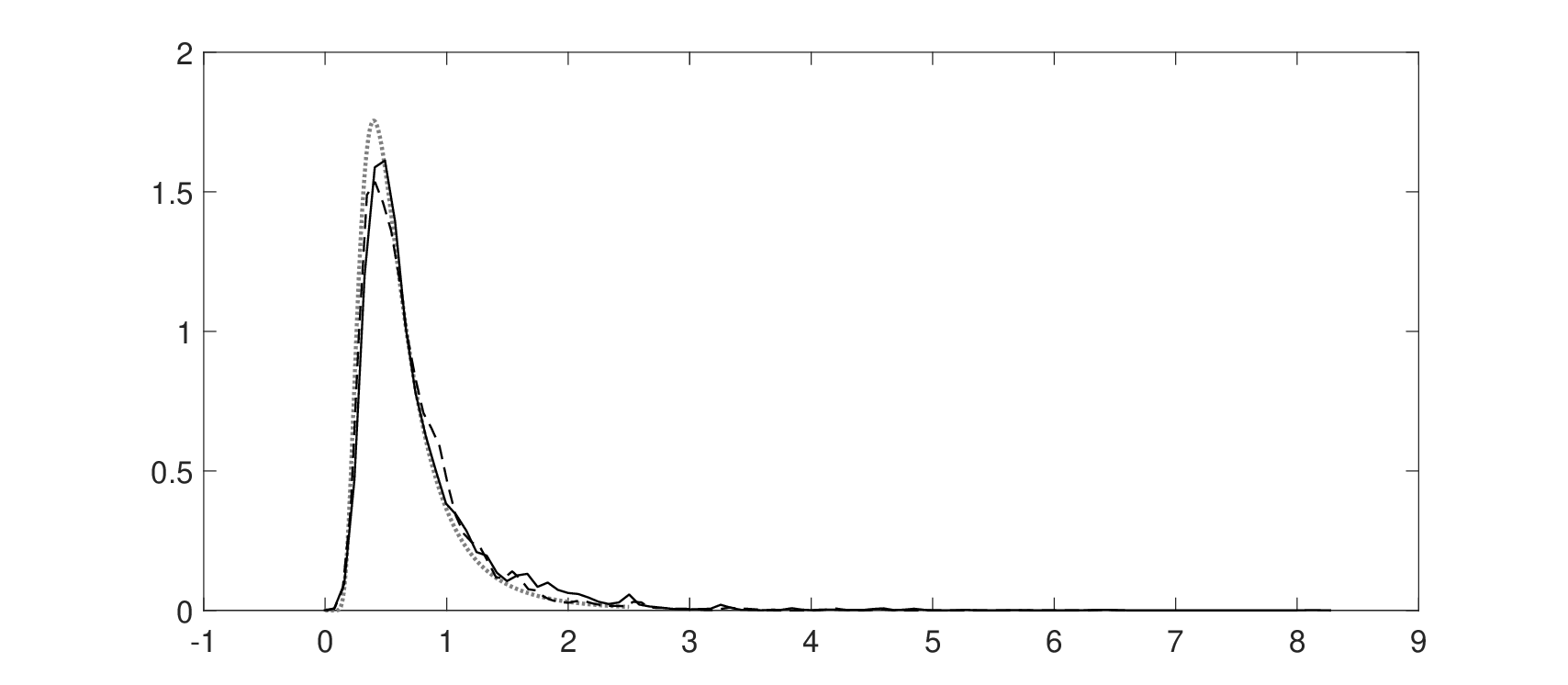}}
\subfloat[$\sigma_{\alpha}$]{
\includegraphics[height=2.9cm,width=5cm]{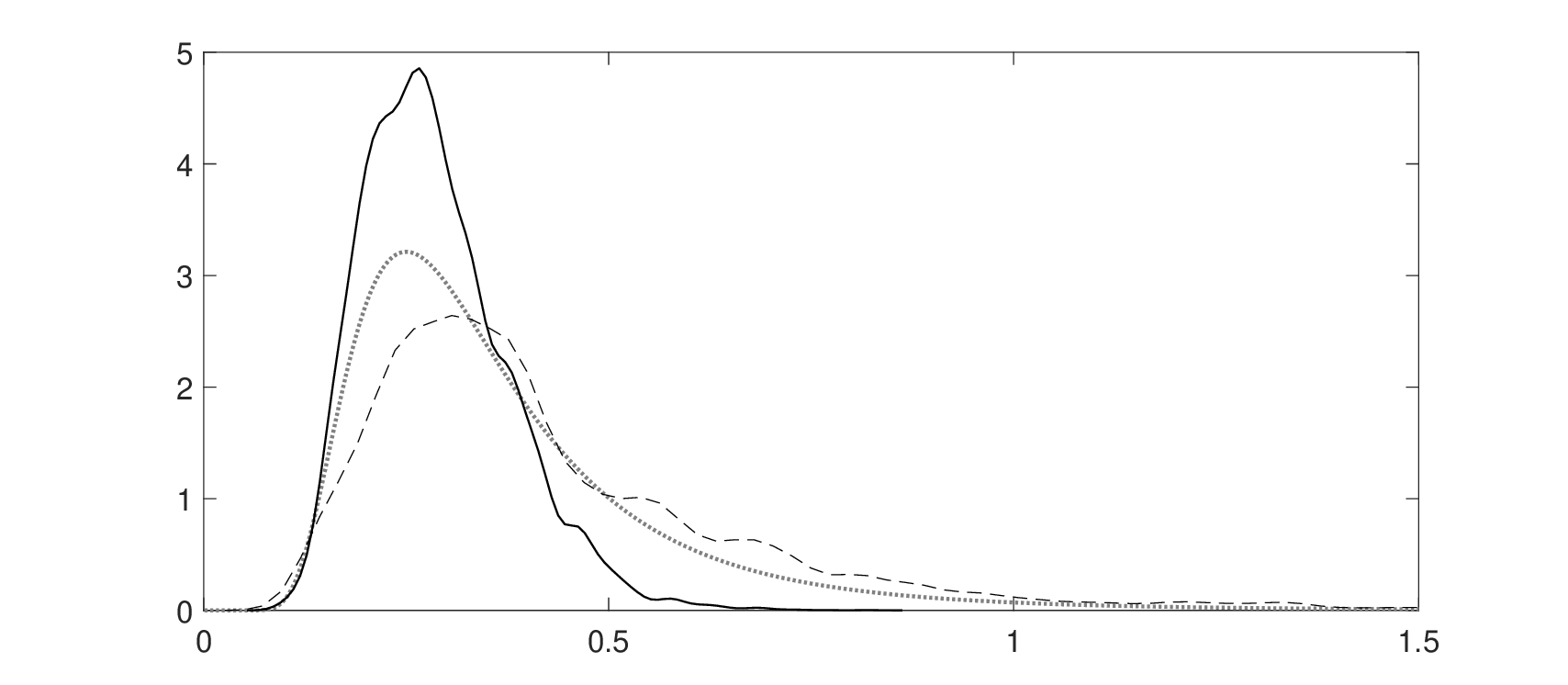}}
\subfloat[$\sigma_{\varepsilon}$]{
\includegraphics[height=2.9cm,width=5cm]{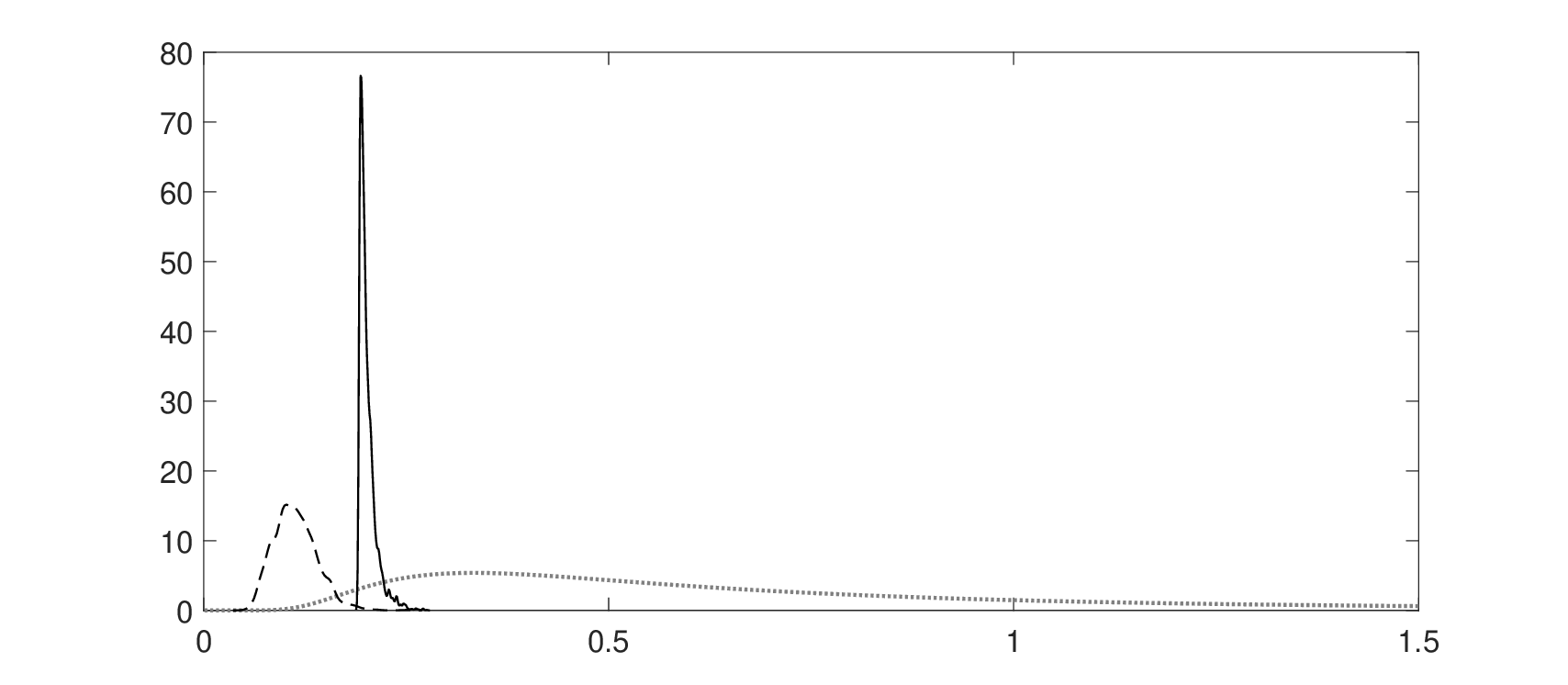}}
\caption{\footnotesize{Treatment group 3, exact posteriors via PMM using the auxiliary particle filter (solid lines), synthetic likelihoods posteriors (dashed) and prior densities (dotted gray). The prior density for $\sigma_{\varepsilon}$ was multiplied by 4 for ease of display.}}
\label{fig:group3marginals}
\end{figure}

\subsection{Results using synthetic likelihoods}\label{sec:results-synlik}

The application of the synthetic likelihood (SL) methodology from section \ref{sec:synlik} relies on a selection of summary statistics. Therefore, before presenting any results, we first describe how our chosen summary statistics were constructed.

\paragraph{Summary statistics:}
We first define the components for each individual (vector) summary $\bm{s}_i:=\bm{s}^{\mathrm{intra}}(\bm{y}_i)$: 
(i) the mean absolute deviation for the repeated measurements $\mathrm{MAD}\{y_{ij}\}_{j=1:n_i}$
(ii) the slope of the line segment connecting the first and the
last observation, $(y_i(t_{n_i})-y_i(t_1))/(t_{n_i}-t_1)$; 
(iii+iv) the values of the first and second measurements $y_{i1}$ and $y_{i2}$; 
(v) the estimated slope of a first order autoregressive fit of the
repeated measurements, that is $\hat{\beta}_{i1}$ from the regression $E(y_{ij})=\beta_{i0}+\beta_{i1}y_{i,j-1}$. 
Note that when fitting model (\ref{e:sde0}) to the control group, the last 
summary statistic was dropped to prevent $\bm{\Sigma}_N(\bm{\theta})$ from becoming
singular (several mice had only two observations so that the second
and fifth summary were perfectly correlated).
Additional inter-individual (population) summary statistics $\bm{s}^{\mathrm{inter}}$ are:
(i) $ \mathrm{MAD} \{y_{i1}\}_{i=1:M}$, the mean absolute deviation
between subjects at the first time point (day 6 for the active
treatment groups and day 1 for the control group); 
(ii) the same as in (i) but for the second time point; (iii) the same as in (i) but for the last time point. Therefore when fitting group 3 ($M=8$ subjects) the total vector of summaries $\bm{s}$ contains 43 features, since we have 5 features per subject plus 3 inter-individuals features. In absence of previous literature considering the construction of summary statistics for SDEMEMs, our custom-made summaries follow common sense intuition. For example it seems reasonable to include into $\bm{s}^{\mathrm{intra}}(\bm{y}_i)$ a robust measure of variability (MAD) for individual trajectories. Also, since the overall behavior of the trajectories is increasing, we believe the slope of the line connecting first and last observations can give insight on the volume growth rate. Similarly, the values of the first two individual measurements could represent an assessment of the initial growth. The first order autocorrelation is a standard measure of information in dynamic models. Similarly, we assess variation between-subjects using $\bm{s}^{\mathrm{inter}}$: to this end we assess the variation between trajectories at several sampling times, in our case by using MAD on measurements from three different times points.

\paragraph{Results using BSL:}
We use $N=3,000$ simulated datasets to construct the synthetic likelihood
approximation at each value of $\bm{\theta}$, and run $R=20,000$ iterations of the BSL algorithm 
\ref{alg:synlikMCMC}, as described in section \ref{sec:synlik}. For group 5, due to the small number of subjects, we doubled $N$ to $N=6,000$ (otherwise we obtain a very variable synthetic likelihood, whose occasional overestimation causes stickiness in the chains). 
For comparability, we adopted the same priors and initial parameter values as in the exact
Bayesian analysis. During the execution of the algorithm we observed an acceptance rate of about $30\%$ and the procedure required about 520 seconds for group 3 ($M=8$ subjects).
We checked the convergence of three chains initialized at the same dispersed values used to assess the convergence of PMM. For each parameter the $\hat{R}$ value was below 1.04, hence the chains are converging.

\begin{figure}[ht]
\centering
\includegraphics[height=3cm,width=6cm]{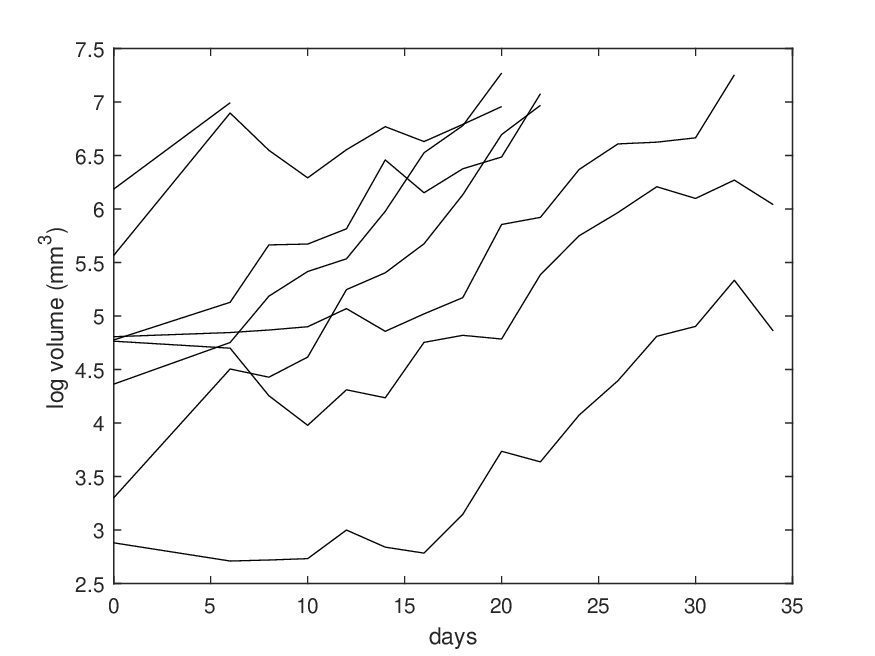}\quad
\includegraphics[height=3cm,width=6cm]{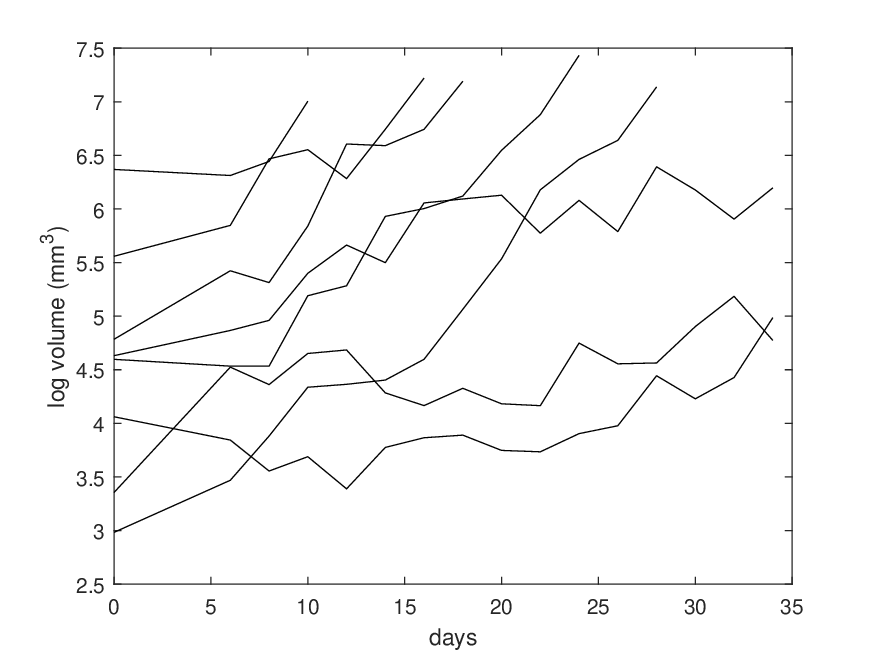}\\
\caption{\footnotesize{Group 3: two realizations from model \eqref{e:sde1} estimated with synthetic likelihoods.}}
\label{fig:group3-synlike-sim1-2}
\end{figure}

Posterior estimates obtained with BSL (discarding a burn-in of 10,000 iterations) are compared with PMM in Table \ref{tab:realdata-estimates},  and the approximate marginals for group 3 are compared in Figure \ref{fig:group3marginals}. Although some differences in posterior means are found between BSL and PMM, these appear to be of minor consequence to the anticipated growth patterns. Figure \ref{fig:group3-synlike-sim1-2} shows simulated growth curves (based on the posterior means estimated with BSL) that are overall similar to Figure \ref{fig:group3-simdata}, which was obtained using the corresponding PMM estimates. 
A more careful evaluation of the BSL model fit is carried out using posterior predictive checks in section \ref{sec:bsl-posterior-predictive} below. 
All of the selected summary statistics were found to be approximately normally distributed (see supplementary material).

Additional analyses reported in the supplementary material show that BSL returns results that are closer to those from PMM, provided that PMM is run with a larger number of particles (say, $L=3,000$) and moreover that BSL is less sensitive to the specific choice of $N$ than PMM is to the choice of $L$. Furthermore we investigated the effect of using less informative priors for $\log\bar{\delta}$, $\sigma_{\beta}$ and $\sigma_{\delta}$. The conclusion is that the considered volume of data is not informative enough for these parameters, that is the information carried by the model is unable to depart from the prior information for $\sigma_{\beta}$ and $\sigma_{\delta}$. The (log-)elimination rate $\log\bar{\delta}$ does depart from its prior, but at the expense of increased variability. 

Comparing computational times between BSL and PMM in a fair way is difficult, since the two algorithms have a completely different structure. Both methods perform similarly in terms of acceptance rate (30\% in both cases) however in terms of raw numbers, PMM is clearly more intensive since at each MCMC iteration the model is simulated $L\times L_2= 10,000$ times, while for BSL the model is simulated only $N=3,000$ times (plus the overhead time needed to compute summary statistics out of each simulated trajectory). Given the above, 1,000 MCMC iterations using data from group 3 require 8.35 minutes with PMM and 0.44 minutes with BSL.

\subsubsection{Posterior predictive checks}\label{sec:bsl-posterior-predictive}

\begin{figure}[ht]
\centering
\includegraphics[width=11cm,height=6.5cm]{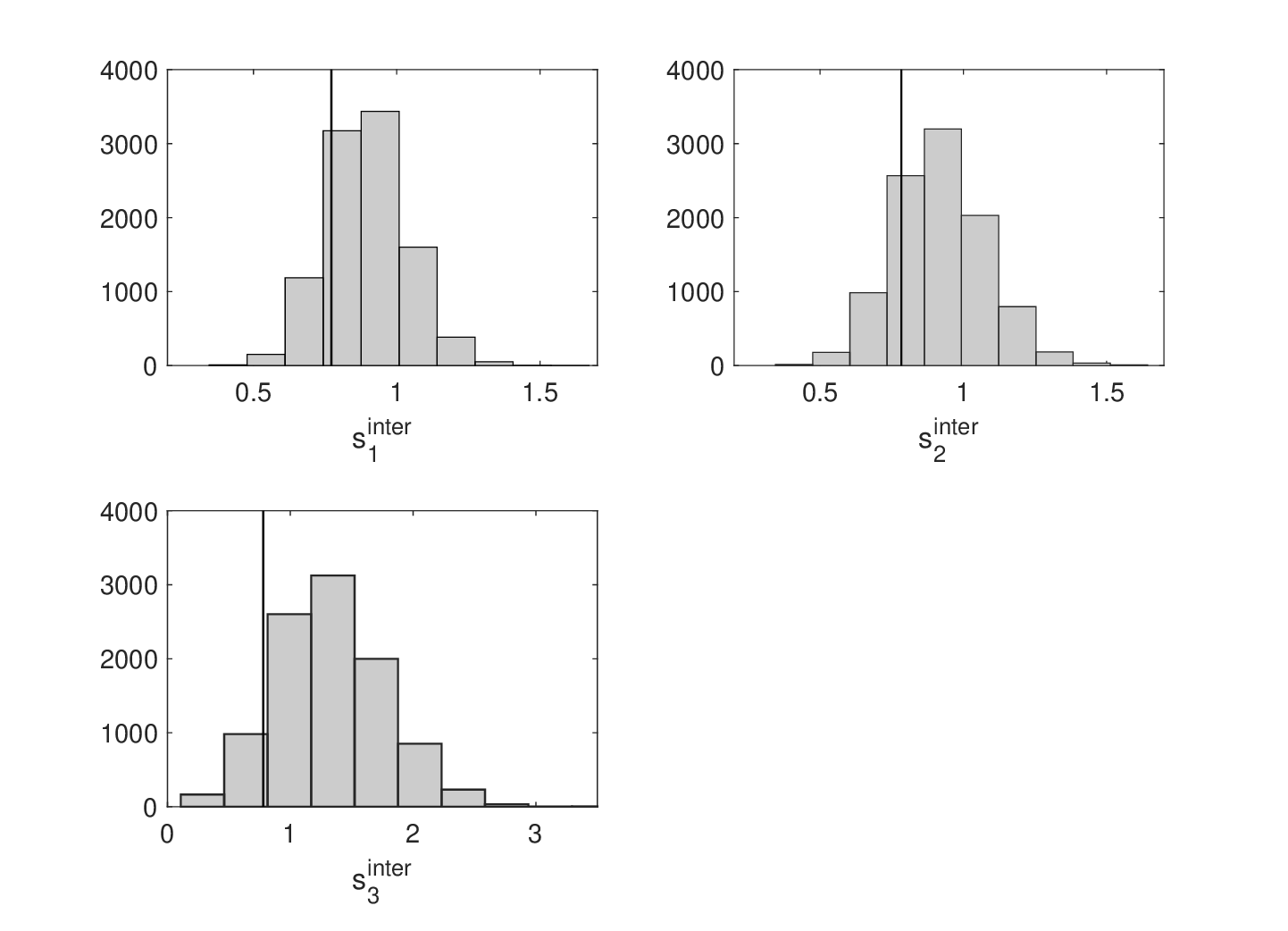}
\caption{\footnotesize{Posterior predictive checks for group 3 generated from BSL. Simulated inter-subjects summaries $s^{\mathrm{inter}}_1$ (top-left), $s^{\mathrm{inter}}_2$ (top-right) and $s^{\mathrm{inter}}_3$ (bottom). Vertical lines mark the values for the corresponding statistics from the observed data.}}
\label{fig:group3-synlike-postpredchecks_S-INTER}
\end{figure}

\begin{figure}[ht]
\centering
\includegraphics[width=11cm,height=10.5cm]{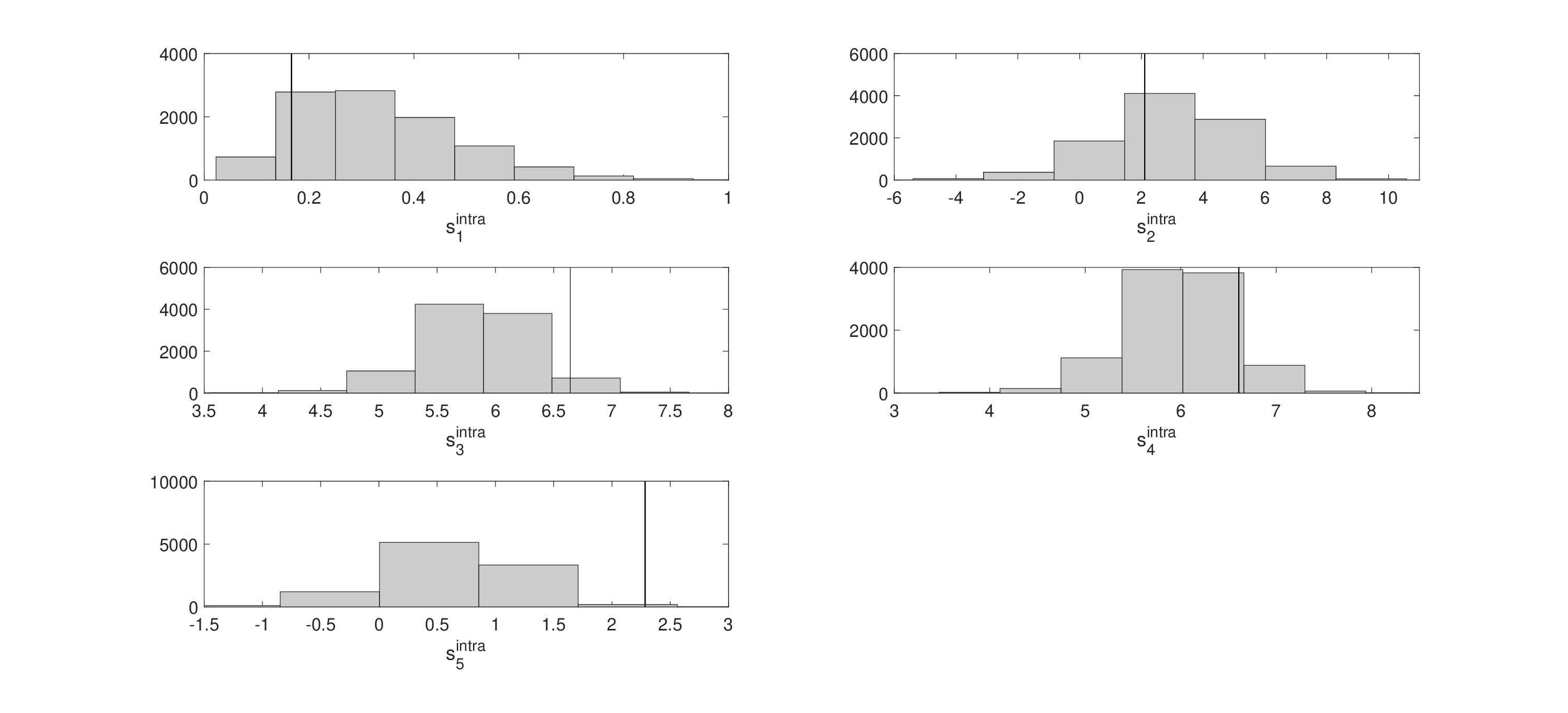}
\caption{\footnotesize{Posterior predictive checks for group 3 generated from BSL. Simulated intra-subjects summaries for subject 1: $s^{\mathrm{intra}}_1$ and $s^{\mathrm{intra}}_2$ (top), $s^{\mathrm{intra}}_3$ and $s^{\mathrm{intra}}_4$ (middle) and $s^{\mathrm{intra}}_5$ (bottom). Vertical lines mark the values for the corresponding statistics from the observed data.}}
\label{fig:group3-synlike-postpredchecks_S-INTRA_subj1}
\end{figure}

Posterior predictive checks were made for data in group 3, following the reasoning and notation in the supplementary material. We used the 10,000 draws produced as output of the BSL algorithm \ref{alg:synlikMCMC} (after burn-in) to simulate corresponding 10,000 independent sets of summaries from the posterior predictive distribution. Next, these were compared with the observed summaries from the experimental data. Since $M=8$ subjects are considered, each $\bm{s}^*$ contains 43 summaries.
Figure \ref{fig:group3-synlike-postpredchecks_S-INTER} shows the 
histograms for the three inter-subject summaries while
the intra-subject summaries for subject 1 are shown in Figure \ref{fig:group3-synlike-postpredchecks_S-INTRA_subj1}. Corresponding plots for the remaining subjects can be found in the supplementary material. Regarding the inter-subjects summaries, the model generates summaries that are comparable to the corresponding observed summaries.
Regarding the intra-subjects summaries for subject 1, the estimated model complies well with the observed summaries, except for ${s}_5^{\mathrm{intra}}$ which seems less plausible. However, for the other subjects, the observed ${s}_5^{\mathrm{intra}}$  is highly probable under $p(\bm{s}^*|\bm{s})$ (see supplementary material).

\section{Comparison to an ODE mixed-effects model}\label{s:odefit}
While SDEMEMs may have a better chance to capture real life within- and between-subjects variability, inference is complicated by the need to make delicate implementation and calibration decisions for the simulations setup, in order to approximate their likelihoods. On the opposite side, ordinary differential equation mixed-effects models (ODEMEMs) have tractable likelihood functions, and can be analyzed with off-the-shelf software utilizing robust MCMC inference via Hamiltonian Monte Carlo \citep{betancourt2017conceptual}, where exploration of the posterior surface is guided by exact gradients.

For a comparison of the two model types we have fitted the ODEMEM \eqref{e:obsmodel}--\eqref{e:ode1} separately to data from treatment groups 1 and 3. Parameters of interest are $\bm{\theta} = (\bar{\beta},\bar{\delta}, \bar{\alpha},\sigma_\beta,\sigma_\delta,\sigma_\alpha)$.
Inference based on 20,000 draws from the true posterior $\pi(\bm{\theta}|\bm{y})$ is presented in Table \ref{tab:realdata-estimates-odemems}
(see supplementary material for details).
We note that the estimated mean growth and decay rates ($\bar{\beta}$ and $\bar{\delta}$ respectively) from the ODEMEMs  are higher compared to those estimated from the SDEMEMs. Importantly, estimated residual variation $\sigma_\varepsilon$ here is 3-4 times larger than for the SDEMEMs.  Also, marginal posteriors are wider for all parameters. In particular, identification of $\bar{\alpha}$ remains elusive in both groups.
In analogy with section \ref{sec:bsl-posterior-predictive} we performed posterior predictive checks to evaluate the fit of the ODEMEM for group 3. Results are shown in the supplementary material. These indicate a better fit for subject 1 than the one obtained using SDEMEMs.

However, larger sample sizes (or more frequent measurements over time) are needed to determine which model performs best in terms of both in-sample and out of sample predictions.  

\begin{table}[ht]
\caption{\footnotesize{Posterior means and 95\% posterior intervals from exact Bayesian inference for the ODEMEMs.}}
\centering \scriptsize
\begin{tabular}{rrr}
\hline
{} & group 1 & group 3\\
\hline\\
$\bar{\beta}$ & 9.93 [6.90,13.30] & 5.71 [3.43,7.48] \\
$\bar{\delta}$& 2.24 [0.68,5.58] & 1.77 [0.63,4.02]\\
$\bar{\alpha}$& 0.42 [0.08,0.80] & 0.46 [0.09,0.82]  \\
$\sigma_{\beta}$& 0.69 [0.23,2.01] & 1.87 [0.47,3.81]\\
$\sigma_{\delta}$& 0.59 [0.22,1.44]  & 0.54 [0.22,1.22] \\
$\sigma_{\alpha}$ & 0.36 [0.14,0.91] & 0.46 [0.16,1.15] \\
$\sigma_{\varepsilon}$& 0.76 [0.56,1.04] & 0.60 [0.51,0.72] \\
\hline
\end{tabular}
\label{tab:realdata-estimates-odemems}
\end{table}

\section{Simulation studies}\label{sec:simulation-study}

We have conducted a small-scale simulation study to investigate the statistical properties of PMM and BSL in the context of the SDEMEM \eqref{e:sde1}. 

\begin{figure}[ht]
\centering
\includegraphics[height=4.5cm,width=9cm]{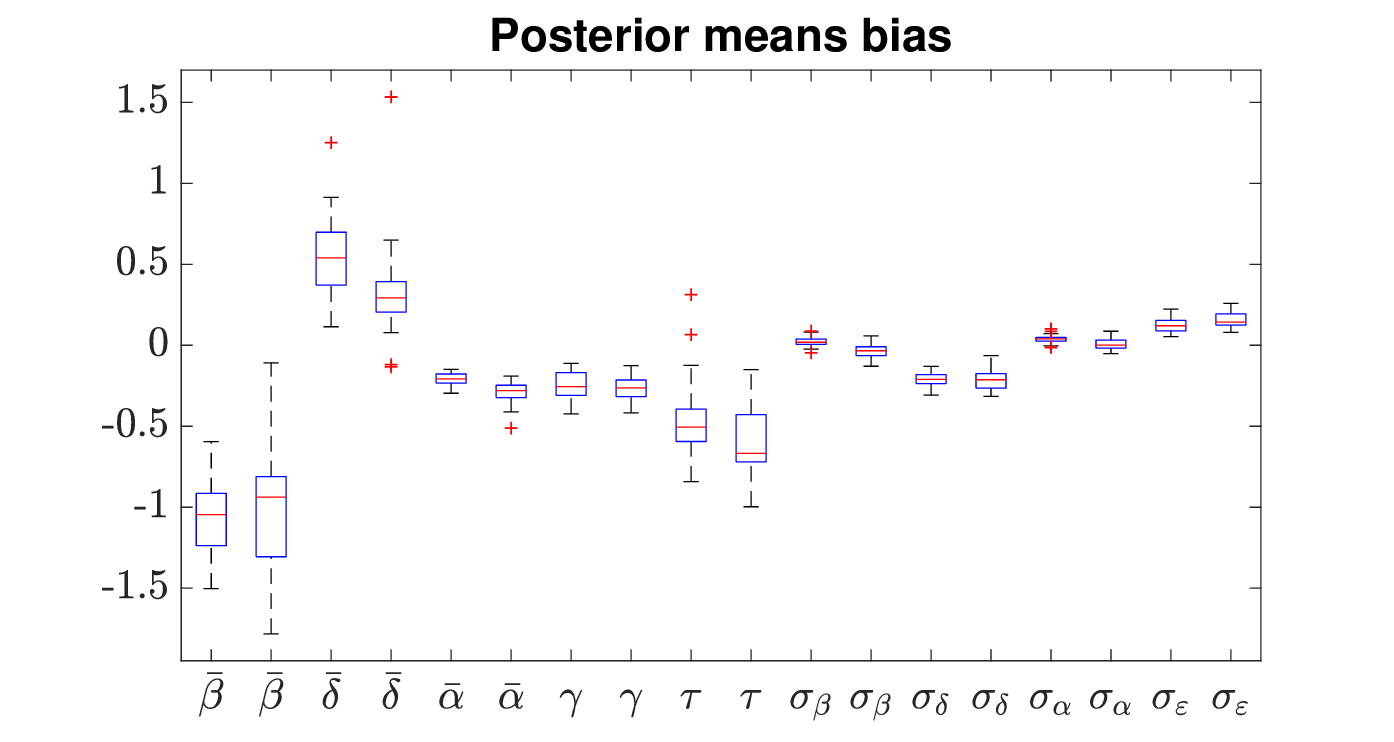}
\caption{\footnotesize{Simulation study with $M=8$ subjects: boxplots of the bias of thirty posterior means obtained with PMM and BSL. Starting from the left side: bias of $\bar{\beta}$ obtained via PMM, then the bias of $\bar{\beta}$ obtained via BSL, and so on.}}\label{fig:boxplots_meansbias_30simulations}
\end{figure}

\begin{figure}[ht]
\centering
\includegraphics[height=4.5cm,width=9cm]{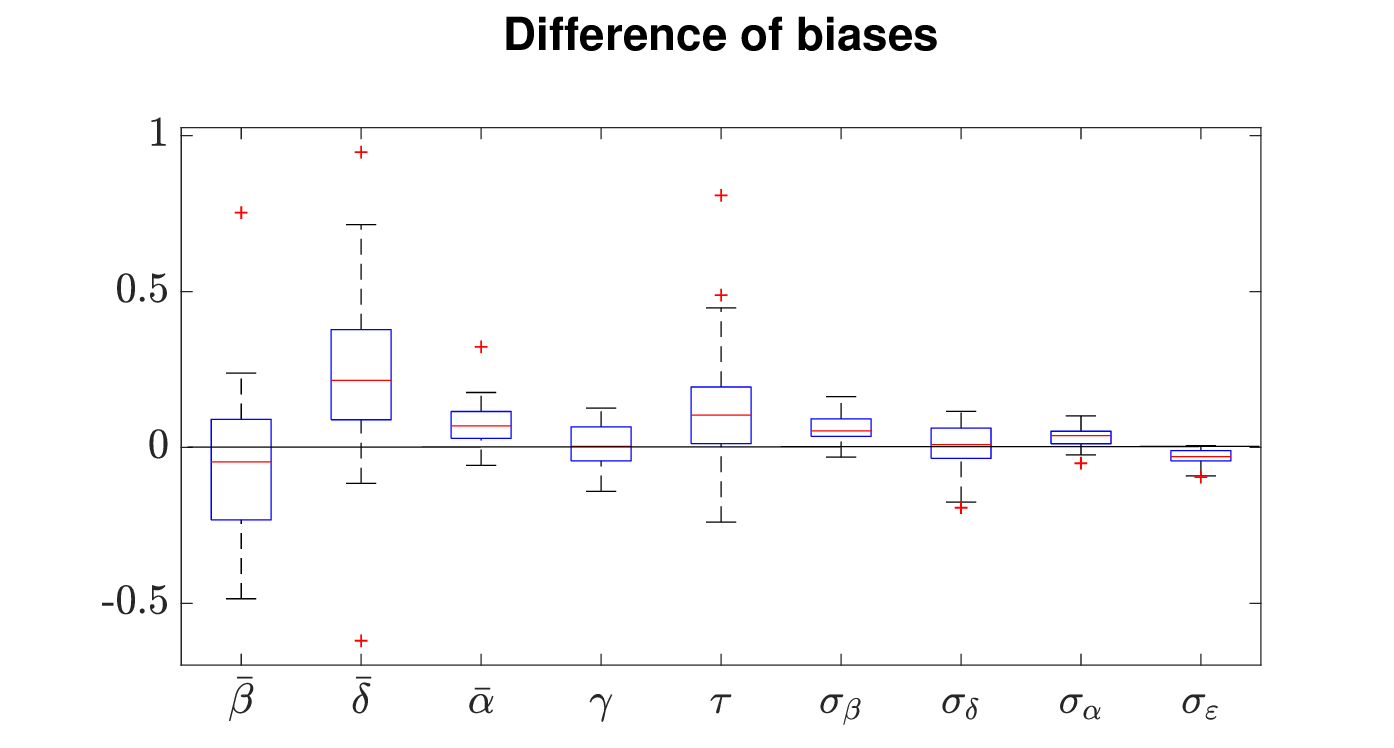}
\caption{\footnotesize{Simulation study with $M=8$ subjects: boxplots of the difference of the biases shown in Figure \ref{fig:boxplots_meansbias_30simulations}, namely bias(PMM)-bias(BSL).  A black line marks a difference of biases equal to zero.}}\label{fig:boxplots_differencebiases_30simulations}
\end{figure}

Thirty datasets were generated independently from the model with ground-truth parameters $\bm{\theta}_0$ set to the posterior means obtained with exact Bayesian methodology on group 3, as found in Table \ref{tab:realdata-estimates} except for $\bar{\alpha}$ which was set to $0.75$ for consistency with the simulation study in section \ref{sec:larger-simulated-data} below. 
Each simulated dataset has measurements for $M=8$ subjects, with observational times identical to those of group 3, and growth curves initiated at the same values $v_{i,0}$ as in the previous sections.
To each of the generated datasets we applied both PMM and BSL, initializing the algorithms at the same starting parameters as in previous analyses and using the same setup as in the case-study, where PMM uses $L=2,000$ particles and $L_2=5$ and the Bayesian synthetic likelihoods approach uses $N=3,000$.
We collected the thirty posterior means $\hat{\bm{\theta}}_{b}$ and computed their median biases
and root mean square errors (RMSE)  $\sqrt{\sum_{b=1}^{30}(\hat{\bm{\theta}}_{b}-\bm{\theta}_0)'(\hat{\bm{\theta}}_{b}-\bm{\theta}_0)/30} $. Results obtained with PMM and BSL are similar, except for $\bar{\delta}$, see Table \ref{tab:simulation-results} and Figure \ref{fig:boxplots_meansbias_30simulations}--\ref{fig:boxplots_differencebiases_30simulations}. Figure \ref{fig:boxplots_differencebiases_30simulations} considers boxplots of the difference of the posterior means biases, so that a positive difference for a given parameter implies that the bias is larger (in absolute value) when PMM is used.
Figure \ref{fig:boxplots_differencebiases_30simulations} seems to suggest a slightly better performance of BSL compared to PMM. However, the number of repetitions $B=30$ is too small to be conclusive. Unfortunately, performing a larger simulation study would be computationally very intensive. Running only thirty simulations required about 41 hours with PMM and 7 hours with BSL. In the next section we explore the effect of increasing the number of subjects for a single experiment.

\begin{table}[ht]
\caption{\footnotesize{Simulation study with $M=8$ subjects: true parameter values ($\theta_0$), median bias and RSME using the pseudo-marginal MCMC method (PMM) and Bayesian synthetic likelihoods (BSL).}}
\label{tab:simulation-results}
\centering \scriptsize
\begin{tabular}{llrrrrrrrrr}
\hline
\\
& & $\bar{\beta}$ & $\bar{\delta}$ & $\bar{\alpha}$ & $\gamma$ & $\tau$ & $\sigma_\beta$ & $\sigma_\delta$ & $\sigma_\alpha$ & $\sigma_\varepsilon$\\
\hline
$\theta_0$ & & 3.33  &  1.14  &  0.75 &   1.09  &  1.82  &  0.51   & 0.76 &   0.29  &  0.20\\
PMM\\
& bias &-1.05 &  -0.54  &  -0.21 &  -0.26  & -0.51  & -0.019 &   -0.211  & 0.039 &   0.121\\
& RMSE & 1.09  &  0.60  &  0.21  &  0.27   & 0.51 &   0.039    & 0.213  &  0.045   & 0.129\\
BSL\\
& bias & -0.93  & 0.29 &  -0.28 &  -0.26 &  -0.67  & -0.039  &  -0.213 &  0.001  &  0.143\\
& RMSE & 1.07  &  0.44  &  0.30   & 0.27  &  0.62  &  0.053   & 0.216 &   0.033 &   0.161\\
\hline
\end{tabular}
\end{table}


\subsection{Results using larger sample sizes}\label{sec:larger-simulated-data}

\begin{figure}[ht]
\centering
\subfloat[$\log\bar{\beta}$]{\includegraphics[height=2.9cm,width=5cm]{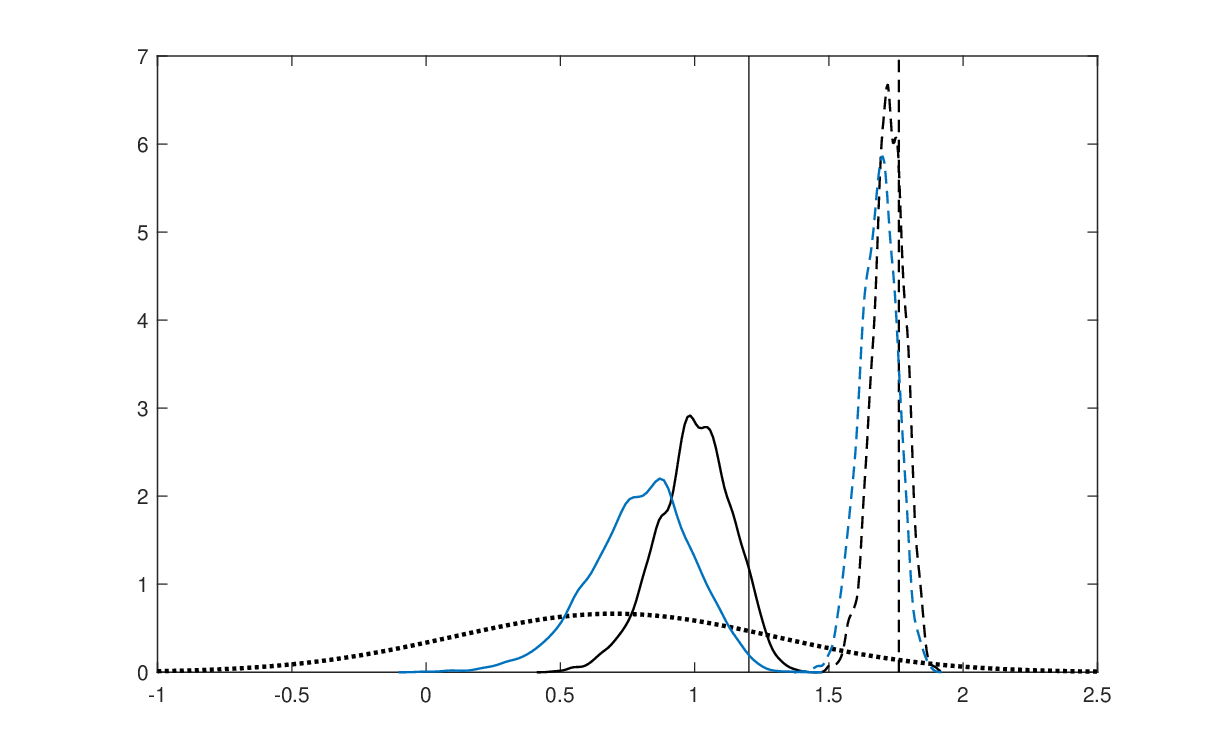}}
\subfloat[$\log\bar{\delta}$]
{\includegraphics[height=2.9cm,width=5cm]{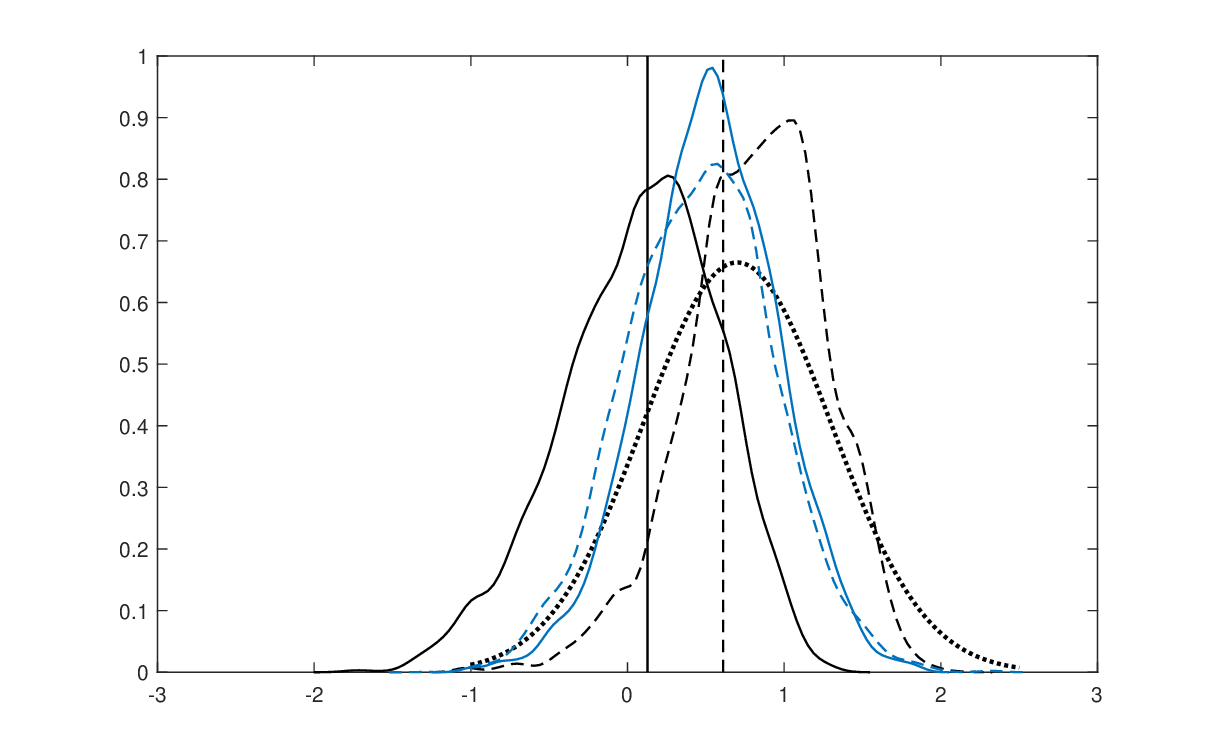}}
\subfloat[$\bar{\alpha}$]{
\includegraphics[height=2.9cm,width=5cm]{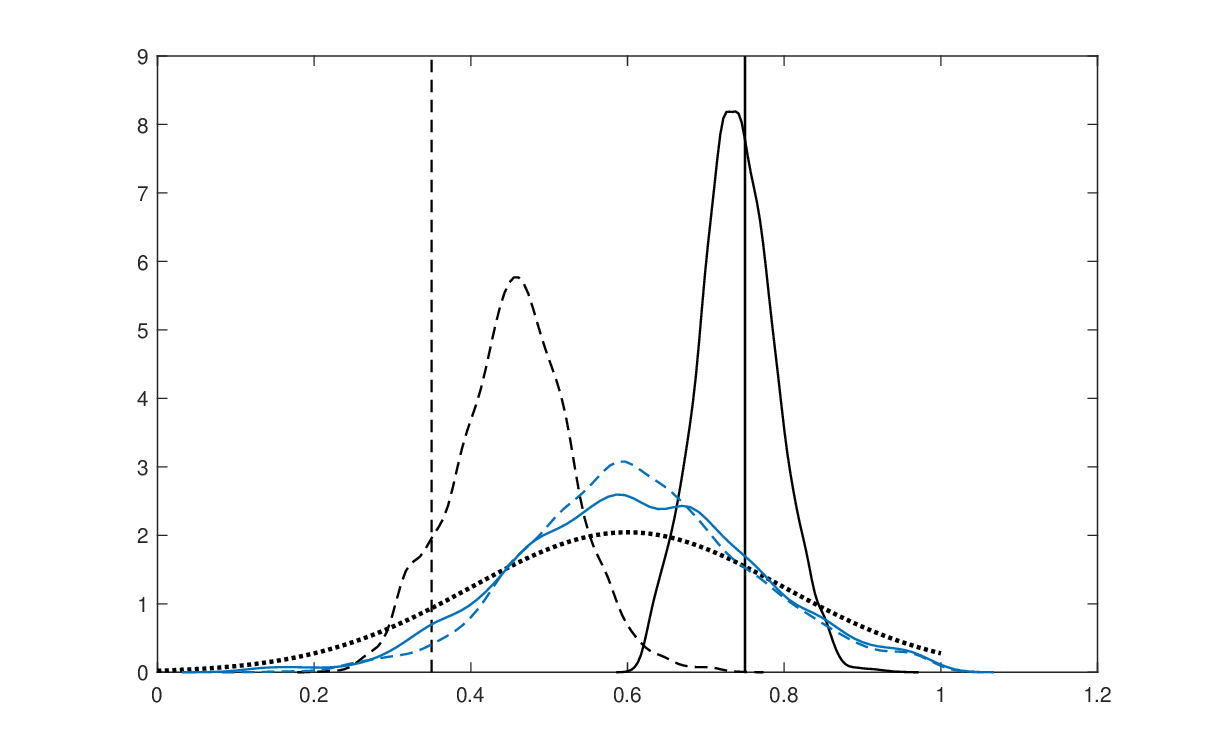}}\\
\subfloat[$\gamma$]{
\includegraphics[height=2.9cm,width=5cm]{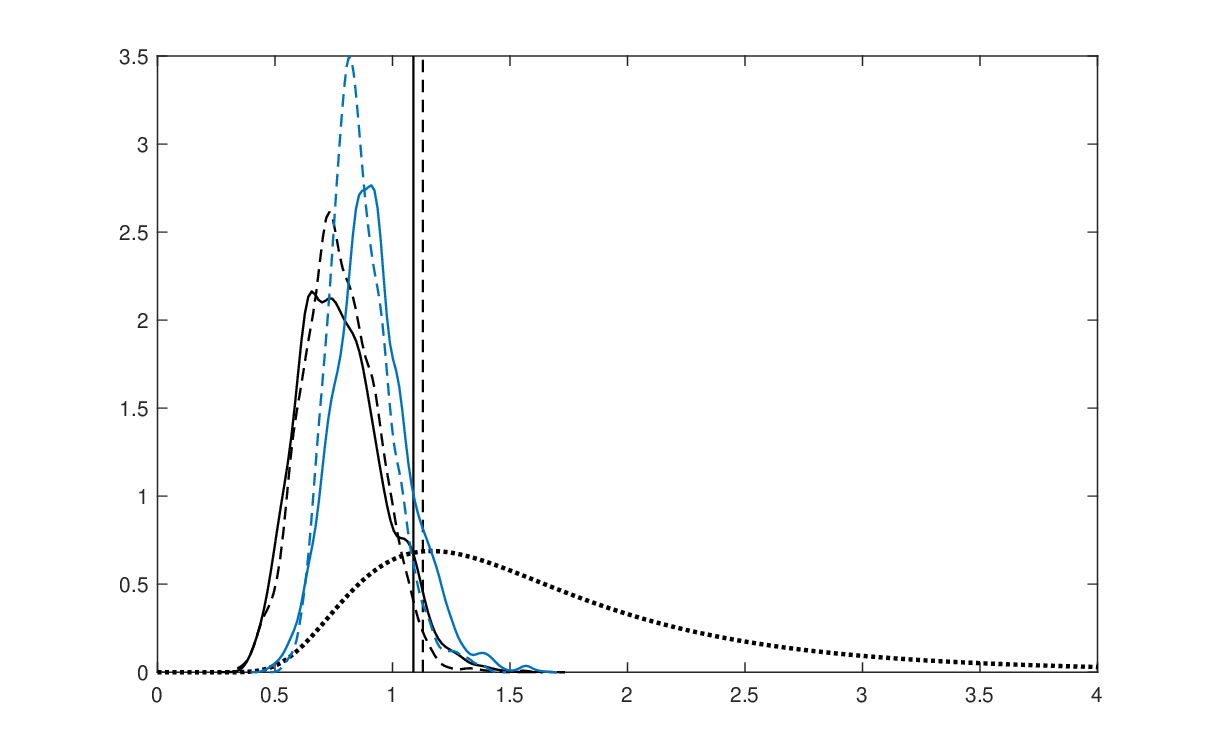}}
\subfloat[$\tau$]{
\includegraphics[height=2.9cm,width=5cm]{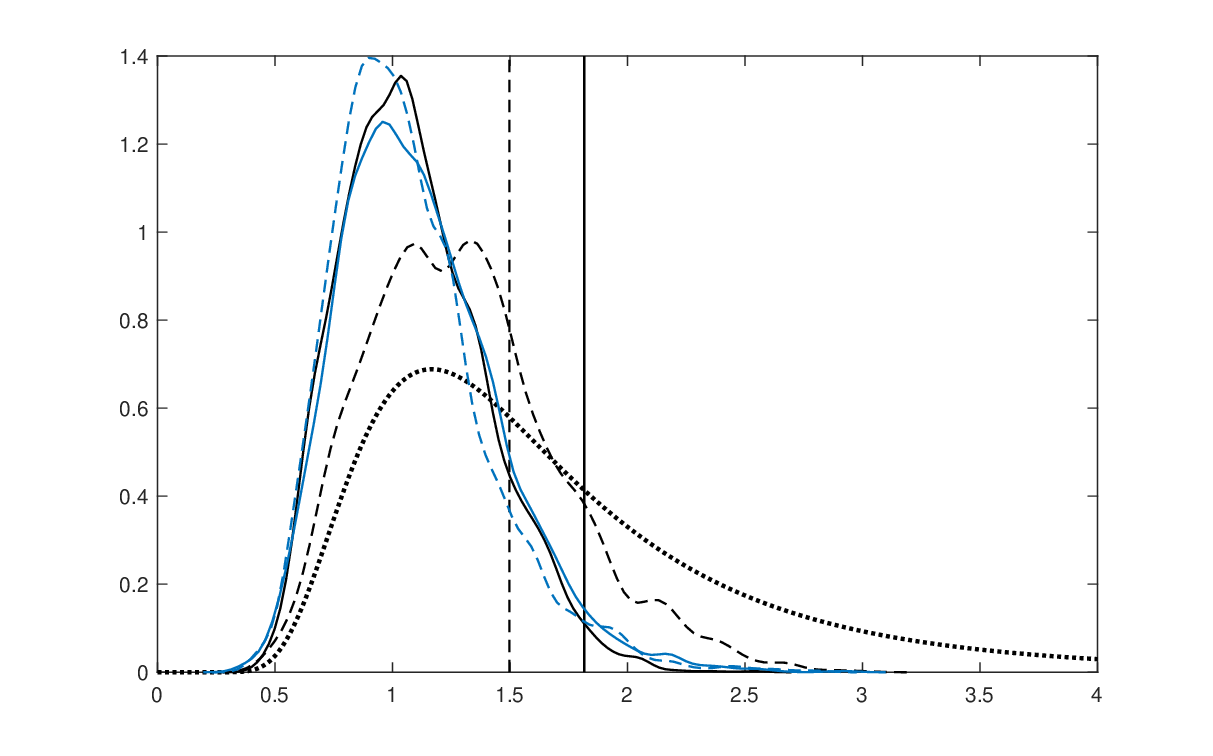}}
\subfloat[$\sigma_{\beta}$]{
\includegraphics[height=2.9cm,width=5cm]{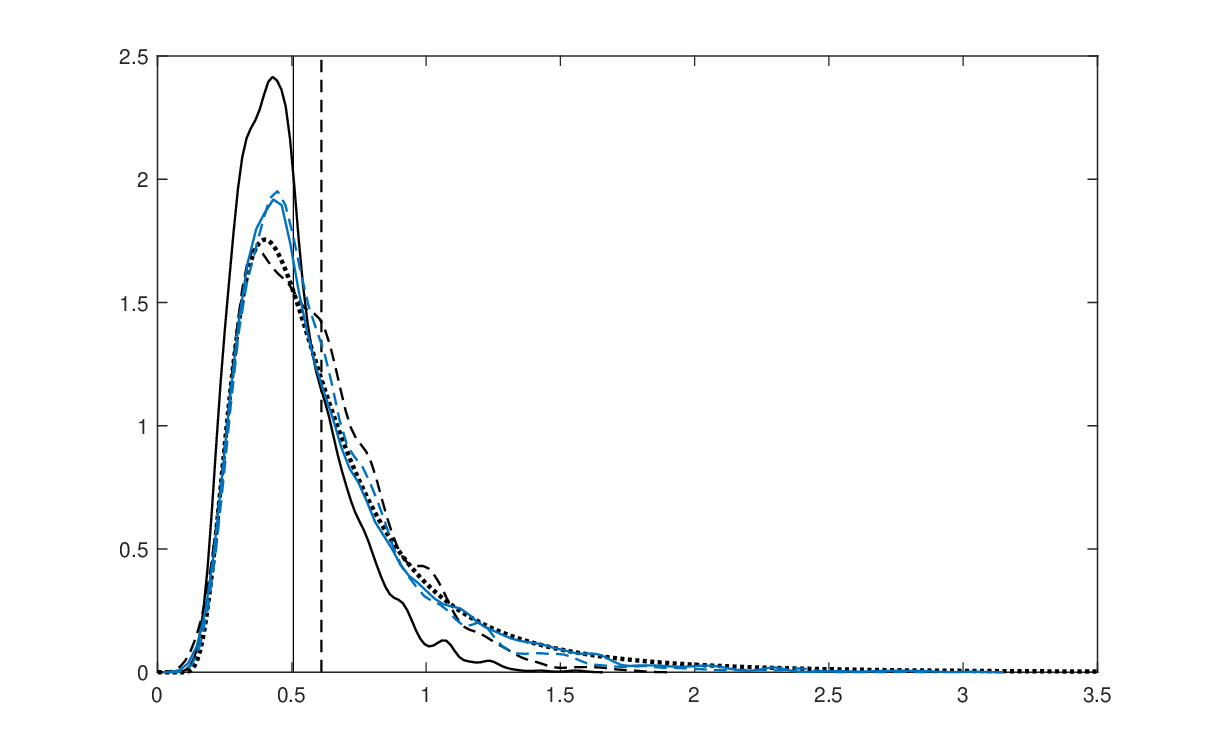}}\\
\subfloat[$\sigma_{\delta}$]{
\includegraphics[height=2.9cm,width=5cm]{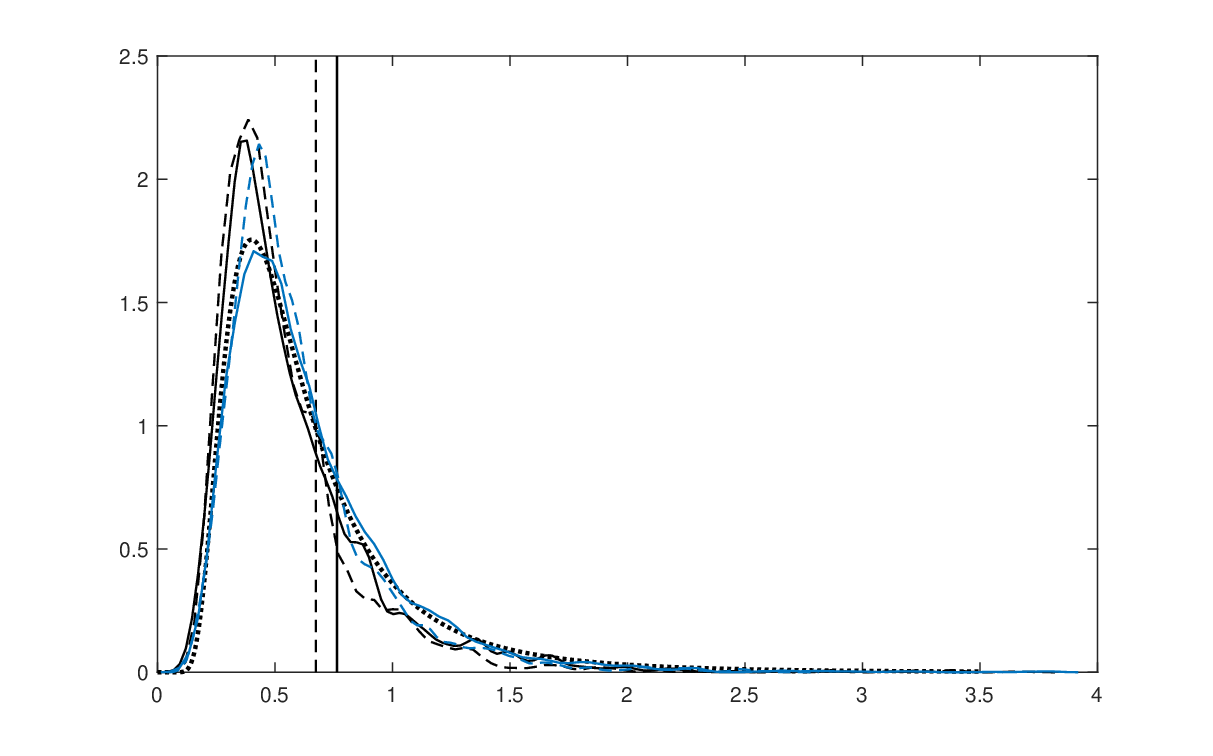}}
\subfloat[$\sigma_{\alpha}$]{
\includegraphics[height=2.9cm,width=5cm]{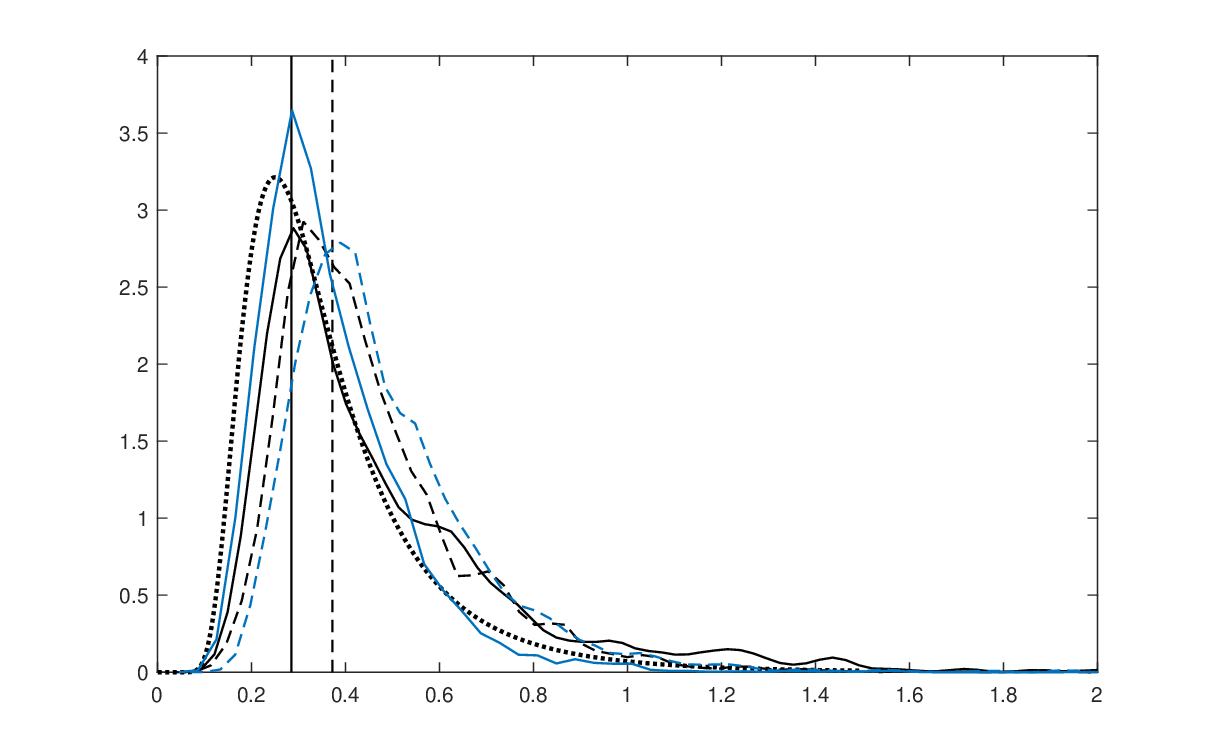}}
\subfloat[$\sigma_{\varepsilon}$]{
\includegraphics[height=2.9cm,width=5cm]{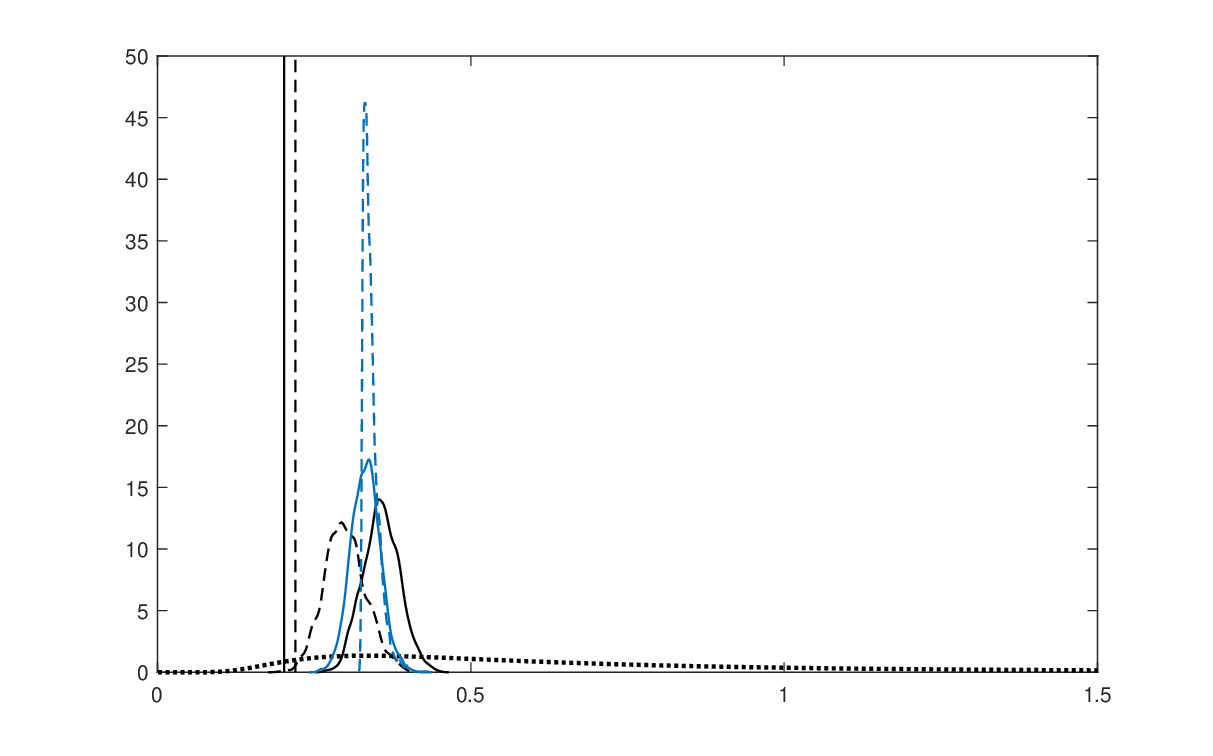}}
\caption{\footnotesize{Posteriors based on simulated data with $M=17$ subjects $\mathcal{D}_1$ (dashed curves) and $\mathcal{D}_2$ (solid lines), obtained with BSL (black) and PMM (blue). Dashed vertical lines are ground-truth parameters for $\mathcal{D}_1$. Solid vertical lines are ground-truth parameters for $\mathcal{D}_2$. Dotted lines are prior densities.}}
\label{fig:simdata-D1D2-17subjects}
\end{figure}

In order to investigate whether the problems we had in identifying the model parameters cease if sample size is increased,
we simulated two datasets corresponding to two groups, each containing $M=17$ subjects but having different treatment efficacies. The model parameters for the first group, $\mathcal{D}_1$, was set to the same values as for the PMM estimates for group 1 in Table \ref{tab:realdata-estimates}, except for $\bar{\alpha}$, here set to $\bar{\alpha} = 0.35$ (low treatment efficacy). The model parameters for the second group, $\mathcal{D}_2$, was set to the same values as for the PMM estimates for group 3 in Table \ref{tab:realdata-estimates}, except for $\bar{\alpha}$, here set to $\bar{\alpha}=0.75$ (high treatment efficacy). We applied both BSL and PMM to analyze the data, initializing the algorithms at the same starting values as in previous sections. Because of the increased sample size (hence a larger spread of data points) we use a larger number of simulations $N=6,000$ with BSL while for PMM we use $L=5,000$ and $L_2=10$. Posteriors are in Figure \ref{fig:simdata-D1D2-17subjects}.

\paragraph{BSL results:}
Figure \ref{fig:simdata-D1D2-17subjects} shows that the mean growth rates $\bar{\beta}$ for the surviving tumor cells are correctly identified, with a higher growth rate for $\mathcal{D}_1$ than for $\mathcal{D}_2$, and the two posteriors for $\bar{\beta}$ are well separated. Compared to the smaller sample size $M=8$ (Figure \ref{fig:group3marginals}) the mean treatment efficacy  $\bar{\alpha}$ is much better identified. Please note that
%, as the posteriors are more concentrated . 
the posterior for $\bar{\alpha}$ in $\mathcal{D}_2$ shows a better identification of the ground truth parameter, than in $\mathcal{D}_1$. This most likely due to the longer trajectories in $\mathcal{D}_2$ (subjects survive longer). 
The separation between the two marginal posteriors suggests that we could obtain more accurate inferences for treatments efficacy in real data using BSL. We further note that the residual variability $\sigma_\varepsilon$ is difficult to identify with high precision. Given that for log-normal data the coefficient of variation is given by $\sqrt{\exp(\sigma^2_\varepsilon)-1}$, and since the four marginals in subfigure \ref{fig:simdata-D1D2-17subjects}(i) suggest an estimate $\hat{\sigma}_\varepsilon\approx 0.3$, we obtain an estimated coefficient of variation of about 0.31. The true coefficient of variation equals 0.20.

\paragraph{PMM results:}
The most notable difference between BSL and PMM is that PMM is  unable to identify $\bar{\alpha}$ both for $\mathcal{D}_1$ and $\mathcal{D}_2$. The posteriors for the remaining model parameters are overall similar. Given that BSL is an approximate methodology, some differences between the two methods are expected.
 We believe that the failure to identify $\bar{\alpha}$ points to the need of improving the way the PMM algorithm is constructed. Currently the random effects $(\log\alpha_i,\log\beta_i,\log\delta_i)$ are simulated from their \textit{unconditional} distributions, e.g. $\log\alpha_i\sim \mathcal{N}_{[0,1]}(\bar{\alpha},\sigma^2_\alpha)$, instead of being simulated from a distribution \textit{conditional} on data. How to construct such distributions is left for future research, but clearly ``blind'' simulation of $\alpha_i$ comes at a cost, as this enters the starting conditions for the dynamics in equation \eqref{e:sde1}. BSL also propagates random effects blindly, but it explicitly encodes information on the population variability via the $\bm{s}^{\mathrm{inter}}$ statistics.

\section{Summary}\label{sec:summary}
We have introduced a new mixed effects model for the analysis of repeated measurements of tumor volumes in mice in a tumor xenograpy study. For each subject the dynamics for the exact, unobservable, tumor volumes are modeled by stochastic differential equations (SDEs), while observed volumes are assumed perturbed with measurement error. The resulting model is a stochastic differential mixed effects model (SDEMEM), which is of state-space type. SDEMEMs provide a very useful representation for repeated measurement data since they are able to distinguish several sources of variability, in the present example: intra-individual temporal variability, biologic variability between subjects, and measurement error (residual variability). We considered two different SDEMEMs: one for unperturbed growth, modelling an untreated control group, and one for tumor (re)growth following an active treatment such as chemo- or radiation therapy. The former is a one-compartment model while the latter is a two compartments model. The two compartments represent the unobserved fraction of tumor cells that has been killed by the treatment ($V^{\mathrm{kill}}$) and the unobserved fraction that has survived the treatment ($V^{\mathrm{surv}}$), respectively. Hence the model extends the classical double exponential model by including random perturbations in the growth dynamics. 

Parameter inference for the SDEMEM is difficult for several reasons. One is the intractability of the likelihood function. Another reason is that model parameters are difficult to identify, since data consist of noisy measurements of the total tumor volume $V=V^{\mathrm{surv}}+V^{\mathrm{surv}}$, not the separate compartments. Finally, most tumor xenograft studies are performed with small sample sizes. We have considered methods for exact and approximate Bayesian inference to overcome this. In particular, we have compared approximate Bayesian inference using the synthetic likelihood (BSL) approach to exact  Bayesian inference using a pseudo-marginal method (PMM). BSL bases inference on the likelihood function of normally distributed summary statistics, instead of the intractable likelihood function of the actual data. The efficiency of the resulting estimator relies on the choice of the summary statistics. For the application to SDEMEMs we advocated the use of subject specific summaries, which can be further comprised over groups by taking averages and computing covariances.  
In an application to a tumor xenography study with two active treatment groups and an untreated control, comprising  data from 5-8 subjects in each group, we found that inference results produced by BSL are similar to those from PMM, indicating that our choice of summary statistics was appropriate. The small sample bias was similar between the two methods. A further advantage of synthetic likelihoods is that, unlike exact particle-based inference, it can be applied to models other than the state-space type, thus inference could be extended to other stochastic growth rate models than state-space SDEMEMs. Also, results obtained with BSL are quite robust to the simulation setup (i.e. to the number of simulated datasets per MCMC iteration), whereas our specific implementation of the PMM seems much more sensitive to changes in the number of considered particles (see supplementary material). Improvements with PMM could be achieved with further research on simulating random effects conditionally to data.

A finding from the case study was that larger sample sizes are needed to identify all model parameters and obtain accurate estimates of the treatment contrasts. 
This was confirmed in a simulation study considering eight and seventeen subjects, where BSL was able to identify treatment contrasts with seventeen subjects in each group while PMM was not.

We have compared the fit of the SDEMEM for the two treatment groups in the case study with that of a ODE mixed-effects model, assuming no within-subject variation in growth and elimination rates over time. The ODE mixed-effects model appears to fit the case study data well, but parameters have larger uncertainties and residual variation is much larger than for the SDEMEM. Unfortunately, the small sample sizes prevent us from determining which model is truly the better. Hence, overall realistic modeling of tumor growth dynamics in response to treatment remains an open question.
Although we recommend larger sample sizes for obtaining valid statistical inference, Bayesian inference may still be used to perform exploratory analyses in small scale experiments. In the latter case, judicious informative priors based on subject matter expertise may compensate for the otherwise too small sample size.

In conclusion, SDEMEMs allow for mechanistic modeling of tumor growth and response to treatment including natural sources of variability. These may be useful for power calculations and optimal design, even in studies where more robust statistical methods are preferred for confirmatory data analysis.

\section*{Acknowledgements}
We are grateful for fruitful comments by three anonymous reviewers and the associate editor.
Research was partially supported by the Swedish Research Council (VR grant 2013-05167). We thank the research team at the Center for Nanomedicine and Theranostics (DTU Nanotech, Denmark) for providing the data for the case study and for introducing us to the problem of making inference from tumor xenography experiments.

\bibliographystyle{abbrvnat}
\bibliography{biblio}

\newpage

\input{supplementary.tex}

\end{document}

%% file: supplementary.tex
\begin{center}
\textbf{SUPPLEMENTARY MATERIAL}
\end{center}

\subsection*{Auxiliary particle filter for mixed-effects state-space models}
In the algorithm below we give a version of the auxiliary particle filter (APF), adapted for mixed-effects state-space models (SSM). APF was initially proposed by \cite{pitt1999filtering} to make inference for the latent state of a SSM. For the purpose of obtaining an unbiased approximation of the likelihood function, we consider \cite{pitt2012some}. For the case where the SSM has dynamics driven by a stochastic differential equation with no closed form solution, an appealing proposal function is given in \cite{golightly2011bayesian}.  Same as for algorithm 2 in the main text, we assume a fixed initial state $x_0$, but otherwise sampling of particles $x_0^l\sim p(x_0)$ should be performed. 

\begin{algorithm*}[ht]
\scriptsize
\caption{Auxiliary particle filter (APF) for mixed-effects state-space models}
\begin{algorithmic}
\State \textbf{Input:} positive integers $L$ and $L_2$, a value for $\bm{\theta}$. Set time $t_{0}=0$. Everything that follows is conditional on the current value of $\bm{\theta}$, which is therefore removed from the notation.\\

\textbf{Output:} all the $\hat{p}(\bm{y}_{ij}|\bm{y}_{i,1:j-1})$, $i=1,...,M$; $j=1,...,n_i$.
\For{$i=1,...,M$}
\State draw $\bm{\phi}_i^l\sim p(\bm{\phi}_i)$ for all $l\in\{1,\ldots,L\}$. Form $\bm{x}_{i0}^l:=\bm{x}_{i0}(\bm{\phi}_i^l)$ accordingly and $\bm{x}_{i0}:=(\bm{x}_{i0}^1,...,\bm{x}_{i0}^{L})$. Set normalised weights $\tilde{w}_{i0}^{l}=1/L$ for all $l\in\{1,...,L\}$.
  \If{$j=1$}
  \State  
  $\bar{\bm{x}}_{i1}$:=\texttt{FirstStagePropagate}($\bm{x}_{i0},L,L_2$)
   \State Compute first stage weights $\omega_{i1}^l =p(\bm{y}_{i1}|\bar{\bm{x}}_{i1}^l)\tilde{w}_{i0}^l$.
      \State Normalization:  $\tilde{\omega}_{i1}^l:={\omega}_{i1}^l/\sum_{l=1}^L {\omega}_{i1}^l$. Interpret $\tilde{\omega}_{i1}^l$ as the probability for index $l_{i1}$ associated to $\bar{\bm{x}}_{i1}^l$.
\State Resampling: sample $L$ times with replacement from the probability distribution $\{l_{i1},\tilde{\omega}_{i1}^l\}$. Denote the sampled indeces with $k^1,...,k^L$.
\State Second propagation: sample $\bm{x}_{i1}^l\sim p(\bm{x}_{i1}|\bm{x}_{i,0}^{k^l})$.
\State Compute second stage weights $w_{i1}^l=p(\bm{y}_{i1}|\bm{x}_{i1}^l)/p(\bm{y}_{i1}|\bar{\bm{x}}_{i1}^l)$.
\State Compute $\hat{p}(\bm{y}_{i1})=\biggl(\frac{\sum_{l=1}^L w_{i1}^l}{L}\biggr)\sum_{l=1}^L\omega_{i1}^l$.
\State Normalise: $\tilde{w}_{i1}^{l}:={w}_{i1}^{l}/\sum_{l=1}^L{w}_{i1}^{l}$. 
  \EndIf
 \For{$j=2,...,n_i$}
\State  
  $\bar{x}_{ij}$:=\texttt{FirstStagePropagate}($\bm{x}_{i,j-1},L,L_2$)
   \State Compute $\omega_{ij}^l =p(y_{ij}|\bar{\bm{x}}_{ij}^l)\tilde{w}_{i,j-1}^l$.
      \State Normalization:  $\tilde{\omega}_{ij}^l:={\omega}_{ij}^l/\sum_{l=1}^L {\omega}_{ij}^l$. Interpret $\tilde{\omega}_{ij}^l$ as the probability for index $l_{ij}$ associated to $\bar{\bm{x}}_{ij}^l$.
\State Resampling: sample $L$ times with replacement from the probability distribution $\{l_{ij},\tilde{\omega}_{ij}^l\}$. Denote the sampled indeces with $k^1,...,k^L$.
\State Second propagation: sample $\bm{x}_{ij}^l\sim p(\bm{x}_{ij}|{\bm{x}}_{i,j-1}^{k^l})$.
\State Compute second stage weights $w_{ij}^l=p(\bm{y}_{ij}|\bm{x}_{ij}^l)/p(\bm{y}_{ij}|\bar{\bm{x}}_{ij}^l)$.
\State Compute $\hat{p}(\bm{y}_{ij}|\bm{y}_{i,1:j-1})=\biggl(\frac{\sum_{l=1}^L w_{ij}^l}{L}\biggr)\sum_{l=1}^L\omega_{ij}^l$.
\State Normalise: $\tilde{w}_{ij}^{l}:={w}_{ij}^{l}/\sum_{l=1}^L{w}_{ij}^{l}$. 
  
\EndFor
\EndFor
\\\hrulefill\\

\textbf{Function $\bar{\bm{x}}_{ij}$:=\texttt{FirstStagePropagate}($\bm{x}_{i,j-1}$,$L,L_2$):}\\
\For{$l=1,...,L$}
\State Sample $\bm{x}_{ij}^{l_2}  \sim p(\bm{x}_{ij}|\bm{x}_{i,j-1}^{l})$, for each $l_2\in\{1,...,L_2\}$.
\State Compute $\bar{\bm{x}}_{ij}^{l}:=\sum_{l_2=1}^{L_2} \bm{x}_{ij}^{l_2}/L_2$.
\EndFor
\State Return $\bar{\bm{x}}_{ij}:=(\bar{\bm{x}}_{ij}^{1},...,\bar{\bm{x}}_{ij}^{L})$.
\end{algorithmic}
\end{algorithm*}

 For \textit{each} of the $L$ particles available at time $t_{i,j-1}$, the function \texttt{FirstStagePropagate} propagates forward $L_2$ particles to the next time $t_{i,j}$, then computes the sample mean from the cloud of $L_2$ particles. Based on the results of this preliminary propagation, a second propagation $x_{ij}^l\sim p(x_{ij}|{x}_{i,j-1}^{k^l})$ simulates particles forward starting from those particles that appear to be promising candidates, according to the preliminary ``exploration'' conducted via the first stage propagation (i.e. the promising particles are those having indeces $k^l$ sampled according the first stage weights $\omega$). The obtained approximate likelihood $\hat{p}(\bm{y}|\bm{\theta})=\prod_{i=1}^M\{\hat{p}(\bm{y}_{i1}|\bm{\theta})\prod_{j=2}^{n_1}\hat{p}(\bm{y}_{ij}|\bm{y}_{i,1:j-1},\bm{\theta})\}$ is unbiased (\citealp{pitt2002smooth}, \citealp{pitt2012some}).
  With reference to model (5) in the main text, the notation in the APF algorithm is as follows: the starting total volume is $V_{i0}:=V_i^{\mathrm{surv}}(0)+V_i^{\mathrm{kill}}(0)$ where $V_i^{\mathrm{surv}}(0)=(1-\alpha_i)v_{i0}$ and $V_i^{\mathrm{kill}}(0)=\alpha_iv_{i0}$. Then we have $x_{ij}:=\log V_{ij}$, hence $\bar{x}_{ij}^l:=\sum_{l_2=1}^{L_2} (\log V_{ij}^{l_2})/L_2$, so that $p(y_{ij}|x_{ij}^l)\equiv \mathcal{N}(y_{ij};x_{ij}^l,\sigma^2_\varepsilon)$ and $p(y_{ij}|\bar{x}_{ij}^l)\equiv \mathcal{N}(y_{ij};\bar{x}_{ij}^l,\sigma^2_\varepsilon)$. In general, for any particle and regardless of whether this is $x_{ij}^l:=\log V_{ij}^l$ or $x_{ij}^{l_2}:=\log V_{ij}^{l_2}$, we have that $V_{ij}^l:=(V_{ij}^{\mathrm{kill}})^l+(V_{ij}^{\mathrm{surv}})^l$ (respectively $V_{ij}^{l_2}:=(V_{ij}^{\mathrm{kill}})^{l_2}+(V_{ij}^{\mathrm{surv}})^{l_2}$), that is the indeces $l$ (resp. $l_2$) obtained when resampling the total volumes are used to select the ``surviving'' and ``killed'' states. Finally note that for APF (and similarly for the bootstrap filter in algorithm 2 of the main text) the ``best'' particles for the total volumes are not necessarily the best particles for $V^{\mathrm{surv}}$ and $V^{\mathrm{kill}}$, when these are considered separately.

\subsection*{Considerations for implementing BSL}

For the implementation of the BSL algorithm 3 in the main text, note that multiplicative constants such as the
$c(k,v)$'s appearing in $\hat{p}(\bm{s}|\bm{\theta})$ are independent of $\bm{\theta}$, hence these cancel-out in the likelihood ratio that defines the acceptance
probability. To prevent the MCMC algorithm from reaching a premature
halt, we recommend to set $\hat{p}(\bm{s}|\bm{\theta}):=0$ whenever the
argument of $\psi(\cdot)$ in equation (10) is not a positive
definite matrix (except for the starting value $\bm{\theta}^*$, of
course).

Notice the literature on synthetic likelihoods does not indicate strategies for the identification of informative summary statistics. Procedures for the construction of informative summaries could be borrowed from the approximate Bayesian computation literature, see \cite{blum2013comparative} and \cite{prangle2015summary} for reviews, but these do not ensure Gaussianity of the resulting summaries. 

\subsection*{Gaussianity of the summary statistics}

\begin{figure}[ht]
\centering
\includegraphics[height=8cm,width=10cm]{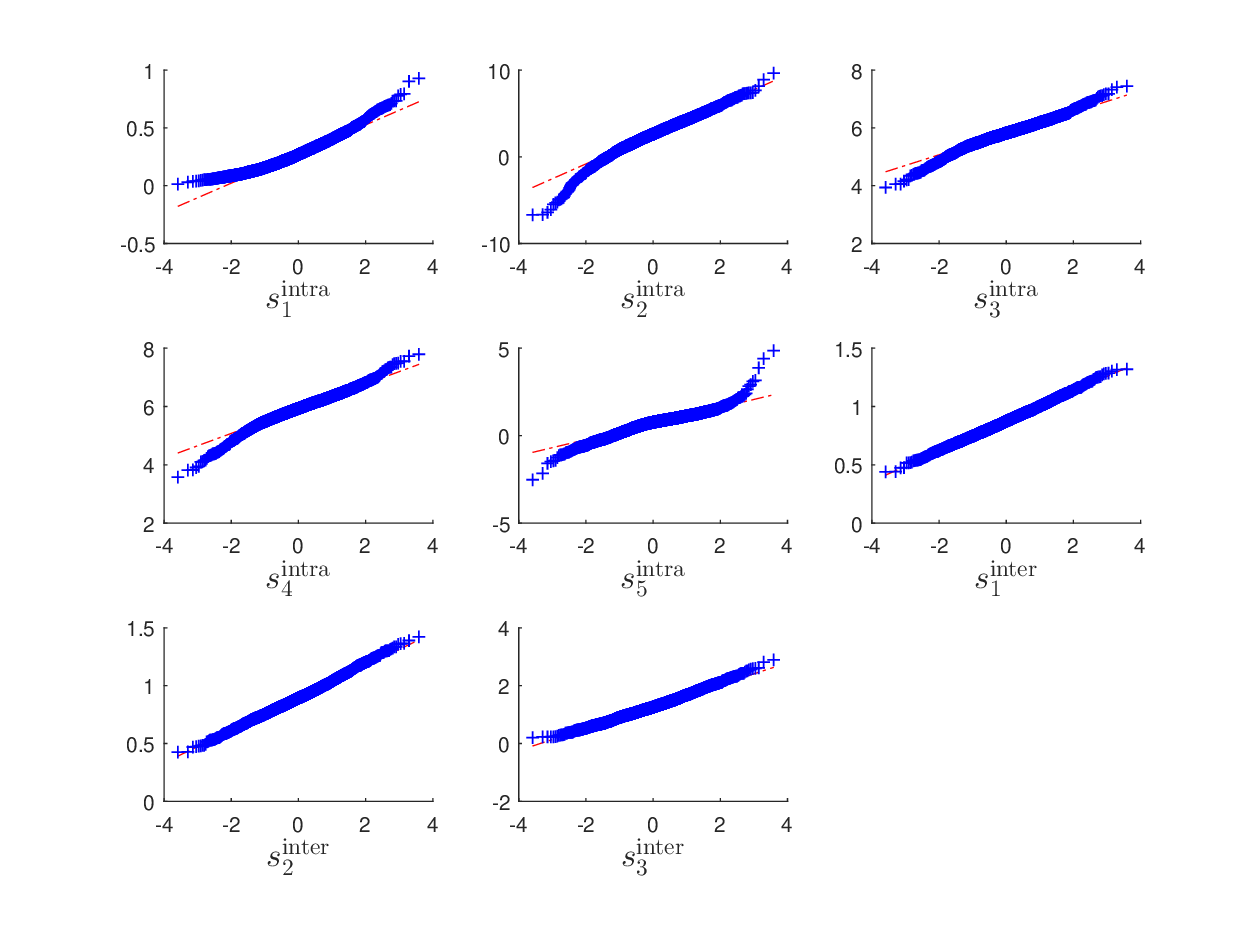}
\caption{\footnotesize{Group 3: normal qq-plots for the intra-individual summary statistics generated for a specific subject from group 3 ($s_1^{\mathrm{intra}},...,s_5^{\mathrm{intra}}$) as well as inter-individual summaries ($s_1^{\mathrm{inter}},,...,s_3^{\mathrm{inter}}$). All summaries have been generated in correspondence of the last simulated parameter draw in the MCMC.}}
\label{fig:group3-synlike-qqplots}
\end{figure}

The following pertains results in section 5.2 of the main text: Figure \ref{fig:group3-synlike-qqplots} gives normal qq-plots of simulated summaries corresponding to the last draw generated with BSL. All summaries appear fairly close to normality.

\subsection*{Wider priors for section 5.2}

\begin{figure}[ht]
\centering
\subfloat[$\log\bar{\beta}$]{\includegraphics[height=2.9cm,width=5cm]{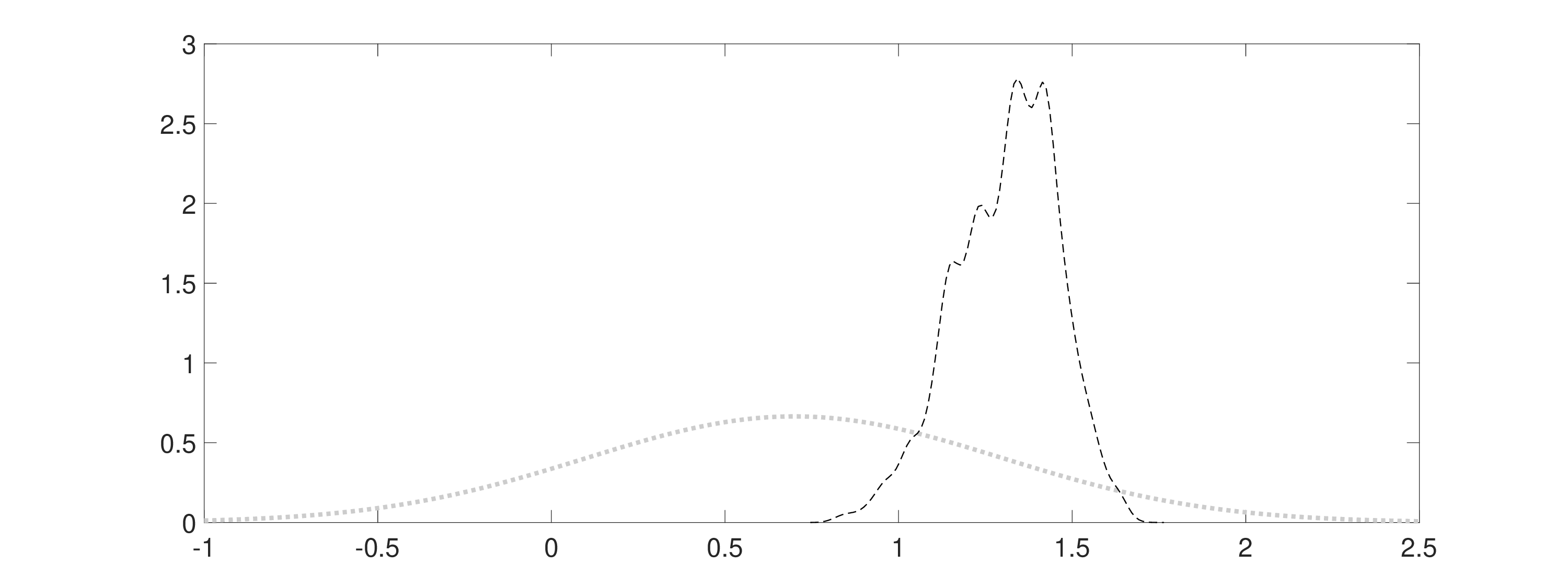}}
\subfloat[$\log\bar{\delta}$]
{\includegraphics[height=2.9cm,width=5cm]{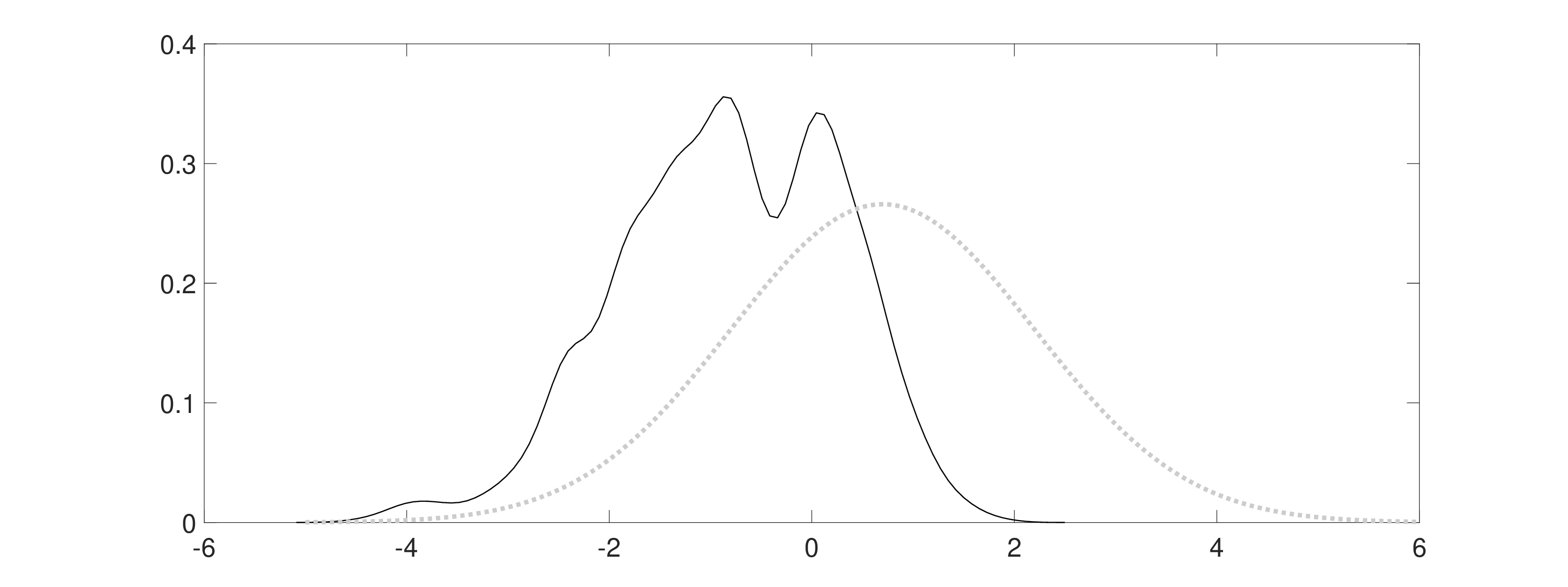}}
\subfloat[$\bar{\alpha}$]{
\includegraphics[height=2.9cm,width=5cm]{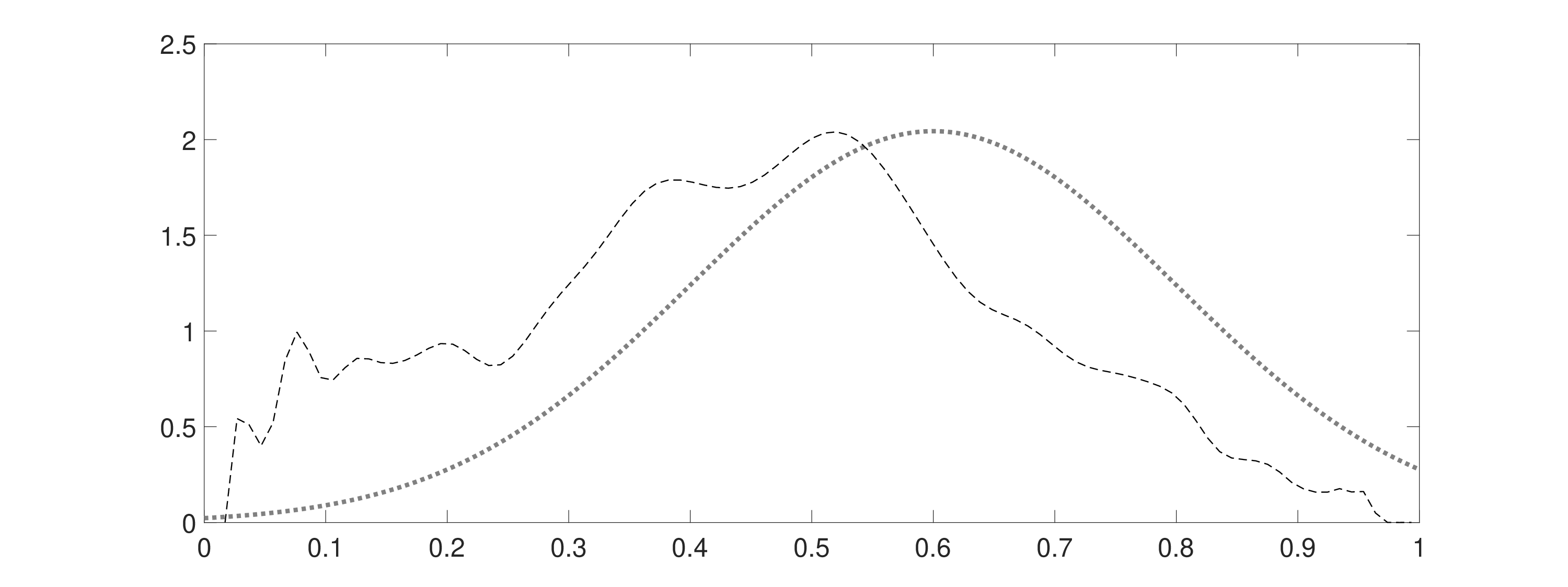}}\\
\subfloat[$\gamma$]{
\includegraphics[height=2.9cm,width=5cm]{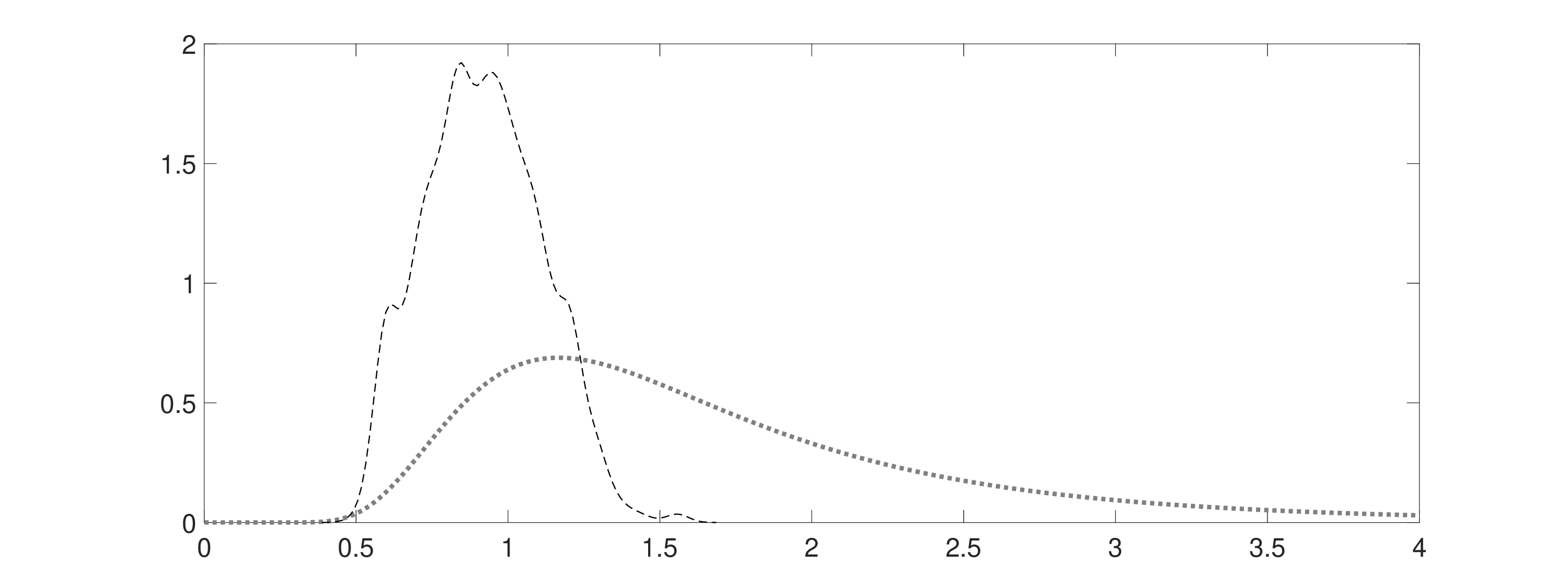}}
\subfloat[$\tau$]{
\includegraphics[height=2.9cm,width=5cm]{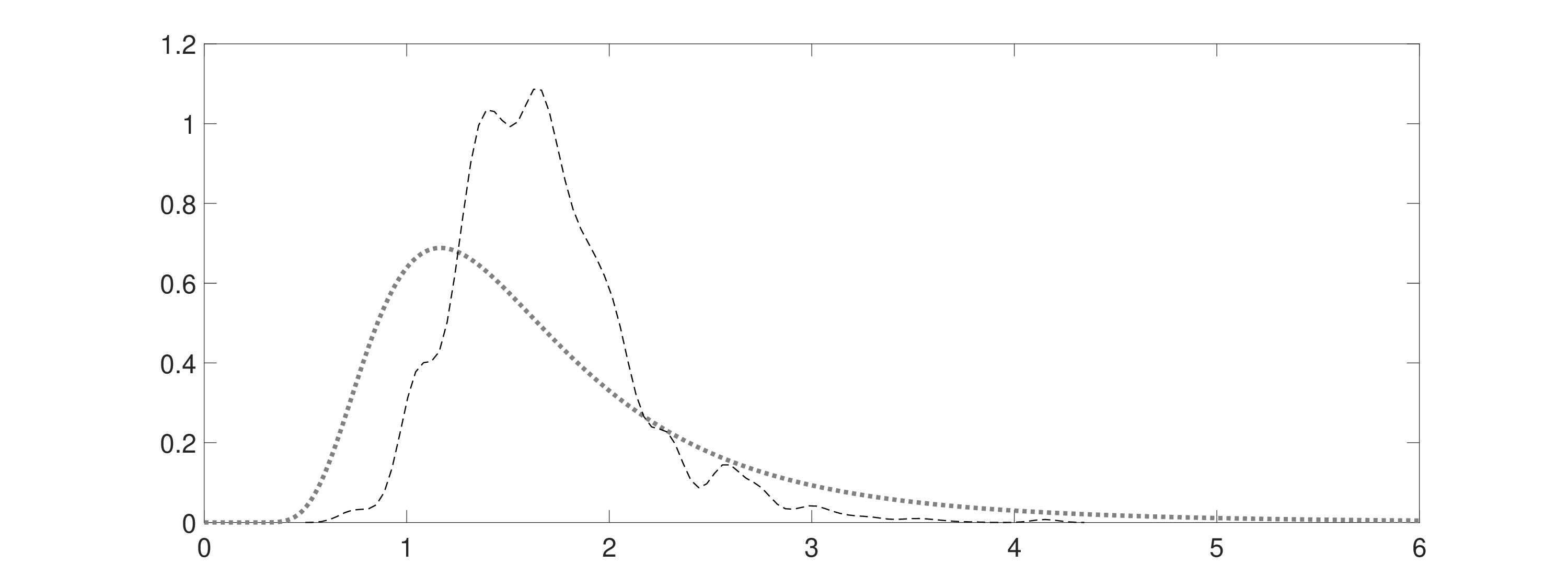}}
\subfloat[$\sigma_{\beta}$]{
\includegraphics[height=2.9cm,width=5cm]{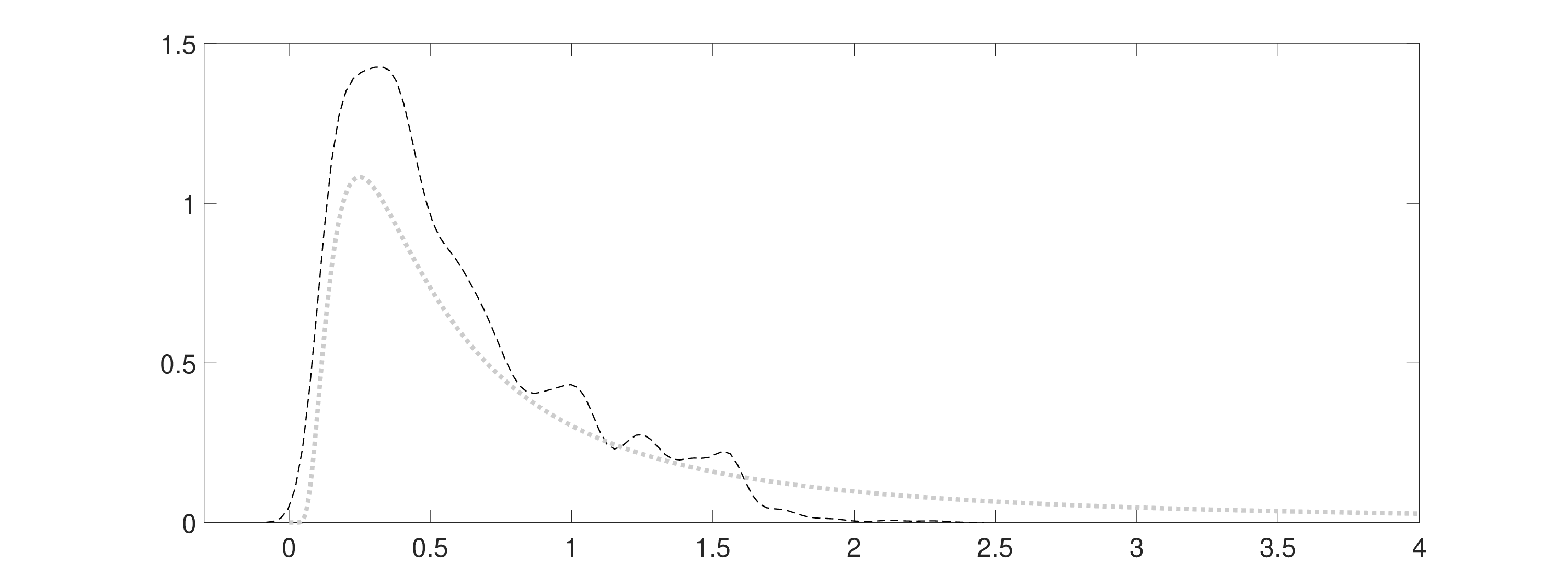}}\\
\subfloat[$\sigma_{\delta}$]{
\includegraphics[height=2.9cm,width=5cm]{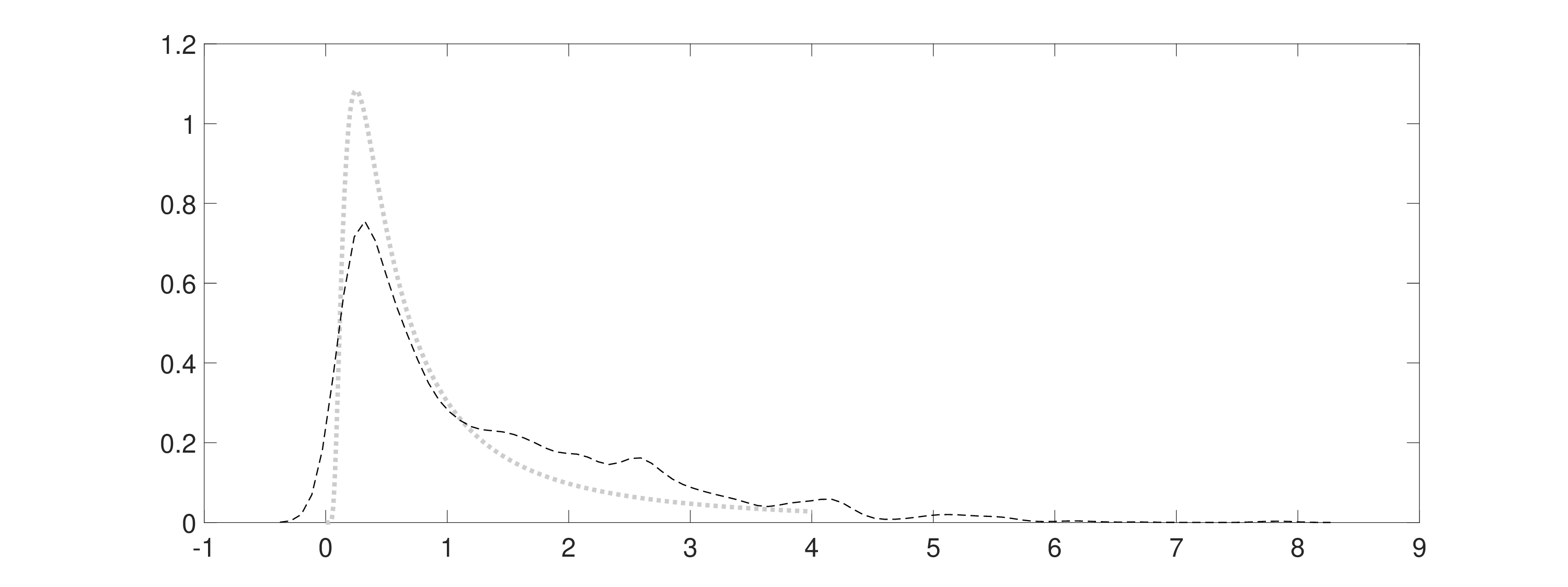}}
\subfloat[$\sigma_{\alpha}$]{
\includegraphics[height=2.9cm,width=5cm]{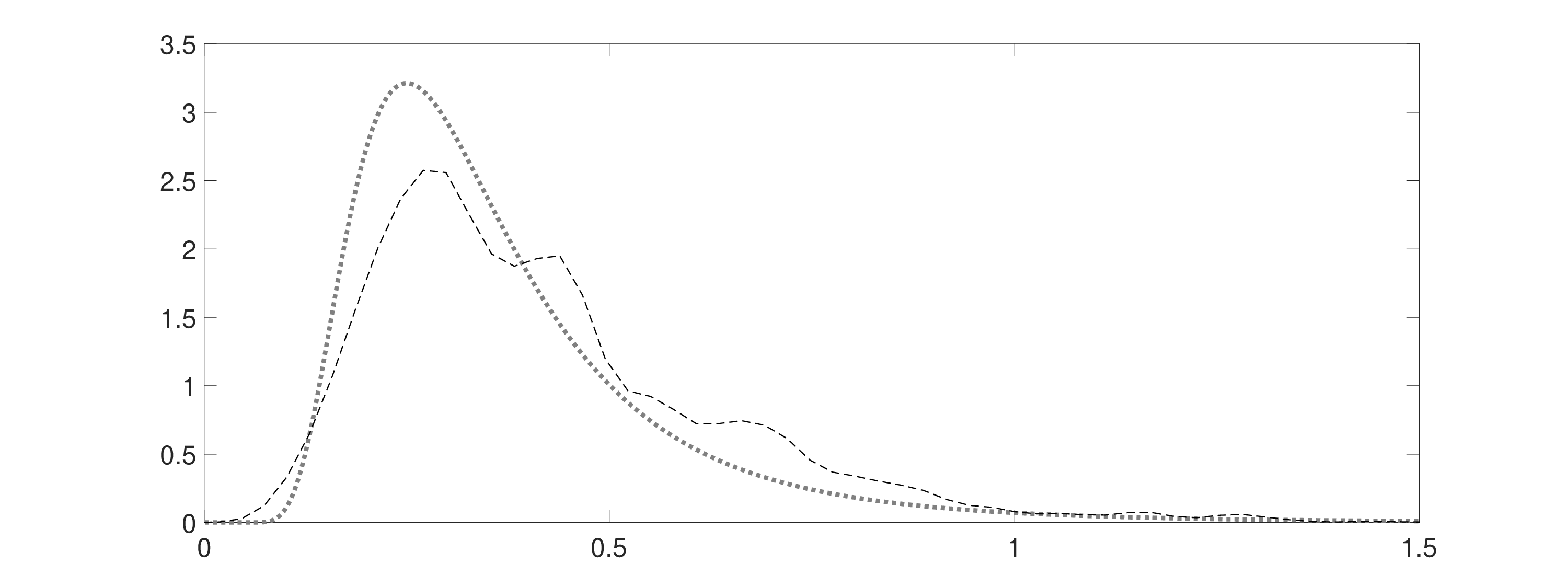}}
\subfloat[$\sigma_{\varepsilon}$]{
\includegraphics[height=2.9cm,width=5cm]{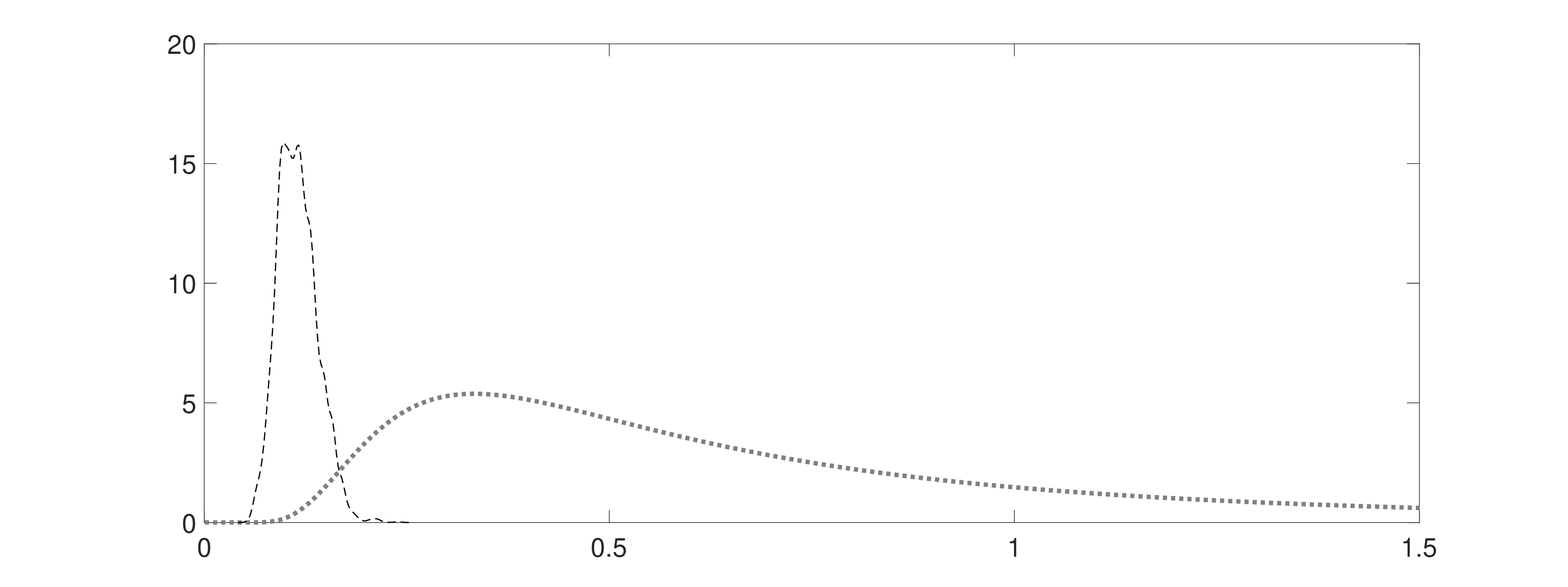}}
\caption{\footnotesize{Treatment group 3 using less informative priors for $\log\bar{\delta}$, $\sigma_{\beta}$ and $\sigma_{\delta}$. Marginal posteriors obtained with synthetic likelihoods (dashed) and prior densities (dotted gray).}}
\label{fig:group3marginals-widerpriors}
\end{figure}

Here we report results obtained by running BSL as in section 5.2, except for considering less informative priors for $\log\bar{\delta}$, $\sigma_{\beta}$ and $\sigma_{\delta}$, namely here we used $\log\bar{\delta}\sim
\mathcal{N}(0.7,1.5^2)$, $\sigma_{\beta}\sim InvGam(1,0.5)$, $\sigma_{\delta}\sim InvGam(1,0.5)$.
All remaining priors are the same as in section 5.2. This way most of the prior mass for $\log\bar{\delta}$ is contained in [-4,4] (vs [-1.0,2.5] in Figure 2 of the main article), and for $\sigma_{\beta}$ and $\sigma_{\delta}$ most of the prior mass is in (0,4] (vs (0,2]).
Results are in Figure \ref{fig:group3marginals-widerpriors}. We can tell that both $\sigma_{\beta}$ and $\sigma_{\delta}$ to some extent follow their priors (similarly to Figure 2 in the main article), that is the available amount of data do not seem to contain enough information to allow estimation of these parameters. Regarding the posterior for the (log-)elimination rate $\log\bar{\delta}$, we notice a major shift towards smaller values (as compared to the counterpart in Figure 2 in the main text) however the spread of the posterior has also increased. It seems that also this parameter is sensitive to its prior, at least in the small sample scenario. 

\section*{Posterior predictive checks}

We discuss posterior predictive checks for both the pseudo-marginal Metropolis algorithm (PMM) and for Bayesian synthetic likelihoods (BSL).
Performing posterior predictive checks when the likelihood is approximated using particle filters, as in the PMM algorithm, is less immediate. Denote with $y^*$ a simulated realization from the hypothesized data-generating model, that is $y^*\sim p(y^*|\theta)$ where $p(y|\theta)$ denotes the likelihood function and $y$ the observed data. The posterior predictive distribution $p(y^*|y)$ (e.g. \citealp{gelman2013bayesian} chapter 6) is given by
\[
p(y^*|y)=\int p(y^*|\theta)\pi(\theta|y)d\theta
\]
where $\pi(\theta|y)$ is the posterior of $\theta$. We would like to first simulate $\theta^*\sim \pi(\theta|y)$ (which may be obtained from the PMM output, after burnin), and next $y^*\sim p(y^*|\theta^*)$.  To compare the predicted distribution of $y^*$ with the observed data $y$, we introduce some summary statistics $T(\cdot)$, and compare $T(y)$ to the distribution of $T(y^*)$ (not to be confused with the summaries used in the synthetic likelihood approach, $s(\cdot)$).

Clearly, since $p(y|\theta)$ is unknown in closed form, it must be approximated, for example using algorithms 2-4 to return $\hat{p}(y|\theta)$. In our case, because of the dependence of data $y$ on unobservables $(X,\phi)$, and because of the multidimensional integral in (7) each likelihood term has an unknown distribution and we cannot sample a $y^*$ from $\hat{p}(y|\theta)$. A possibility, which we leave for the interested reader, is to sample for the generic subject $i$ an $X^*_i$ from the filtering distribution $p(X_i|y_i;\theta^*)$, where $\theta^*$ is a draw obtained via PMM, then form $y^*_{ij}=g(X^*_{ij},\varepsilon^*_{ij})$, with $\varepsilon_{ij}\sim N(0,{\sigma^*}^2_{ij})$ following the notation in model (6). Here $X^*_i$ is a trajectory that it is possible to obtain as a by-product of either algorithm 2 or 4, by sampling a single index $l'$ from the cloud of particles obtained at the last time point $t_{in_i}$ then follow the genealogy of $l'$ backwards up to time $t_0=0$. The sequence of ancestors of the particle $l'$ provides a single path from $p(X_i|y_i;\theta^*)$.

For BSL the approach is much simpler as we can consider $T(y)\equiv s(y)$. In this case we have 
\[
p(s^*|s)=\int p(s^*|\theta)\pi(\theta|s)d\theta
\]
where $s:=s(y)$. To sample from $p(s^*|s)$ we plug  a draw $\theta^*\sim \pi(\theta|s)$ into our model simulator to obtain a corresponding  $y^*\sim p(y|\theta^*)$, and finally take $s^*=s(y^*)$, where clearly $s^*\sim p(s|\theta^*)$. If we repeat the procedure for all the posterior draws $\theta^*$ returned by BSL, we can then produce e.g. histograms from the ensemble of all drawn $s^*$, thus obtaining an approximation to $p(s^*|s)$.

Notice in particular that $p(s|\theta)$ is the true (albeit analytically unknown) likelihood of the summary statistics, not its BSL approximation. This is because here we are interested in evaluating the performance of our assumed data-generating model (conditionally on posterior draws obtained via BSL), and not in testing BSL itself.

\subsection*{Additional posterior predictive checks for section 5.2.1}

\begin{figure}[ht]
\centering
\includegraphics[width=14cm,height=10cm]{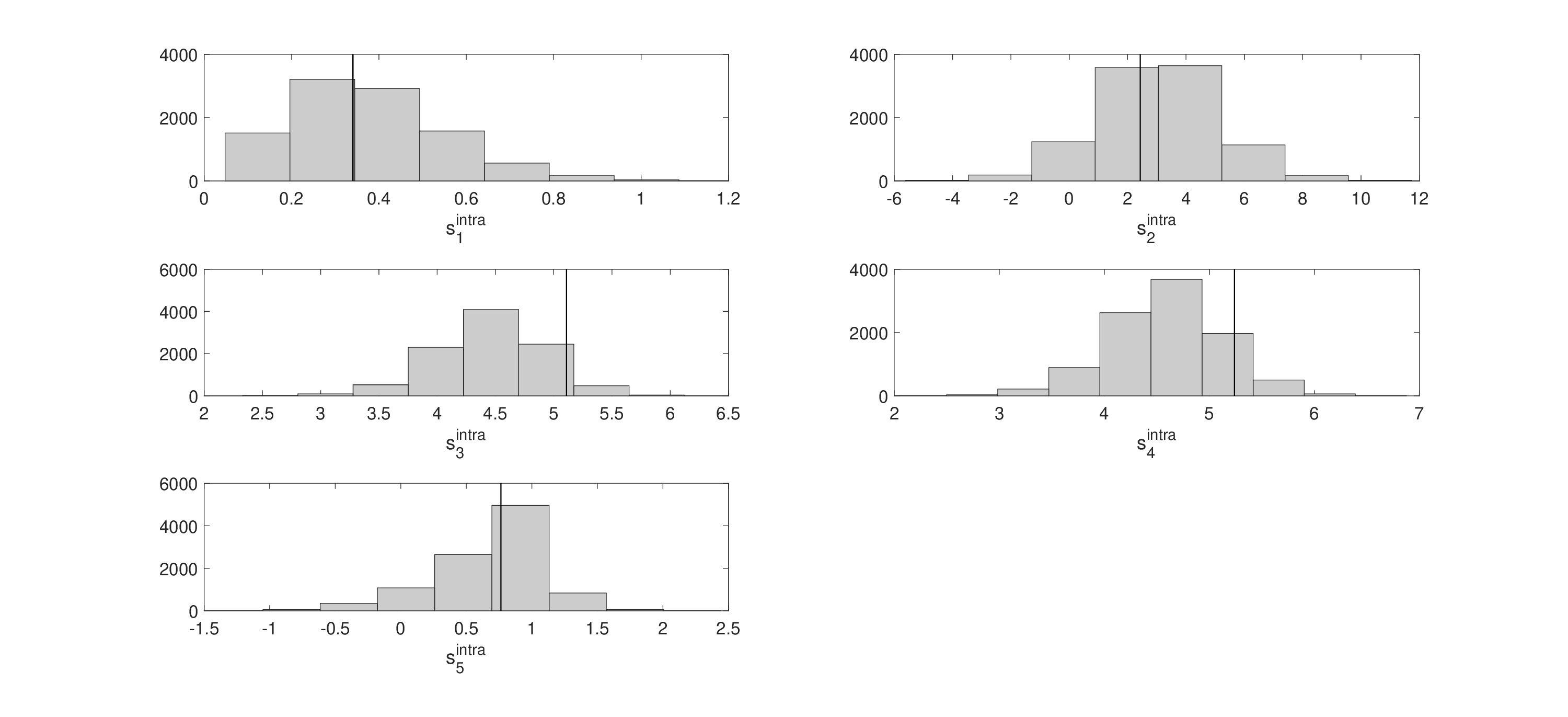}
\caption{\footnotesize{Posterior predictive checks for group 3 generated using draws from BSL. Distribution of the simulated statistics for the intra-subjects variability for subject 2: $s^{\mathrm{intra}}_1$ and $s^{\mathrm{intra}}_2$ (top), $s^{\mathrm{intra}}_3$ and $s^{\mathrm{intra}}_4$ (middle) and $s^{\mathrm{intra}}_5$ (bottom). Vertical lines mark the values for the corresponding statistics from the observed data.}}
\label{fig:group3-synlike-postpredchecks_S-INTRA_subj2}
\end{figure}

\begin{figure}[ht]
\centering
\includegraphics[width=14cm,height=10cm]{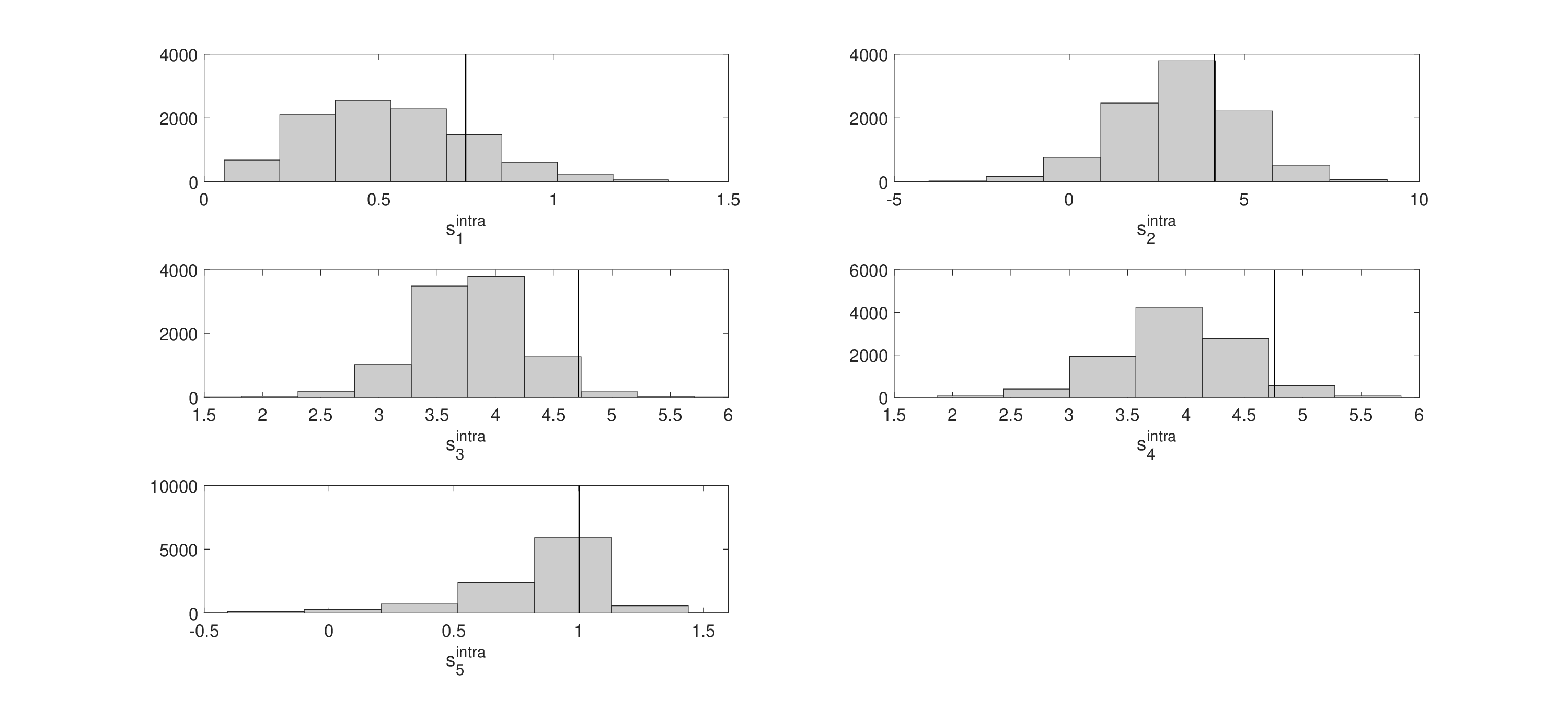}
\caption{\footnotesize{Posterior predictive checks for group 3 generated using draws from BSL. Distribution of the simulated statistics for the intra-subjects variability for subject 3: $s^{\mathrm{intra}}_1$ and $s^{\mathrm{intra}}_2$ (top), $s^{\mathrm{intra}}_3$ and $s^{\mathrm{intra}}_4$ (middle) and $s^{\mathrm{intra}}_5$ (bottom). Vertical lines mark the values for the corresponding statistics from the observed data.}}
\label{fig:group3-synlike-postpredchecks_S-INTRA_subj3}
\end{figure}

Here we consider further plots for posterior predictive checks (PPC) produced when using BSL on group 3, see section 5.2.1 in the manuscript. There we have reported the PPC for inter-subjects variability (Figure 6 in the manuscript) and the individual intra-subject variability pertaining to subject 1. Here we report further plots for intra-subject variability for two additional subjects, namely subject 2 and 3. See Figures \ref{fig:group3-synlike-postpredchecks_S-INTRA_subj2}--\ref{fig:group3-synlike-postpredchecks_S-INTRA_subj3}, showing that the observed summaries are plausible according to the estimated model. Also in this case, most observed summaries are consistent with those produced by the prior predictive distribution.

\subsection*{Small study on varying $L$ and $N$}

We reconsider the real data for group 3 analysed in section 5, to assess the sensitivity of the inference to variations in the number of simulated particles $L$, when using the auxiliary particle filter pseudo marginal method (PMM), and to variations in the number of simulated datasets $N$ when using Bayesian synthetic likelihoods (BSL). For PMM we always keep the number $L_2$ of particles propagated in the ``first stage'' constant to $L_2=5$. See algorithm 4 in the main text for details. Results of this study are in Table \ref{tab:realdata-varyingsimulations} and are all produced using $R=20,000$ MCMC iterations (first 6,000 discarded as burnin). The first column in Table \ref{tab:realdata-varyingsimulations} reports the results from the corresponding column in Table 3 from the main text, that were obtained with $(L,N)=(2000,3000)$. Then, in the second column we increase by 50\% the values of $(L,N)$. Finally, in the third column we reduce by 50\% the values of $(L,N)$ from the first column.

We notice that PMM returns different results when varying the value of $L$. This is not only affecting the posterior variability, but also the location of the mean. See in particular $\bar{\beta}$, $\bar{\delta}$, $\tau$. Results from BSL are much more stable to changes in $N$. In particular, it is reassuring that the value of $N$ used to produce results in section 5 ($N=3,000$) does not produce substantially different results when increased to $N=4,500$. Instead, using $N=1,500$ is not enough to produce a chain for $\bar{\alpha}$ that is able to reach apparent stationarity (plot not reported), not even if we use $R=40,000$ iterations; hence $N=1,500$ would be too small in this case. Finally, for the middle column we note a stronger similarity between results across methods, compared to results from the other columns.

\begin{table}[ht]
\caption{\footnotesize{Posterior means and 95\% posterior intervals: for each parameter we first report exact Bayesian inference using the auxiliary particle filter PMM and then BSL. PMM always uses $L_2=5$ for all cases.}}
\label{tab:realdata-varyingsimulations}
\centering
\begin{tabular}{rrrr}
\hline
{} & $(L,N)=(2000,3000)$ & $(L,N)=(3000,4500)$ & $(L,N)=(1000,1500)$\\
\hline\\
$\bar{\beta}$ & 3.33 [2.07,4.64] & 3.75 [2.76,4.95] & 2.58 [1.33,3.81]\\
& 3.93 [2.93,5.04] & 3.92 [2.92,5.00] & 4.11 [3.22,5.16]\\ 
$\bar{\delta}$& 1.14 [0.40,2.32] & 1.59 [0.60,3.31] & 1.72 [0.54,4.16]\\
& 1.52 [0.43,3.68] & 1.50 [0.50,3.56] & 1.45 [0.51,3.13]\\
$ \bar{\alpha}$ & 0.60 [0.31,0.91] & 0.55 [0.22,0.86] & 0.59 [0.34,0.88]\\
& 0.47 [0.17,0.84] & 0.44 [0.12,0.84] & 0.69 [0.47,0.91]\\
$\gamma$& 1.09 [0.70,1.52] & 1.02 [0.67,1.41] & 1.26 [0.89,1.68]\\
& 0.92 [0.56,1.32] & 0.97 [0.61,1.36] & 0.95 [0.64,1.45]\\
$\tau$& 1.82 [1.02,2.63]  & 2.06 [1.28,3.00] & 2.28 [1.37,3.45]\\
& 1.75 [1.03,2.64] & 1.70 [0.99,2.61] & 1.61 [0.95,2.39]\\
$\sigma_{\beta}$& 0.51 [0.19,1.67] & 0.68 [0.23,1.74] & 0.59 [0.22,1.42]\\
&  0.59 [0.23,1.28] & 0.54 [0.21,1.22] & 0.55 [0.24,1.15]\\
$\sigma_{\delta}$ & 0.76 [0.26,2.23] & 0.55 [0.22,1.22] & 0.60 [0.23,1.37]\\
& 0.71 [0.25,1.91] & 0.75 [0.25,2.17] & 0.76 [0.24,2.33]\\
$\sigma_{\alpha}$ & 0.29 [0.15,0.48] & 0.40 [0.17,0.88] & 0.41 [0.18,0.90]\\
& 0.43 [0.14,1.16] & 0.47 [0.14,1.50] & 0.47 [0.18,1.14]\\
$\sigma_{\varepsilon}$& 0.20 [0.19,0.23] & 0.20 [0.19,0.22] & 0.20 [0.19,0.21]\\
& 0.11 [0.07,0.17] & 0.11 [0.07,0.16] & 0.12 [0.07,0.18]\\
\hline
\end{tabular}
\end{table}

\subsection*{Comparison between the bootstrap filter and the auxiliary particle filter}

\begin{figure}[ht]
\centering
\includegraphics[scale=0.8]{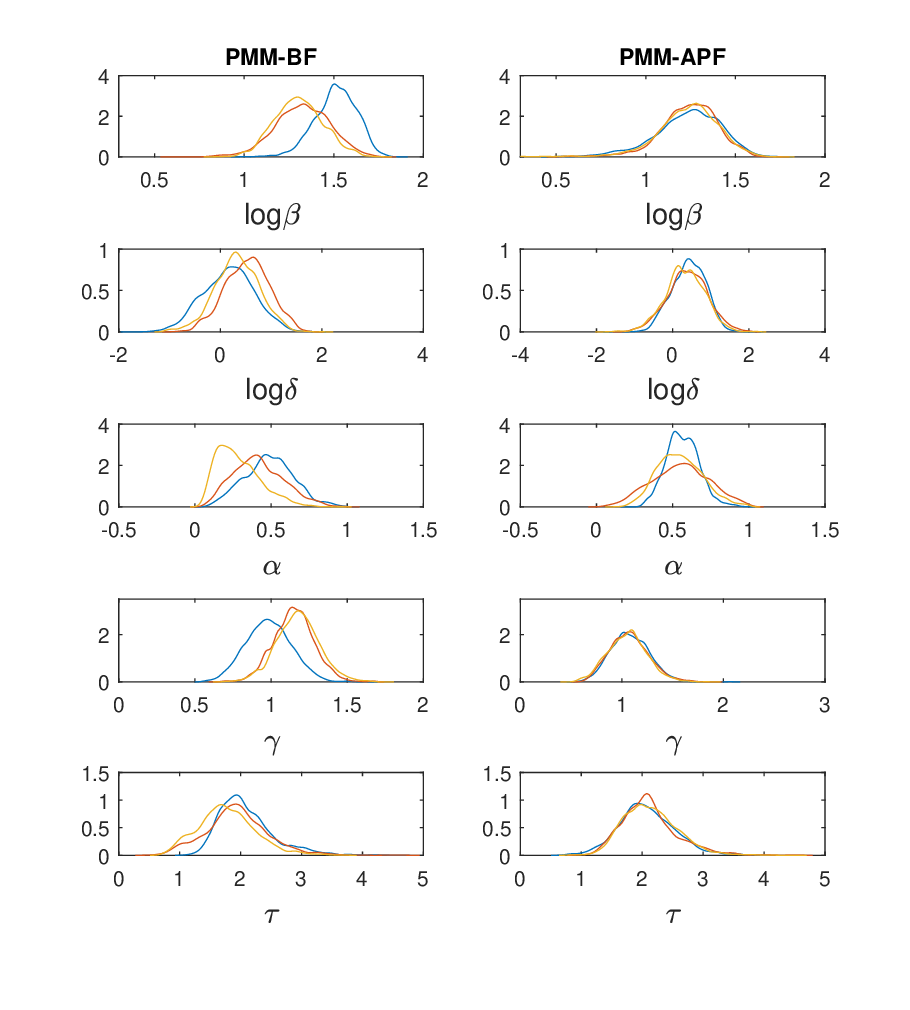}
\caption{\footnotesize{Posterior marginals for each of three chains obtained using PMM-BF (left) and PMM-APF (right). Fitted data are those from group 3.}}
\label{fig:marginals-BFvsAPF}
\end{figure}

In the main text we introduced both the bootstrap filter (BF, algorithm 2) and the auxiliary particle filter (APF, algorithm 4 in the appendix). In section 5.1 we claim that ``the bootstrap filter [...] is known to degenerate when the measurements noise is  very small, as in our case with $\sigma_\varepsilon$ more than an order of magnitude smaller than log-volumes. With a small $\sigma_\varepsilon$ it is difficult for particles propagated blindly to ``hit'' the narrow support of the density function for the next observation, hence the use of the auxiliary particle filter.''
In this section we compare results obtained using two pseudo-marginal methods (PMM): the first PMM employs the BF (PMM-BF) and the second one uses the APF (PMM-APF). Both methods are applied to data from experimental group 3 (i.e. the same data analyzed in section 5.1). In the interest of the comparison, we use the following setup: for both PMM-BF and PMM-APF we run three MCMC chains, where each chain is initialized at a different seed for the pseudo-random numbers generation and at different starting values for the parameters. Of course comparison between PMM-BF and PMM-APF is consistent, i.e. we use the same seeds and parameter starting values for the two methods.
PMM-BF uses $L=2,000$ particles while PMM-APF uses $L=2,000$ and $L_2=5$, just as in section 5.1. The three sets of parameter starting values are in Table \ref{tab:starting-parameters-BFvsAPF}.
\begin{table}[ht]
\caption{\footnotesize{Three sets of parameter starting values used for comparing PMM-BF and PMM-APF.}}
\centering 
\begin{tabular}{lccccccccc}
\hline
{} & $\log\bar{\beta}$ &   $\log\bar{\delta}$ &  $\log\bar{\alpha}$ &   $\log\gamma$  &  $\log\tau$ &  $\log\sigma_{\beta}$ &  $\log\sigma_{\delta}$  & $\log\sigma_{\alpha}$ &   $\log\sigma_{\varepsilon}$\\
set1 & -1    &    -2.20    &   -0.69    &  -3.5    &    -3   &   -2.3      &        -3       &          -1.35       &      -1.39\\
set2 & 1.6   &     1.6     &   -0.36     &   0     &    0    &   -0.7      &        -0.7     &          -2.3        &        0   \\
set3 & 0     &     1     &     -0.1     &    -1     &   -1   &    -1       &        -1       &           -1.5       &        -0.5\\
\hline\end{tabular}
\label{tab:starting-parameters-BFvsAPF}
\end{table}
As an illustration of our several comparisons, Figure \ref{fig:marginals-BFvsAPF} reports the marginal posteriors separately for each of the three chains and (in the interest of space) for the first five parameters only ($\log\bar{\beta}$, $\log\bar{\delta}$,  $\bar{\alpha}$,   $\gamma$,  $\tau$). These are based on the last 10,000 MCMC draws obtained for each chain. It is clear that when the APF is employed within PMM results are more stable across simulations, with the exception of $\bar{\alpha}$ which, once more, seems to be the most difficult parameter to capture for the given data.

\subsection*{Exact Bayesian inference for ODE mixed-effects models}

Here we fit an ordinary differential equations mixed-effects model (ODEMEM) separately to data from groups 1 and 3, using exact Bayesian inference. These data have already been analyzed in section 5 of the main text using SDEMEMs. The ODEMEM is given as equations (1)--(3) in the main text. Same as for the SDEMEM case, we assume a deterministic initial state $v_{i,0}$ for all subjects. Hence, parameters of interest are $\theta=(\bar{\alpha},\bar{\beta},\bar{\delta},\sigma_{\alpha},\sigma_{\beta},\sigma_{\delta},\sigma_\varepsilon)$. The ODEMEM does not involve latent stochastic processes, hence the likelihood function is available in closed-form, as measurements (observed log-volumes) arise as independent random samples from the following model
\begin{equation} 
Y_{ij} \sim \mathcal{N}(\log v_{i,0}+\log((1-\alpha_i)\exp(\beta_i t_{ij})+\alpha_i\exp(-\delta_i t_{ij})), \sigma^2_\varepsilon). \label{eq:gaussian-likelihood}
\end{equation} 
As such, for given observations on all $M$ subjects $y=(y_1,...,y_M)$, the likelihood function for $\theta$ is given by $p(y|\theta)=\prod_{i=1}^M p(y_i|\theta)$, where each $p(y_i)$ is written as $p(y_i|\theta)=\prod_{j=1}^{n_i} p(y_{ij};a_{ij},\sigma^2_\varepsilon)$, with $p(y_{ij};a_{ij},\sigma^2_\varepsilon)$ the Gaussian density function corresponding to \eqref{eq:gaussian-likelihood} and evaluated at $y_{ij}$, with mean $a_{ij}=\log v_{i,0}+\log((1-\alpha_i)\exp(\beta_i t_{ij})+\alpha_i\exp(-\delta_i t_{ij}))$. The corresponding posterior distribution is proportional to $p(y|\theta)\pi(\theta)$ where we use the same priors $\pi(\theta)$ as considered for the SDEMEM case (of course here we do not have priors on $\tau$ and $\gamma$ which are not part of the ODEMEM). This makes the ODEMEM case study simple to fit using reliable off-the-shelf statistical libraries such as \texttt{Stan} \citep{carpenter2017stan}. We used the \texttt{Rstan} interface to \texttt{Stan} and the code is available as supplementary material. We ran 10,000 iterations for 4 chains in parallel. The obtained \texttt{Rhat} equals 1 for each parameter, this diagnosing apparent convergence. The results reported in Table 3 in the main text are obtained from 5,000 post-burnin draws for each chain, hence inference is based on 20,000 draws.

\begin{figure}[ht]
\centering
\includegraphics[width=12cm,height=6cm]{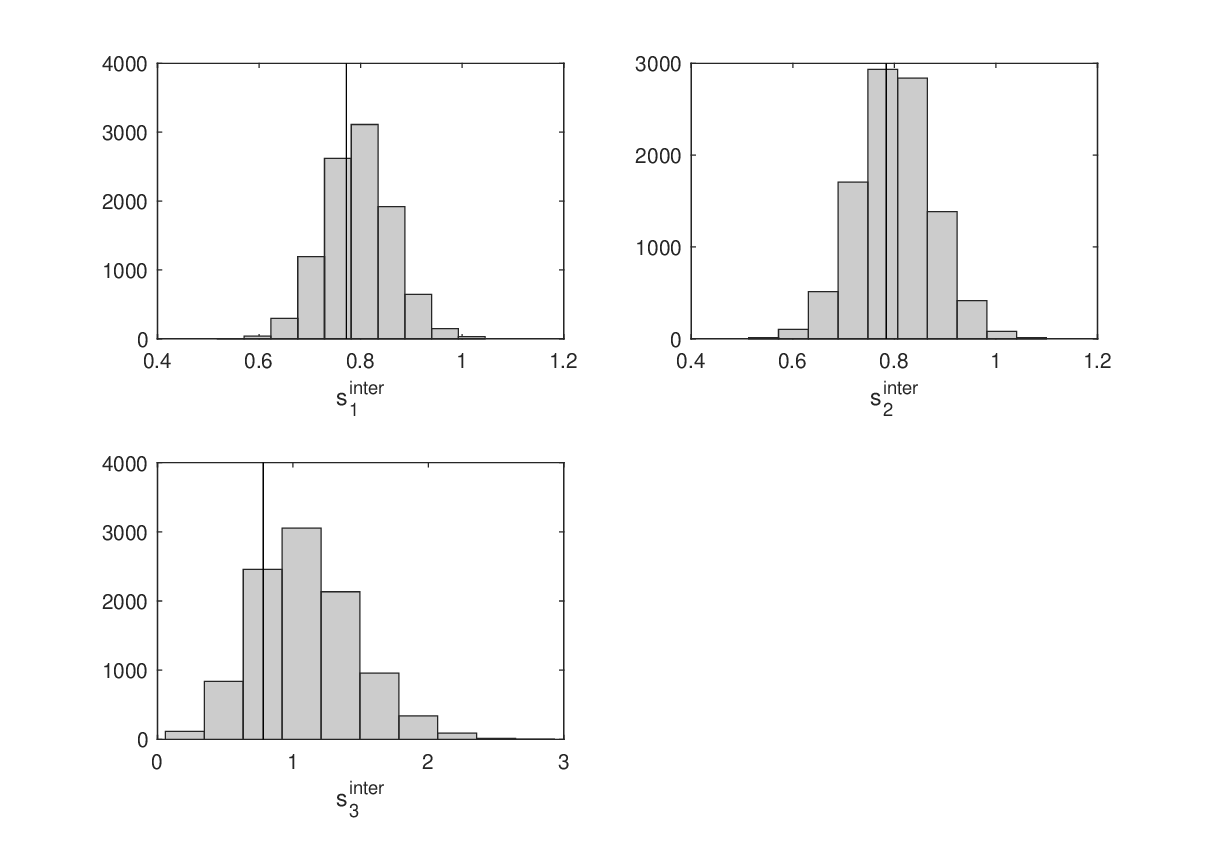}
\caption{\footnotesize{Posterior predictive checks from ODEMEM fitting of group 3. Distribution of the simulated statistics for the inter-subjects variability $s^{\mathrm{inter}}_1$ (top-left), $s^{\mathrm{inter}}_2$ (top-right) and $s^{\mathrm{inter}}_3$ (bottom). Vertical lines mark the values for the corresponding statistics from the observed data.}}
\label{fig:group3-odemem-postpredchecks_S-INTER}
\end{figure}

\begin{figure}[ht]
\centering
\includegraphics[width=10cm,height=7cm]{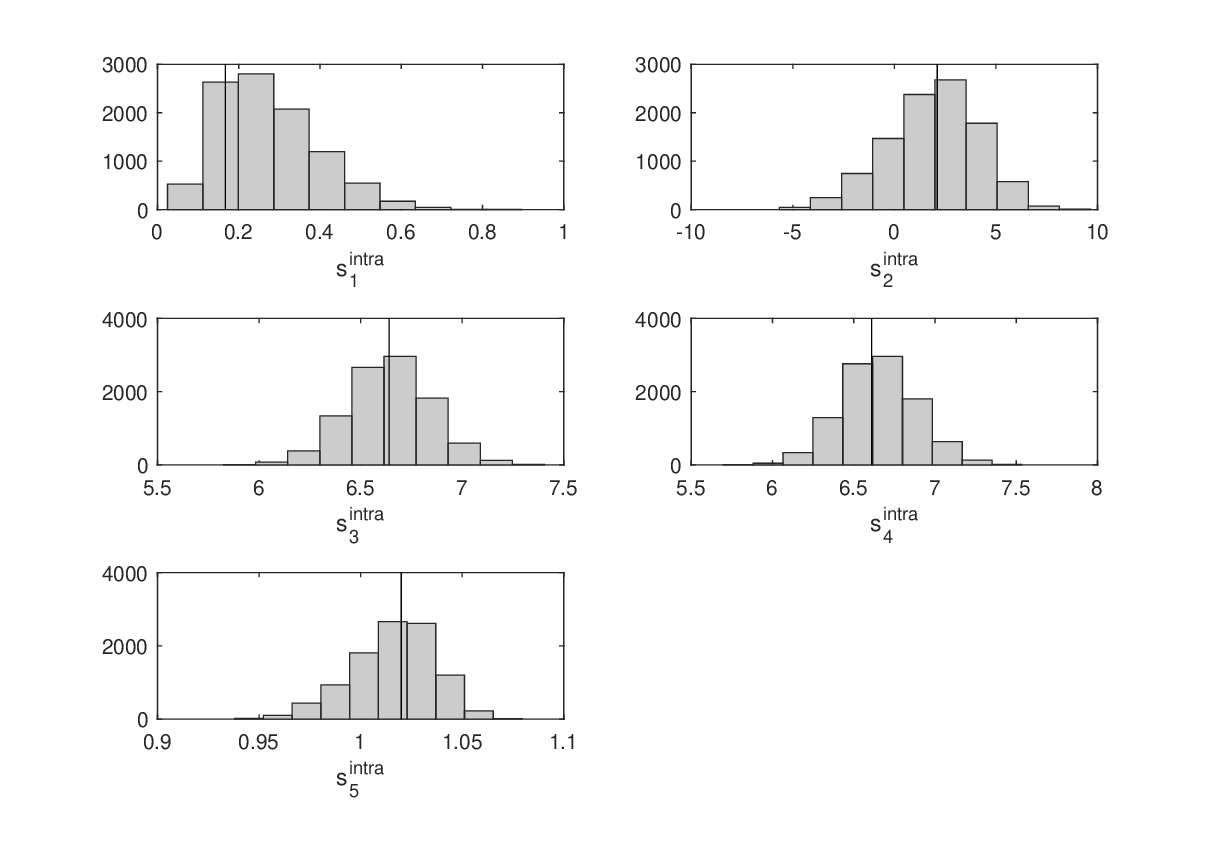}
\caption{\footnotesize{Posterior predictive checks from ODEMEM fitting of group 3. Distribution of the simulated statistics for the intra-subjects variability for subject 1: $s^{\mathrm{intra}}_1$ (top-left), $s^{\mathrm{intra}}_2$ (top-right), $s^{\mathrm{intra}}_3$ (middle-left), $s^{\mathrm{intra}}_4$ (middle-right) and $s^{\mathrm{intra}}_5$ (bottom). Vertical lines mark the values for the corresponding statistics from the observed data.}}
\label{fig:group3-odemem-postpredchecks_S-INTRA_subj1}
\end{figure}

We now report the corresponding posterior predictive checks (PPC). These are simply obtained by plugging the 20,000 posterior draws into the ODEMEM, and used to simulate corresponding 20,000 synthetic datasets. As a visual aid for the comparison with the observed data, we use the summary statistics employed for inference via BSL.
For group 3, Figure \ref{fig:group3-odemem-postpredchecks_S-INTER} gives the PPC for the inter-subjects variability, while Figure \ref{fig:group3-odemem-postpredchecks_S-INTRA_subj1} gives the PPC for the intra-subject variability for subject 1 (as an example). There is a minor discrepancy in the way we compute $s_5^{\mathrm{intra}}$, compared to the one used for BSL inference. In the present case, where we use the R software to compute the PPC, the function \texttt{ar.ols} returning coefficients for autoregression of order one is documented to have issues, when an intercept term is considered in the model and the regression is fitted without taking differences from the mean of the data (i.e. \texttt{demean} is set to \texttt{FALSE}). We indeed experienced computational issues and turned to estimating the coefficient $\beta_1$ of an autoregression without intercept term. Hence here $s_5^{\mathrm{intra}}$ is given by $\hat{\beta}_1$ from a model fitted without intercept. PPC seems to show that the ODEMEM using exact inference is performing satisfactorily. In fact it performs better (in terms of PPC) than the SDEMEM using BSL, however notice that we did not obtain PPC for the SDEMEM estimated using PMM. Compare with Figures 6--7 in the main text showing PPC for SDEMEMs obtained  using BSL, and we can tell that while the inter-subjects checks are similar to the ODEMEM case, instead $s_3^{intra}$, $s_4^{intra}$ and $s_5^{intra}$ are much more precise for the ODEMEM. However we still do not know which model is best, since posterior inference for ODEMEMs parameters is much more variable than for SDEMEMs, and estimated residual variability $\sigma_\varepsilon$ is 3-4 times larger for ODEMEMs.

\subsection*{BSL inference for ODE mixed-effects models}

Here we conduct inference for the ODEMEM fitted to group 3 using BSL (see the previous section for details on the model). With this model we are required to use a much larger number of model simulations for each MCMC iteration, in order to reach stationarity and good mixing for the chains. Interestingly, it seems that inference via the introduction of summary statistics is here more challenging than in the SDEMEMs case, as if by removing systemic noise (and corresponding stochastic intensities $\gamma$ and $\tau$), the resulting synthetic likelihood is more variable than before. In fact using the usual $N=3,000$ simulated datasets produces very nonstationary and badly mixing chains. By using $N=20,000$ we solve this issues, but not the quality of the final inference. In fact, results below show that the use of inference based on summary statistics is unable to capture most notably $\sigma_\delta$. We have the following posterior means and 95\% posterior intervals for subjects in group 3: $\bar{\beta}= 3.14$ [2.33,4.10],   $\bar{\delta}= 2.06$ [0.62,5.55],    $\bar{\alpha}=0.65$ [0.27,0.94],    $\sigma_\beta=0.58$ [0.27,1.23],    $\sigma_\delta=4.60$ [2.95,6.93],    $\sigma_\alpha=0.51$ [0.22,1.03],    $\sigma_\varepsilon=0.23$ [0.13,0.31]. These should be compared with the second column in Table 3 in the main text. Since the only difference between the fitting of the SDEMEM and the fitting of the ODEMEM is that in the latter we have $\gamma=\tau=0$, we can only deduce that  approximate inference via BSL is facilitated by the richer stochastic model.

%% file: main.bbl
\begin{thebibliography}{46}
\providecommand{\natexlab}[1]{#1}
\providecommand{\url}[1]{\texttt{#1}}
\expandafter\ifx\csname urlstyle\endcsname\relax
  \providecommand{\doi}[1]{doi: #1}\else
  \providecommand{\doi}{doi: \begingroup \urlstyle{rm}\Url}\fi

\bibitem[An et~al.(2018)An, Nott, and Drovandi]{an2018robust}
Z.~An, D.~J. Nott, and C.~Drovandi.
\newblock Robust {B}ayesian synthetic likelihood via a semi-parametric
  approach.
\newblock \emph{arXiv preprint arXiv:1809.05800}, 2018.

\bibitem[Andrieu and Roberts(2009)]{andrieu2009pseudo}
C.~Andrieu and G.~Roberts.
\newblock The pseudo-marginal approach for efficient {M}onte {C}arlo
  computations.
\newblock \emph{The Annals of Statistics}, pages 697--725, 2009.

\bibitem[Beaumont(2003)]{beaumont2003estimation}
M.~Beaumont.
\newblock Estimation of population growth or decline in genetically monitored
  populations.
\newblock \emph{Genetics}, 164\penalty0 (3):\penalty0 1139--1160, 2003.

\bibitem[Betancourt(2017)]{betancourt2017conceptual}
M.~Betancourt.
\newblock A conceptual introduction to {H}amiltonian {M}onte {C}arlo.
\newblock \emph{arXiv preprint arXiv:1701.02434}, 2017.

\bibitem[Blum et~al.(2013)Blum, Nunes, Prangle, Sisson,
  et~al.]{blum2013comparative}
M.~G. Blum, M.~A. Nunes, D.~Prangle, S.~A. Sisson, et~al.
\newblock A comparative review of dimension reduction methods in approximate
  {B}ayesian computation.
\newblock \emph{Statistical Science}, 28\penalty0 (2):\penalty0 189--208, 2013.

\bibitem[Capp{\'e} et~al.(2006)Capp{\'e}, Moulines, and
  Ryd{\'e}n]{cappe2006inference}
O.~Capp{\'e}, E.~Moulines, and T.~Ryd{\'e}n.
\newblock \emph{Inference in hidden {M}arkov models}.
\newblock Springer Science \& Business Media, 2006.

\bibitem[Carpenter et~al.(2017)Carpenter, Gelman, Hoffman, Lee, Goodrich,
  Betancourt, Brubaker, Guo, Li, and Riddell]{carpenter2017stan}
B.~Carpenter, A.~Gelman, M.~D. Hoffman, D.~Lee, B.~Goodrich, M.~Betancourt,
  M.~Brubaker, J.~Guo, P.~Li, and A.~Riddell.
\newblock Stan: A probabilistic programming language.
\newblock \emph{Journal of Statistical Software}, 76\penalty0 (1), 2017.

\bibitem[Del~Moral(2004)]{del2004genealogical}
P.~Del~Moral.
\newblock \emph{Feynman-Kac formulae: genealogical and interacting particle
  systems with applications}.
\newblock New York: Springer, 2004.

\bibitem[Delattre and Lavielle(2013)]{delattre2013coupling}
M.~Delattre and M.~Lavielle.
\newblock Coupling the {SAEM} algorithm and the extended {K}alman filter for
  maximum likelihood estimation in mixed-effects diffusion models.
\newblock \emph{Statistics and its interface}, 6\penalty0 (4):\penalty0
  519--532, 2013.

\bibitem[Demidenko(2006)]{demidenko2006assessment}
E.~Demidenko.
\newblock The assessment of tumour response to treatment.
\newblock \emph{Journal of the Royal Statistical Society: Series C (Applied
  Statistics)}, 55\penalty0 (3):\penalty0 365--377, 2006.

\bibitem[Demidenko(2010)]{demidenko2010three}
E.~Demidenko.
\newblock Three endpoints of in vivo tumour radiobiology and their statistical
  estimation.
\newblock \emph{International journal of radiation biology}, 86\penalty0
  (2):\penalty0 164--173, 2010.

\bibitem[Demidenko(2013)]{demidenko2013mixed}
E.~Demidenko.
\newblock \emph{Mixed models: theory and applications with R}.
\newblock John Wiley \& Sons, 2013.

\bibitem[Donnet and Samson(2013)]{donnet2013review}
S.~Donnet and A.~Samson.
\newblock A review on estimation of stochastic differential equations for
  pharmacokinetic/pharmacodynamic models.
\newblock \emph{Advanced Drug Delivery Reviews}, 65\penalty0 (7):\penalty0
  929--939, 2013.

\bibitem[Donnet and Samson(2014)]{donnet2014using}
S.~Donnet and A.~Samson.
\newblock Using {PMCMC} in {EM} algorithm for stochastic mixed models:
  theoretical and practical issues.
\newblock \emph{Journal de la Soci{\'e}t{\'e} Fran{\c{c}}aise de Statistique},
  155\penalty0 (1):\penalty0 49--72, 2014.

\bibitem[Donnet et~al.(2010)Donnet, Foulley, and Samson]{donnet2010bayesian}
S.~Donnet, J.~Foulley, and A.~Samson.
\newblock Bayesian analysis of growth curves using mixed models defined by
  stochastic differential equations.
\newblock \emph{Biometrics}, 66\penalty0 (3):\penalty0 733--741, 2010.

\bibitem[Doucet et~al.(2015)Doucet, Pitt, Deligiannidis, and
  Kohn]{doucet2015efficient}
A.~Doucet, M.~Pitt, G.~Deligiannidis, and R.~Kohn.
\newblock Efficient implementation of {M}arkov chain {M}onte {C}arlo when using
  an unbiased likelihood estimator.
\newblock \emph{Biometrika}, 2015.
\newblock \doi{doi: 10.1093/biomet/asu075}.

\bibitem[Fasiolo et~al.(2018)Fasiolo, Wood, Hartig, and
  Bravington]{fasiolo2016extended}
M.~Fasiolo, S.~Wood, F.~Hartig, and M.~Bravington.
\newblock An extended empirical saddlepoint approximation for intractable
  likelihoods.
\newblock \emph{Electronic Journal of Statistics}, 12\penalty0 (1):\penalty0
  1544--1578, 2018.

\bibitem[Fuchs(2013)]{fuchs2013inference}
C.~Fuchs.
\newblock \emph{Inference for Diffusion Processes: With Applications in Life
  Sciences}.
\newblock Springer Science \& Business Media, 2013.

\bibitem[Gelman and Rubin(1992)]{gelman1992inference}
A.~Gelman and D.~Rubin.
\newblock Inference from iterative simulation using multiple sequences.
\newblock \emph{Statistical Science}, pages 457--472, 1992.

\bibitem[Gelman et~al.(2013)Gelman, Stern, Carlin, Dunson, Vehtari, and
  Rubin]{gelman2013bayesian}
A.~Gelman, H.~S. Stern, J.~B. Carlin, D.~B. Dunson, A.~Vehtari, and D.~B.
  Rubin.
\newblock \emph{Bayesian Data Analysis}.
\newblock Chapman and Hall/CRC, third edition, 2013.

\bibitem[Ghurye and Olkin(1969)]{ghurye1969unbiased}
S.~Ghurye and I.~Olkin.
\newblock Unbiased estimation of some multivariate probability densities and
  related functions.
\newblock \emph{The Annals of Mathematical Statistics}, pages 1261--1271, 1969.

\bibitem[Golightly and Wilkinson(2011)]{golightly2011bayesian}
A.~Golightly and D.~Wilkinson.
\newblock Bayesian parameter inference for stochastic biochemical network
  models using particle {M}arkov chain {M}onte {C}arlo.
\newblock \emph{Interface Focus}, 1\penalty0 (6):\penalty0 807--820, 2011.

\bibitem[Gordon et~al.(1993)Gordon, Salmond, and Smith]{gordon1993novel}
N.~Gordon, D.~Salmond, and A.~Smith.
\newblock Novel approach to nonlinear/non-{G}aussian {B}ayesian state
  estimation.
\newblock In \emph{Radar and Signal Processing, IEE Proceedings F}, volume 140,
  pages 107--113, 1993.

\bibitem[Haario et~al.(2001)Haario, Saksman, and Tamminen]{haario2001adaptive}
H.~Haario, E.~Saksman, and J.~Tamminen.
\newblock An adaptive {M}etropolis algorithm.
\newblock \emph{Bernoulli}, pages 223--242, 2001.

\bibitem[Heitjan et~al.(1993)Heitjan, Manni, and
  Santen]{heitjan1993statistical}
D.~Heitjan, A.~Manni, and R.~Santen.
\newblock Statistical analysis of in vivo tumor growth experiments.
\newblock \emph{Cancer Research}, 53\penalty0 (24):\penalty0 6042--6050, 1993.

\bibitem[Kitagawa(1996)]{kitagawa1996monte}
G.~Kitagawa.
\newblock Monte {C}arlo filter and smoother for non-{G}aussian nonlinear state
  space models.
\newblock \emph{Journal of computational and graphical statistics}, 5\penalty0
  (1):\penalty0 1--25, 1996.

\bibitem[Kong and Yan(2011)]{kong2011modeling}
M.~Kong and J.~Yan.
\newblock Modeling and testing treated tumor growth using cubic smoothing
  splines.
\newblock \emph{Biometrical Journal}, 53\penalty0 (4):\penalty0 595--613, 2011.

\bibitem[Laajala et~al.(2012)Laajala, Corander, Saarinen, M{\"a}kel{\"a},
  Savolainen, Suominen, Alhoniemi, M{\"a}kel{\"a}, Poutanen, and
  Aittokallio]{laajala2012improved}
T.~Laajala, J.~Corander, N.~Saarinen, K.~M{\"a}kel{\"a}, S.~Savolainen,
  M.~Suominen, E.~Alhoniemi, S.~M{\"a}kel{\"a}, M.~Poutanen, and
  T.~Aittokallio.
\newblock Improved statistical modeling of tumor growth and treatment effect in
  preclinical animal studies with highly heterogeneous responses in vivo.
\newblock \emph{Clinical Cancer Research}, 18\penalty0 (16):\penalty0
  4385--4396, 2012.

\bibitem[Marin et~al.(2012)Marin, Pudlo, Robert, and
  Ryder]{marin2012approximate}
J.~Marin, P.~Pudlo, C.~Robert, and R.~Ryder.
\newblock Approximate {B}ayesian computational methods.
\newblock \emph{Statistics and Computing}, 22\penalty0 (6):\penalty0
  1167--1180, 2012.

\bibitem[McFadden(1989)]{mcfadden1989smm}
D.~McFadden.
\newblock A method of simulated moments for estimation of discrete response
  models without numerical integration.
\newblock \emph{Econometrica}, 57\penalty0 (5):\penalty0 995--1026, 1989.

\bibitem[P{\'e}ron et~al.(2016)P{\'e}ron, Buyse, Ozenne, Roche, and
  Roy]{peron2016}
J.~P{\'e}ron, M.~Buyse, B.~Ozenne, L.~Roche, and P.~Roy.
\newblock An extension of generalized pairwise comparisons for prioritized
  outcomes in the presence of censoring.
\newblock \emph{Statistical methods in medical research}, 2016.
\newblock \doi{10.1177/0962280216658320}.

\bibitem[Picchini(2018)]{picchini2016likelihood}
U.~Picchini.
\newblock Likelihood-free stochastic approximation {EM} for inference in
  complex models.
\newblock \emph{Communications in Statistics-Simulation and Computation}, 2018.
\newblock \doi{10.1080/03610918.2017.1401082}.

\bibitem[Pitt(2002)]{pitt2002smooth}
M.~Pitt.
\newblock Smooth particle filters for likelihood evaluation and maximisation.
\newblock Technical Report 651, 2002.

\bibitem[Pitt and Shephard(1999)]{pitt1999filtering}
M.~Pitt and N.~Shephard.
\newblock Filtering via simulation: Auxiliary particle filters.
\newblock \emph{Journal of the American statistical association}, 94\penalty0
  (446):\penalty0 590--599, 1999.

\bibitem[Pitt et~al.(2012)Pitt, dos Santos~Silva, Giordani, and
  Kohn]{pitt2012some}
M.~Pitt, R.~dos Santos~Silva, P.~Giordani, and R.~Kohn.
\newblock On some properties of {M}arkov chain {M}onte {C}arlo simulation
  methods based on the particle filter.
\newblock \emph{Journal of Econometrics}, 171\penalty0 (2):\penalty0 134--151,
  2012.

\bibitem[Plummer et~al.(2006)Plummer, Best, Cowles, and Vines]{coda}
M.~Plummer, N.~Best, K.~Cowles, and K.~Vines.
\newblock {CODA}: Convergence diagnosis and output analysis for {MCMC}.
\newblock \emph{R News}, 6\penalty0 (1):\penalty0 7--11, 2006.

\bibitem[Prangle(2015)]{prangle2015summary}
D.~Prangle.
\newblock Summary statistics in approximate {B}ayesian computation.
\newblock \emph{arXiv:1512.05633}, 2015.

\bibitem[Price et~al.(2017)Price, Drovandi, Lee, and Nott]{price2016bayesian}
L.~Price, C.~Drovandi, A.~Lee, and D.~Nott.
\newblock Bayesian synthetic likelihood.
\newblock \emph{Journal of Computational and Graphical Statistics}, 2017.
\newblock \doi{10.1080/10618600.2017.1302882}.

\bibitem[Sherlock et~al.(2015)Sherlock, Thiery, Roberts, and
  Rosenthal]{sherlock2015efficiency}
C.~Sherlock, A.~Thiery, G.~Roberts, and J.~Rosenthal.
\newblock On the efficiency of pseudo-marginal random walk {M}etropolis
  algorithms.
\newblock \emph{The Annals of Statistics}, 43\penalty0 (1):\penalty0 238--275,
  2015.

\bibitem[Stuschke et~al.(1990)Stuschke, Budach, Bamberg, and
  Budach]{stuschke1990methods}
M.~Stuschke, V.~Budach, M.~Bamberg, and W.~Budach.
\newblock Methods for analysis of censored tumor growth delay data.
\newblock \emph{Radiation research}, 122\penalty0 (2):\penalty0 172--180, 1990.

\bibitem[Whitaker et~al.(2017)Whitaker, Golightly, Boys, and
  Sherlock]{whitaker2015bayesian}
G.~Whitaker, A.~Golightly, R.~Boys, and C.~Sherlock.
\newblock Bayesian inference for diffusion driven mixed-effects models.
\newblock \emph{Bayesian Analysis}, 12, 2017.

\bibitem[Wood(2010)]{wood2010statistical}
S.~Wood.
\newblock Statistical inference for noisy nonlinear ecological dynamic systems.
\newblock \emph{Nature}, 466\penalty0 (7310):\penalty0 1102--1104, 2010.

\bibitem[Wu(2011)]{wu2011confidence}
J.~Wu.
\newblock Confidence intervals for the difference of median failure times
  applied to censored tumor growth delay data.
\newblock \emph{Statistics in Biopharmaceutical Research}, 3\penalty0
  (3):\penalty0 488--496, 2011.

\bibitem[Wu and Houghton(2009)]{wu2009assessing}
J.~Wu and P.~Houghton.
\newblock Assessing cytotoxic treatment effects in preclinical tumor xenograft
  models.
\newblock \emph{Journal of biopharmaceutical statistics}, 19\penalty0
  (5):\penalty0 755--762, 2009.

\bibitem[Xia et~al.(2013)Xia, Wu, and Liang]{xia2013model}
C.~Xia, J.~Wu, and H.~Liang.
\newblock Model tumor pattern and compare treatment effects using
  semiparametric linear mixed-effects models.
\newblock \emph{Journal of Biometrics \& Biostatistics}, 2013, 2013.

\bibitem[Zhao et~al.(2011)Zhao, Morgan, Parsels, Maybaum, Lawrence, and
  Normolle]{zhao2011bayesian}
L.~Zhao, M.~Morgan, L.~Parsels, J.~Maybaum, T.~Lawrence, and D.~Normolle.
\newblock Bayesian hierarchical changepoint methods in modeling the tumor
  growth profiles in xenograft experiments.
\newblock \emph{Clinical Cancer Research}, 17\penalty0 (5):\penalty0
  1057--1064, 2011.

\end{thebibliography}
